\documentclass[preprint,authoryear,11pt]{elsarticle}
\usepackage{dcolumn}
\usepackage{graphicx}
\usepackage{color}
\usepackage{bm}
\usepackage{amsmath}
\usepackage{hyperref}

\hypersetup{colorlinks,
  linkcolor=blue,
  filecolor=blue,
  urlcolor=blue,
  citecolor=blue,
  breaklinks=true}

\usepackage{breakcites}

\journal{Physics Reports}
\topmargin = -\topskip
\advance\topmargin by -\headsep
\begin{document}
\begin{frontmatter}

\title{The Intrinsic Alignment of Galaxies and its Impact on Weak Gravitational Lensing in an Era of Precision Cosmology}
\author{M. A. Troxel$^{a,b}$}
\ead{michael.troxel@manchester.ac.uk}
\author{Mustapha Ishak$^{a}$}
\ead{mishak@utdallas.edu}
\address{$^{a}$Department of Physics, The University of Texas at Dallas, Richardson, TX 75080, USA}
\address{$^{a,b}$Jodrell Bank Centre for Astrophysics, University of Manchester, Manchester M13 9PL, UK}
%
%
\begin{abstract}
The wealth of incoming and future cosmological observations will allow us to map out the structure and evolution of the observable universe to an unprecedented level of precision. Among these observations is the weak gravitational lensing of galaxies, e.g., cosmic shear that measures the minute distortions of background galaxy images by intervening cosmic structure. Weak lensing and cosmic shear promise to be a powerful probe of astrophysics and cosmology, constraining models of dark energy, measuring the evolution of structure in the universe, and testing theories of gravity on cosmic scales. However, the intrinsic alignment of galaxies -- their shape and orientation before being lensed -- may pose a great challenge to the use of weak gravitational lensing as an accurate cosmological probe, and has been identified as one of the primary physical systematic biases in cosmic shear studies. Correlations between this intrinsic alignment and the lensing signal can persist even for large physical separations, and isolating the effect of intrinsic alignment from weak lensing is not trivial. A great deal of work in the last two decades has been devoted to understanding and characterizing this intrinsic alignment, which is also a direct and complementary probe of structure formation and evolution in its own right. In this review, we report in a systematic way the state of our understanding of the intrinsic alignment of galaxies, with a particular emphasis on its large-scale impact on weak lensing measurements and methods for its isolation or mitigation. We begin with an introduction to the use of cosmic shear as a probe for cosmology and describe the various physical contributions by intrinsic alignment to the shear or convergence 2- and 3-point correlations. We then review developments in the modeling of the intrinsic alignment signal, including a trend toward attempting to incorporate more accurate nonlinear and single halo effects. The impact on cosmological constraints by the intrinsic alignment of galaxies is also outlined based on these models. We then summarize direct measurements of the large-scale intrinsic alignment signal in various surveys and discuss their constraints on models of intrinsic alignment, as well as progress in utilizing numerical simulations of structure formation to further our understanding of intrinsic alignment. Finally, we outline the development of a variety of mitigation techniques for reducing the impact of the intrinsic alignment contamination on weak lensing signals both within a galaxy data set and between complementary probes of gravitational lensing. The methodology and projected impact of these techniques are discussed for both 2- and 3-point correlations. We conclude by presenting a summary and outlook on the state of intrinsic alignment study and its impact on ongoing and planned weak lensing surveys.
\end{abstract} 


\end{frontmatter}


\tableofcontents

\newpage
\section{Introduction}\label{intro}

One of the most promising probes of the universe is weak gravitational lensing and in particular that due to large-scale structure (cosmic shear). This effect is measured in galaxy surveys as very weak distortions of the intrinsic shapes of galaxies. Weak lensing is an ideal probe to map distributions of dark matter in the universe in the form of large-scale cosmic structure. It is also an excellent probe of the nature of dark energy, as it can trace a large volume of cosmic space. It is sensitive to both the growth rate of large-scale structure and the expansion history of the Universe, and can lead to significant constraints on cosmological parameters such as the matter density, the matter fluctuation amplitude, and the dark energy equation of state. Similarly, gravitational lensing is a powerful probe for testing the nature of gravity on cosmic scales. The weak lensing signal is even more powerful when combined with the galaxy density-shear cross-correlation and galaxy density-density autocorrelation. For a detailed review of weak gravitational lensing and its applications, we refer the reader to the previous review articles by \cite{SEF,Mellier1999,BSWLReview2001,Wittman2002,Refregier2003,VanwaerbekeMellier2003,Schneider2006,HoekstraJain2008,MunshiEtAl2008,Heavens2009,Bartelmann2010,Huterer2010,MasseyEtAl2010,WeinbergEtAl2013} and references therein. Further detail is also provided in Sec.~\ref{formalisms} below.

Weak lensing or cosmic shear offers multiple statistical measures for cosmological analyses. In addition to the 2-point correlation statistic (i.e., the power spectrum), the inclusion of the shear 3-point correlation statistic (i.e., the bispectrum) is particularly important for upcoming surveys, which will have the statistical power to successfully measure the 3-point correlations at high significance. Indeed, when combined with the power spectrum, the bispectrum has been shown to probe additional physics like primordial non-Gaussianity and to break degeneracies in parameter constraints that are present for the power spectrum alone and thus provides further significant improvements on parameter constraints (e.g., the review by \cite{MunshiEtAl2008} and references therein). The promise of weak lensing as a cosmological probe has been identified by the scientific community (see for example, \cite{DETFReport2006}), and has already provided complementary cosmological constraints (see specific results as discussed in Sec.~\ref{cosmicweak}). These have driven the development of much larger and more precise galaxy weak lensing surveys (e.g., the Dark Energy Survey\footnote{http://www.darkenergysurvey.org/} (DES), the Euclid mission\footnote{http://www.euclid-ec.org/}, the Hyper Suprime-Cam\footnote{http://www.naoj.org/Projects/HSC/} (HSC), the Kilo-Degree Survey\footnote{http://kids.strw.leidenuniv.nl/} (KIDS), the Large Synoptic Sky Telescope\footnote{http://www.lsst.org/lsst/} (LSST), the Square Kilometer Array\footnote{https://www.skatelescope.org/} (SKA), and the Wide-Field Infrared Survey Telescope\footnote{http://wfirst.gsfc.nasa.gov/} (WFIRST)), which will produce unprecedented weak lensing measurements in the coming decades.

However, weak lensing measurements in galaxy surveys are limited in precision by several systematic effects which must be accounted for in order to make full use of the potential of ongoing and planned weak lensing surveys (see for example, the reviews listed above and Sec.~\ref{systematics} below). These systematic effects include challenges to measuring the shape of galaxies, which are smeared or distorted due to atmospheric, camera, or reduction effects, calibration biases, difficulties in accurately determining the redshifts of such large ensembles of objects, and fundamental limits to our current understanding of the matter power spectrum and its nonlinear evolution (e.g., due to the effects of baryons) for both standard and nonstandard cosmologies. It is extremely important to understand and control these and other systematic effects of weak lensing in order to fully explore its potential as a precision cosmological probe.

One of the most serious physical systematic effects of weak lensing is the presence of the correlated intrinsic alignment of galaxies that contaminate shear correlations, and these intrinsic alignments or shapes of galaxies are the subject of this review. The intrinsic alignment of galaxies is due to a variety of physical processes including the structure formation scenarios and primordial potentials in which the galaxies formed, evolution of the galaxies and nearby structures, and particularly at late times, baryonic physics, galaxy mergers, and accretion. These correlated alignments can initially be driven by stretching or compression of initially spherically collapsing mass distributions in some gravitational gradient (e.g., \cite{CatelanKamionkowskiBlandford2001}) or by the mutual acquisition of angular momentum through tidal torquing \citep{Sciama1955,Peebles1969,Doroshkevich1970,White1984} of aspherical protogalactic mass distributions during galaxy formation. This galaxy ellipticity or angular momentum alignment and the potential for finding correlations in the alignments of galaxies has been extensively studied; see for example \cite{Djorgovski1987} and references therein for an early review of the topic. Such searches have been largely inconclusive until recent years and originally focused on correlated alignments in high-density regions as probes of structure formation. In the past decade or so, the focus has instead shifted to searching for large-scale correlations in galaxy alignments as contaminants to the weak gravitational lensing signal, and to developing methods to isolate its impact from that of weak lensing. While early work only considered such correlations between the intrinsic alignment of galaxies (labeled the $II$ correlation), \cite{HirataSeljak2004} later identified a cross-correlation between the intrinsic alignment and the lensing signal (labeled the $GI$ correlation), which has turned out to be the stronger and more problematic signal to mitigate in large weak lensing surveys that probe to high redshift. These correlations are defined in Sec.~\ref{backIA} and a more detailed discussion of the search for correlations in the intrinsic alignment in galaxies can be found in Sec.~\ref{detections}.

The intrinsic alignment of galaxies acts as a nuisance signal to the cosmic shear correlation and can strongly bias its constraints on cosmological parameters. It has been shown, for example, that constraints of the equation of state of dark energy can be biased by 50\% or more when intrinsic alignment is ignored. Similarly, the amplitude of matter fluctuations can be biased by up to 30\%. For more on these impacts and specific references, we refer the reader to Sec.~\ref{impacts}. The mitigation of these biases due to the intrinsic alignment of galaxies and development of methods to isolate or measure the intrinsic alignment signal are thus essential for future precision cosmological measurements from weak lensing surveys. Indeed, the science goals of these surveys are dependent upon an effective approach for mitigating the impact of intrinsic alignment on cosmological constraints. This is not a trivial task, but it is a manageable one toward which significant progress has been made over the last decade (see Sec.~\ref{mitigation}).

It is worth emphasizing, though, that isolating the intrinsic alignment signal has a double advantage. The first is to clean the lensing signal from this systematic effect toward its use as a precise -- and more importantly accurate -- cosmological tool. The second is that the intrinsic alignment signal itself, once isolated, provides valuable information that reflects the formation and evolution of galaxies in their respective environments, and which could help in understanding the structure formation scenarios that generated them. This intrinsic alignment signal also provides complementary cosmological information to the weak lensing signal. 

In this review, we aim to provide an overview of the intrinsic alignment of galaxies focusing on their large-scale aspects and their contamination to weak gravitational lensing. We review progress toward the understanding, measurement, and mitigation of the intrinsic alignment of galaxies as it impacts precision weak lensing science and cosmology.

The review is organized as follows. We first provide a general discussion of cosmic shear as a cosmological probe and the necessary formalisms in Sec.~\ref{formalisms}. The last two decades have seen the development of a basic understanding of the intrinsic alignment of galaxies as a contaminant to the precise shear measurement goals of planned surveys, which we will review in Sec.~\ref{backIA}. A large amount of work has already been devoted to identifying and measuring the effects of the large-scale correlated intrinsic alignment signal in various weak lensing surveys to date, and the methodologies and results of this work are discussed in Sec.~\ref{detections}. We also report progress on measuring and constraining the effects of large-scale correlations of intrinsic alignment in numerical simulations in Sec.~\ref{sims}. Finally, a variety of methodologies have been developed to help address mitigating and isolating the intrinsic alignment correlations in weak lensing surveys, and these are presented in Sec.~\ref{mitigation}. We conclude with a summary and future outlook in Sec.~\ref{summary}.

              \section{Gravitational lensing and cosmology}\label{formalisms}

               \subsection{The standard cosmological model}\label{lcdm} 

The standard model of cosmology is based on the theory of general relativity, where dynamics in the universe are described by Einstein's field equations with a cosmological constant $\Lambda$ (for brevity, we will assume units such that $c=1$ throughout),
\begin{equation}
G_{\mu\nu}+\Lambda g_{\mu\nu}=8\pi G T_{\mu\nu}.\label{eq:efe}
\end{equation}
where $G_{\mu\nu}\equiv R_{\mu\nu}-\frac{1}{2}g_{\mu\nu}R$ is the Einstein tensor representing the curvature of spacetime, $R_{\mu\nu}$ is the Ricci tensor, and $R$ the Ricci scalar. The matter content is represented by the energy momentum tensor of a perfect fluid given by  
\begin{equation}
T_{\mu\nu}=(\rho+p)u_{\mu}u_{\nu}-pg_{\mu\nu},\label{eq:stress}
\end{equation}
where $\rho$ is the mass-energy density, $p$ is the isotropic pressure, $u^{\mu}$ is the tangent velocity vector of the cosmic fluid, and $g_{\mu\nu}$ is the metric, which describes the geometry of the spacetime.

On very large scales, it is assumed that the universe can be described by a metric that is globally isotropic and thus homogeneous. Its geometry is represented by the metric of Friedmann-Lemaitre-Robertson-Walker (FLRW) that can be written from the line element
\begin{equation}
ds^2=-dt^2+a^2(t)\left(\frac{dr^2}{1-kr^2}+r^2(d\theta^2+\sin^2\theta d\phi^2)\right).\label{eq:flrw}
\end{equation}
The scale factor $a(t)$ represents the time-dependent evolution of the spatial part of the metric (surfaces of constant $t$), and $k\in\{-1,0,+1\}$ determines the geometry of these spatial sections: negatively curved, flat, or positively curved, respectively.

Equation (\ref{eq:flrw}) for the FLRW metric and the energy-momentum tensor of Eq.~(\ref{eq:stress}) give the Friedmann equation from their time-time components
\begin{equation}
\frac{\dot{a}^2}{a^2}=H(t)^2=\frac{8\pi G}{3}\rho +\frac{\Lambda}{3}-\frac{k}{a^2},
\label{eq:FriedmannEq1}
\end{equation}
where $\dot{a}$ denotes a derivative of $a$ with respect to the time coordinate, and we have defined the Hubble parameter $H(t)^2\equiv \big{(}\frac{\dot{a}(t)}{a(t)}\big{)}^2$. From the combination of the space-space component and the time-time component, one can write an acceleration/deceleration equation (or second Friedmann equation) 
\begin{equation}
\frac{\ddot{a}}{a}\, = \frac{4\pi G}{3}\left(\rho\, +\, 3p \right)\, +\, \frac{\Lambda}{3}.
\label{eq:FriedmannEq2}
\end{equation}
The current day ($t=t_0$) Hubble constant is denoted $H_0=H(t_0)$, and we normalize the expansion such that $a_0=a(t_0)\equiv 1$. The redshift is then related to $a$ by $a=1/(1+z)$.

The Friedmann equations represent the global, homogeneous evolution of the universe and serve as a basis for distance measurements such as the angular diameter distance given as a function of redshift $z$ by 
\begin{equation}
D_A(z)=\frac{\sin_k(\chi)}{1+z},
\end{equation}
where 
\begin{equation}
\sin_k(\chi)=\left\{ \begin{array}{lcl}
k^{-1/2}\sin (k^{1/2}\chi) & & k>0 \\
\chi & & k=0 \\
|k|^{-1/2}\sinh (|k|^{1/2}\chi) & & k<0 \end{array} \right.,
\end{equation}
and the comoving distance $\chi$ is
\begin{equation}
\chi(z)=\frac{1}{H_0}\int_0^z \frac{dz'}{\sqrt{\Omega_m(1+z')^3+\Omega_k(1+z')^2+\Omega_{\Lambda}}}.
\end{equation}
We require that $1=\Omega_m+\Omega_k+\Omega_{\Lambda}$, according to $\Omega_x=\rho_x/\rho_{cr}$ being a fractional energy density relative to the critical density $\rho_{cr}=3H^2/8\pi G$. The luminosity distance is then just $D_L(z)=(1+z)^2D_A$. On large scales, the universe has been reasonably well constrained to be consistent with the concordance $\Lambda$CDM ($\Lambda$ cold dark matter) model. This model describes a flat ($k=0$) universe containing a dominant $\Lambda$ component, which causes the acceleration of the observed expansion of the universe, and some cold dark matter component, which together with baryons make up the observed matter density $\Omega_m=\Omega_{dm}+\Omega_{b}$. This model has been supported by a wide range of observations, including the cosmic microwave background (CMB) \citep{FowlerEtAl2010,DasEtAl2011,KeislerEtAl2011,ReichardtEtAl2012,AdeEtAl2014,
HinshawEtAl2013}, baryon acoustic oscillations (BAO) \citep{BeutlerEtAl2011,AndersonEtAl2012,BlakeEtAl2012,PadmanabhanEtAl2012}, constraints on $H_0$ \citep{RiessEtAl2011,FreedmanEtAl2012}, type Ia supernovae (SNe Ia) \citep{GuyEtAl2010,ConleyEtAl2011,SullivanEtAl2011,SuzukiEtAl2012}, and weak gravitational lensing \citep{BaconRefregierEllis2000,VanwaerbekeEtAl2000,RhodesRefregierGroth2001,
HoekstraEtAl2002,VanwaerbekeEtAl2002,BrownEtAl2003,JarvisEtAl2003,Heymansetal2004c,
MasseyEtAl2005,BenjaminEtAl2007,MasseyEtAl2007,SchrabbackEtAl2010,JeeEtAl2013,
KilbingerEtAl2013,heymans,KitchingEtAl2014,FuEtAl2014,HuffEtAl2014}.

The universe at smaller scales is rather lumpy and full of cosmic structures. This is represented in the standard approach by linear perturbations of the Einstein equations. This is done by replacing the spatially flat FLRW metric by the perturbed metric in, for example, the Newtonian gauge as
\begin{equation}
ds^2=-(1+2\psi)dt^2+a(t)^2(1-2\phi)dx^idx_i,
\label{eq:FLRWpert}
\end{equation}
where the $x_i$'s are comoving coordinates, and $\phi$ and $\psi$ are scalar potentials describing the scalar mode of the metric perturbations. In the case of matter domination (i.e., no shear stress) and working in Fourier $k$-space, the first-order perturbed Einstein equations  give
\begin{align}
k^2\phi &= -4\pi G a^2 \rho_m \delta_m
\label{eq:Poissonsimp}\\
\phi&=\psi,
\end{align}
where the overdensity perturbation relative to the mean density of the space $\bar{\rho}_m$ is just
\begin{equation}
\delta_m = \frac{\rho_m-\bar{\rho}_m}{\rho_m}.\label{eq:densitypert}
\end{equation}

These linear density perturbations can be shown to evolve as (e.g., \cite{peebles})
\begin{equation}
\ddot{\delta}_m+2H(t)\dot{\delta}_m-4\pi G \rho_m \delta_m=0.\label{eq:growth1}
\end{equation}
For small $\delta_m$, these perturbations evolve without moving in comoving coordinates, and have a solution that can be decomposed into a linear superposition of growing ($D_1$) and decaying ($D_2$) modes. Since the standard model assumes that density perturbations have grown from early times, we consider only the growing mode, which we will refer to simply as the growth factor $D$. In a matter dominated universe, $D\propto t^{2/3}\propto a$. Eq.~(\ref{eq:growth1}) can be rewritten for $D$ as
\begin{equation}
\ddot{D}+2H\dot{D}-\frac{3}{2}\Omega_m^0 H_0^2 (1+z)^3D=0.\label{eq:growth2}
\end{equation}
For some growth factor $D$ and times $t>t_1$, the density perturbation grows simply as
\begin{equation}
\delta_m(x,t)=\delta_m(x,t_1)\frac{D(t)}{D(t_1)}.
\end{equation}
This growth factor, which describes the growth rate of large-scale structure, is used both to write the matter power spectrum and in attempts to build an analytical description of the intrinsic alignment signal below.

The resulting cosmological structure of the $\Lambda$CDM model, as shown in the results of large-scale dark matter simulations like the Millennium Simulation \citep{SpringelEtAl2005} evolved to low redshift, could easily be described as a massive, cosmic web. In this picture of large-scale structure, small perturbations in the homogeneous density field sow the seeds for the eventual development of the clusters and super-clusters of galaxies that form nodes in a web of connecting filaments and sheets of mass. Along this cosmic web, smaller halos and galaxies form, merge, and eventually are drawn toward the clusters at the nodes forming the intersection of the filaments and sheets. This filamentary structure and the massive halos that form its nodes will provide the basis for some studies of intrinsic alignment discussed below.

\begin{figure*}
\begin{center}
\includegraphics[width=\columnwidth]{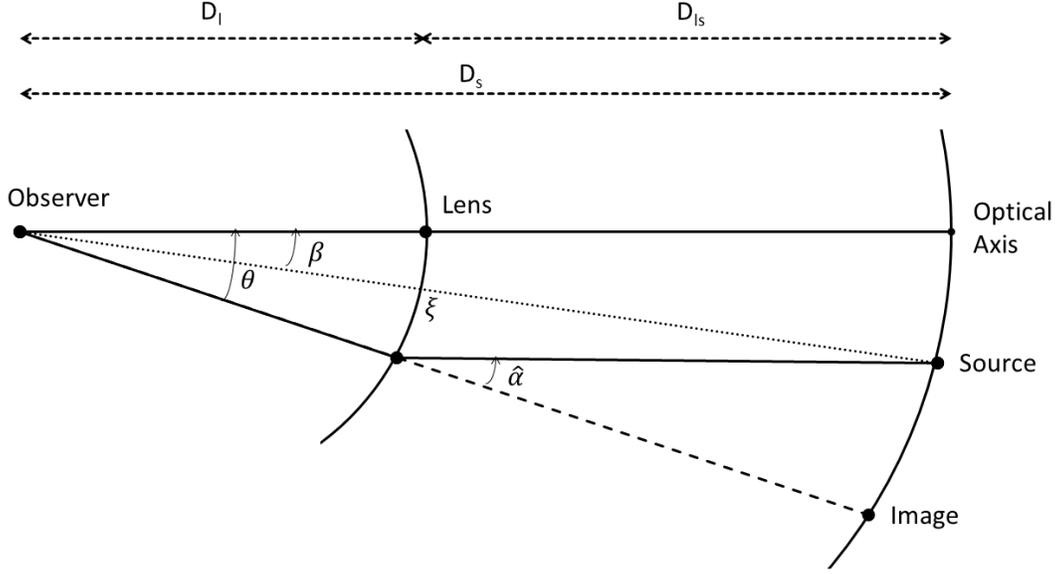}
\caption{\label{fig:lensgeo} 
The geometry of the lens equation. The observer, lens, source, and image positions are shown. The angle between the lens and image is given by $\bm{\theta}$, the angle between the lens and source by $\bm{\beta}$, and the deflection angle by $\bm{\hat{\alpha}}$. The angular diameter distance from the observer to the lens is $D_l$, from the observer to the source is $D_s$, and from the lens to the source is $D_{ls}$. For small angles, the relationship $\bm{\theta} D_{s}= \bm{\beta} D_{s}+\bm{\hat{\alpha}} D_{ls}$ holds. The impact parameter $\xi$ is also shown and can be approximated as a straight-line distance for small $\bm{\theta}$.}
\end{center}
\end{figure*}

\subsection{Gravitational lensing formalism}

A comprehensive review of gravitational lensing has been previously explored by several authors, and we will only briefly introduce the formalism of gravitational lensing here to provide a framework for discussing the impact of galaxy intrinsic alignment on weak gravitational lensing. For further information, we refer the reader to these previous reviews and early papers (e.g., \cite{Miralde1991,Kaiser1992,Blandford1992,SEF,Mellier1999,BSWLReview2001,Wittman2002,Refregier2003,VanwaerbekeMellier2003,Schneider2006,HoekstraJain2008,MunshiEtAl2008,Heavens2009,Bartelmann2010,Huterer2010,MasseyEtAl2010,WeinbergEtAl2013}) and references therein. General relativity predicts a bending angle for light in the neighborhood of some compact, spherically symmetric mass $M$ of 
\begin{equation}
\hat{\alpha}=\frac{4GM}{\xi},\label{eq:bending}
\end{equation}
where $\xi$ is the minimum distance from the path of the light ray to the lensing mass, such that $\xi\gg R_{_S}$, the Schwarzschild radius of the mass. The Born approximation is typically assumed, under which the light ray is represented as a straight line in the neighborhood of the lensing mass with a single discrete bend in its path at the moment it passes closest to the lensing mass. This is valid if the actual deflection is small and the source is spherically symmetric and compact relative to the distances involved. In the weak gravitational field limit and linearized general relativity theory, the total deflection can be considered as the sum of the deflection angles due to some ensemble of lensing masses. The thin lens approximation is then used, where the deflection of the actual light ray is small compared to the typical scales over which the lensing mass distribution changes significantly. The lensing mass can then be represented as a projected surface density. The thin lens approximation is assumed to be valid for most relevant astrophysical lensing systems. Eq.~(\ref{eq:bending}) is then rewritten as the two-vector
\begin{equation}
\bm{\hat{\alpha}}(\bm{\xi})=4G\int d^2\xi'\Sigma(\bm{\xi}')\frac{\bm{\xi}-\bm{\xi'}}{|\bm{\xi}-\bm{\xi'}|^2},\label{eq:bending2}
\end{equation}
where 
\begin{equation}
\Sigma(\bm{\xi})=\int dz\rho(\xi_1,\xi_2,z)
\label{eq:Sigma}
\end{equation}
is the surface mass density, $\bm{\xi}=(\xi_1,\xi_2)$ is the impact two-vector in the lens plane, and $z$ is the perpendicular coordinate to the lens plane, which is not generally the redshift.

In an astrophysical context, the lens geometry can be depicted as in Fig.~\ref{fig:lensgeo}, which under the assumption of small angles gives the relation
\begin{equation}
\bm{\theta} D_{s}= \bm{\beta} D_{s}+\bm{\hat{\alpha}} D_{ls}.\label{eq:rellens1}
\end{equation}
$D_s$ is the angular diameter distance from the observer to the source, $D_{ls}$ is from the lens to the source, and $D_l$ is from the observer to the lens. Eq.~\ref{eq:rellens1} can be written as the familiar lens equation
\begin{equation}
\bm{\beta}=  \bm{\theta}  -\bm{\hat{\alpha}} \frac{D_{ls}}{D_{s}},
\label{eq:lensequation}
\end{equation}
or simply 
\begin{equation}
\bm{\beta}= \bm{\theta} - \bm\alpha
\end{equation}
where $\bm\alpha\equiv \bm{\hat{\alpha}}D_{ls}/D_{s}$ is the scaled deflection angle. 

One can now define the dimensionless surface mass density or convergence
\begin{equation}
\kappa(\bm{\theta})=\Sigma(D_l\bm{\theta})/\Sigma_{cr},
\label{eq:kappa}
\end{equation}
where
\begin{equation}
\Sigma_{cr}=\frac{1}{4\pi G}\frac{D_s}{D_lD_{ls}}
\end{equation}
is the critical surface mass density. $\Sigma_{cr}$ effectively defines the crossover for a given $\Sigma(D_l\bm{\theta})$ from strong to weak lensing such that $\kappa\ge1$ is a sufficient condition for lensing with multiple images. 

The scaled deflection angle can now be expressed in terms of the convergence as 
\begin{equation}
\bm{\alpha}(\bm\theta)=\frac{1}{\pi}\int d^2\theta'\kappa(\bm{\theta'})\frac{\bm{\theta}-\bm{\theta'}}{|\bm{\theta}-\bm{\theta'}|^2},
\label{eq:scaledalpha}
\end{equation}
or through the use of a 2-dimensional deflection potential 
\begin{equation}
\Psi(\bm\theta)=\frac{1}{\pi}\int d^2\theta'\kappa(\bm{\theta'}) \ln|{\bm{\theta}-\bm{\theta'}}|,
\label{eq:scaledalpha2}
\end{equation}
 such that 
\begin{equation}
\bm{\alpha}(\bm{\theta})=\nabla\Psi(\bm{\theta}).
\end{equation}
$\Psi$ then satisfies the 2-dimensional Poisson equation 
\begin{equation}
\nabla^2\Psi(\bm\theta)=2\kappa(\bm{\theta}).
\end{equation} 
The lens mapping can be locally linearized for sufficiently compact sources, which leads to a Jacobian matrix
\begin{align}
\bm{\mathcal{A}}(\bm{\theta}) =\left(\delta_{ij}-\frac{\partial^2\Psi(\bm{\theta})}{\partial\theta_i\partial\theta_j}\right)=
\left(
\begin{array}{cc}
1-\kappa-\gamma_1 & -\gamma_2 \\
-\gamma_2 & 1-\kappa+\gamma_1
\end{array}
\right)\,\,
\label{eq:Jacobian}
\end{align}
where the complex shear is often expressed as $\gamma=\gamma_1+i\gamma_2=|\gamma|e^{2i\phi}$. If the background universe is homogeneous and isotropic, the shape of a source is distorted only by the gravitational tidal field, described by the shear $\gamma$, with no contribution from shear due to the metric. The shape is also magnified both by isotropic focusing due to the convergence $\kappa$ and anisotropic focusing due to $\gamma$. This magnification is expressed from Eq.~(\ref{eq:Jacobian}) as 
\begin{equation} 
\mu=\frac{1}{\rm{det} \mathcal{A}}=\frac{1}{(1-\kappa^2)-|\gamma|^2}.
\label{eq:magnification}
\end{equation}
In more general (non-FLRW) inhomogeneous or anisotropic cosmologies, however, it is often more useful to relate the Jacobian matrix in Eq.~(\ref{eq:Jacobian}) and the lensing convergence and shear directly to the Ricci and Weyl focusing of the spacetime (e.g., \cite{SEF,seitz,clarkson,fanizza,TroxelIshakPeel2014}). 

The complex shear (and thus convergence in the weak limit) is measured in practice as a function of the elliptical galaxy shape. The components of the observed ellipticity ($e=e_1+i e_2$) can be related to the major ($a$) and minor ($b$) axes of the galaxy and its orientation ($\phi$) by
\begin{align}
e=\frac{a-b}{a+b}e^{2i\phi}.
\end{align}
The observed ellipticity is then related to the shear and intrinsic ellipticity $e_I$ by (e.g., \cite{SchrammKayser1995,SeitzSchneider1997})
\begin{align}
e=\frac{e_I+\gamma}{1+\gamma^{*}e_I^{*}},\label{eq:efrome}
\end{align}
where ${}^{*}$ is the complex conjugate. For an unbiased galaxy sample with $\langle e_i\rangle=0$ and $\gamma<1$, the complex shear is estimated as
\begin{align}
\gamma=\langle e\rangle.
\end{align}
Alternately, as used in some intrinsic alignment measurements in Sec. \ref{detections}, the ellipticity can be defined as
\begin{align}
e=\frac{a^2-b^2}{a^2+b^2}e^{2i\phi}, 
\end{align}
with a 'shear responsivity' factor \cite{KaiserSquiresBroadhurst1995,BernsteinJarvis2002}
\begin{align}
\gamma=\frac{\langle e\rangle}{2\mathcal{R}}.
\end{align}

                 \subsection{Weak gravitational lensing power spectra and bispectra}\label{psbs}

We can now generalize the formalism to weak shear or convergence in a $\Lambda$CDM model, where we consider small linear matter density perturbations $\delta_m$ (e.g., Sec.~\ref{lcdm}) with an associated Newtonian potential $\Phi$. These are related by the 3D Poisson equation
\begin{equation}
\nabla^2\Phi=\frac{3}{2}H_0^2\Omega_m\frac{\delta_m}{a}.
\end{equation}
The convergence is then written
\begin{equation}
\kappa(\bm{\theta},\chi)=\frac{3}{2}H_0^2\Omega_m\int_0^{\chi}d\chi'\frac{\sin_k(\chi')\sin_k(\chi-\chi')}{\sin_k(\chi)}\frac{\delta_m(\sin_k(\chi')\bm{\theta},\chi')}{a(\chi')}.\label{eq:keff1}
\end{equation}
For an explicit, normalized distribution of sources in co-moving distance $f(\chi)$, we simply average Eq.~(\ref{eq:keff1}) over the normalized distribution and can write the source-distance weighted effective convergence as
\begin{align}
\kappa(\bm{\theta},\chi)&=\int_0^{\chi_1}d\chi W(\chi)\delta_m(\sin_k(\chi)\bm{\theta},\chi),\label{eq:keff2}
\end{align}
where
\begin{align}
W(\chi)&=\frac{3}{2}H_0^2\frac{\Omega_m}{a(\chi)}\int_{\chi}^{\chi_1}d\chi' f(\chi')\sin_k(\chi)\frac{\sin_k(\chi'-\chi)}{\sin_k(\chi')}.\label{eq:weighting}
\end{align}
$W(\chi)$ is related to the weighted lens efficiency. We have explicitly chosen as the upper bound in the integral, $\chi_1$, the horizon distance, which corresponds to the co-moving distance at infinite redshift, or the edge of the observable universe.

When we study cosmic shear, it is preferable instead to consider the statistical properties of the convergence, which is possible through the correlation function or corresponding spectrum in the harmonic space of the field. Since we have assumed that the space is isotropic and homogeneous, the field $\delta_m$ is also isotropic and homogeneous. With the assumption of Gaussian randomness in $\delta_m$, we can use Limber's approximation \citep{Limber1954,Kaiser1992} to relate the 2D convergence power spectrum and bispectrum to the 3D matter power spectrum and bispectrum. Equation (\ref{eq:keff2}) was written is such a way that it will be immediately recognizable as a weighted projection, and for the 2- and 3-point correlations of $\kappa$, we can write from Limber's approximation the convergence power spectrum and bispectrum
\begin{align}
P_{\kappa}(\ell)&=\int_0^{\chi_1}d\chi\frac{W^2(\chi)}{\sin_k^2(\chi)}P_{\delta}(k=\frac{\ell}{\sin_k(\chi)};\chi)\label{eq:spec}\\
B_{\kappa}(\ell_1,\ell_2,\ell_3)&=\int_0^{\chi_1}d\chi\frac{W^3(\chi)}{\sin_k^4(\chi)}B_{\delta}(k_1=\frac{\ell_1}{\sin_k(\chi)},k_2=\frac{\ell_2}{\sin_k(\chi)},k_3=\frac{\ell_3}{\sin_k(\chi)};\chi).\label{eq:bspec}
\end{align}
where $P_{\delta}(k=\frac{\ell}{\sin_k(\chi)};\chi)$ and $B_{\delta}(k_1=\frac{\ell_1}{\sin_k(\chi)},k_2=\frac{\ell_2}{\sin_k(\chi)},k_3=\frac{\ell_3}{\sin_k(\chi)};\chi)$ are the 3D matter power spectrum and bispectrum, respectively. One can also alternatively write the information in Eqs. (\ref{eq:spec}) \& (\ref{eq:bspec}) in terms of the correlation function or aperture mass statistic \citep{Kaiser1994,SchneiderKilbingerLombardi1995,Schneider1996,PenEtAl2003,JarvisBernsteinJain2004}. For the power spectrum, this is
\begin{align}
\xi_{+}(\theta)=&\frac{1}{2\pi}\int_0^{\infty}d\ell \ell P_{\kappa}(\ell)J_{0}(\ell\theta)\label{eq:corr2ps}\\
\langle M^2_{ap}\rangle(\theta)=&\int d\ell \frac{\ell}{2\pi}P_{\kappa}(\ell)\tilde{U}^2(\theta\ell)\label{eq:apmstat},
\end{align}
where $J_0$ is a zeroth order Bessel function of the first kind and $\tilde{U}$ is the Fourier transform of a filter function $U$ with characteristic smoothing scale $\theta$.

The linear matter power spectrum can be expressed for $k=\ell/\sin_k(\chi)$ as
\begin{align}
P_{\delta}(k,z)=A \frac{T^2(k,z)}{a^2}\frac{D^2(z)}{D^2(0)}k^{n_s},\label{eq:mps}
\end{align}
where $n_s$ is the spectral index, $T$ is a transfer function that modifies the primordial power spectrum, and $A$ is a normalization parameter that can be fixed in relation to $\sigma_8$, which measures the amplitude of matter fluctuations on scales of 8 $h^{-1}$ Mpc. The linear matter power spectrum under-predicts power on small scales, and is often modified to the nonlinear matter power spectrum $P_{nl}$ to include nonlinear effects on small scales (see for example, \cite{smith03}). 

For the matter bispectrum, there are contributions both from primordial non-Gaussianity and the nonlinear clustering of matter. The bispectrum due to nonlinear clustering is often estimated through the fitting formulae of \cite{ScoccimarroCouchman2001}. The bispectrum is related to the power spectrum through second-order perturbation theory \citep{Fry1984,BernardeauEtAl2002}
\begin{align}
B_{nl}(\bm{k_1},\bm{k_2},\bm{k_3},z)=2F_2^{eff}(\bm{k_1},\bm{k_2})P_{nl}(k_1,z)P_{nl}(k_2,z)+2\textrm{ perm.},\label{eq:bfit}
\end{align}
where the $\bm{k}_i$ form a closed triangle and the effective kernel $F_2^{eff}$ is
\begin{align}
F_2^{eff}(\bm{k_1},\bm{k_2})=&\frac{5}{7}a(n,k_1)b(n,k_2)+\frac{1}{2}\frac{\bm{k_1}\cdot\bm{k_2}}{k_1k_2}\left(\frac{k_1}{k_2}+\frac{k_2}{k_1}\right)b(n,k_1)b(n,k_2)\\
&+\frac{2}{7}\left(\frac{\bm{k_1}\cdot\bm{k_2}}{k_1k_2}\right)^2c(n,k_1)c(n,k_2)\nonumber.
\end{align}
The functions $a$, $b$, and $c$ were fit by \cite{ScoccimarroCouchman2001} to numerical simulations. These reduce to $a=b=c=1$ on large scales ($k\ll k_{nl}$), and the perturbation theory prediction is recovered. This fitting function for $B_{nl}$ is only an approximation. \cite{GilMarinEtAl2012} recently presented an improved fitting formula, which modifies the functions $a$, $b$, and $c$ to reflect the ability of more recent simulations to constrain the bispectrum. These improved fitting functions are reported to have an accuracy of typically within $5\%$ when compared to simulation results. 

\begin{figure}
\center
\includegraphics[height=\textwidth,angle=270]{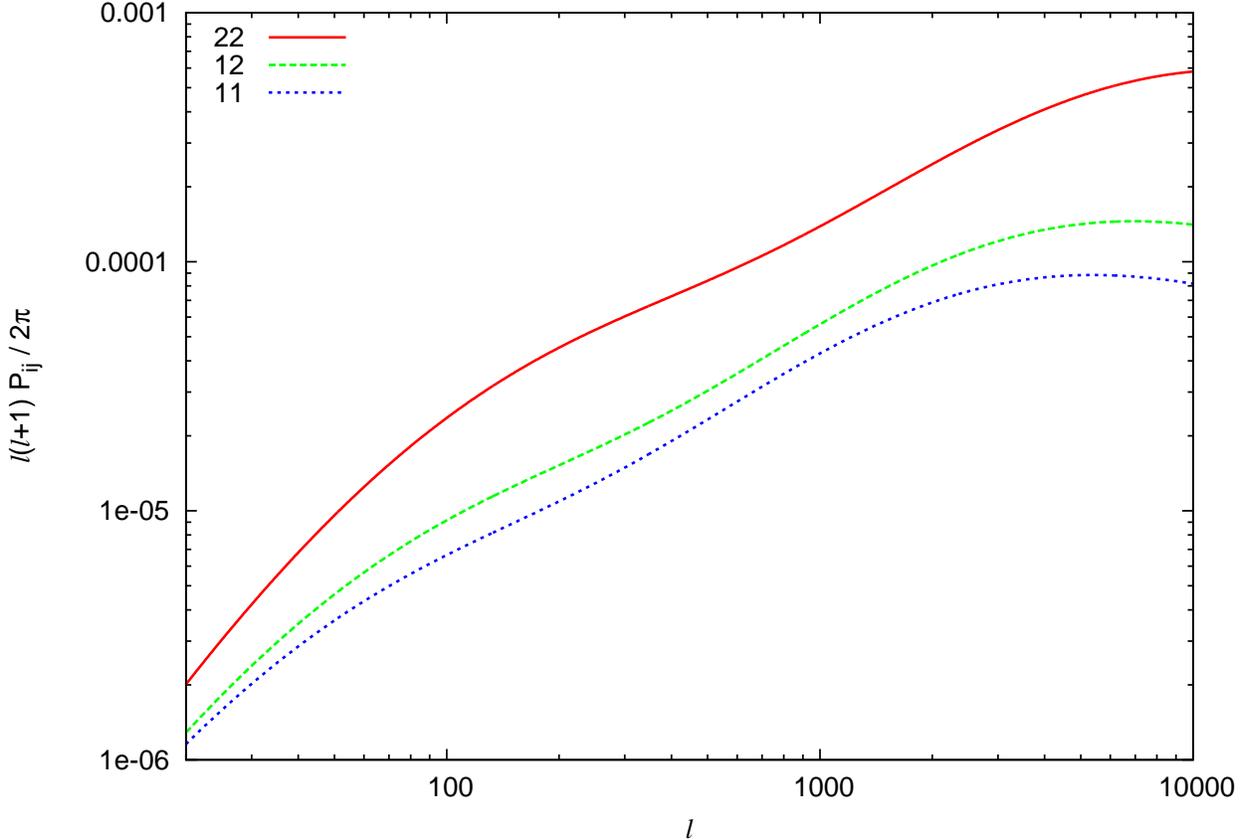}
\caption{The convergence auto- and cross-power spectra in Eq.~(\ref{eq:pstomo}) are shown for the base $\Lambda$CDM model with a cosmological constant, $\Omega_m=0.27$, $\Omega_{\Lambda}=0.73$, $n_s=0.96$, and $\sigma_8=0.84$. The redshift distribution given in Eq.~(\ref{eq:fz}) is split into two bins with boundary $z=0.8$, the median of the distribution. The auto-spectra of the lower and higher redshift bins are labeled `11' and `22', respectively, while the cross-spectrum is labeled `12'.}\label{fig:tomography}
\end{figure}
 
\begin{figure}
\center
\includegraphics[height=\textwidth,angle=270]{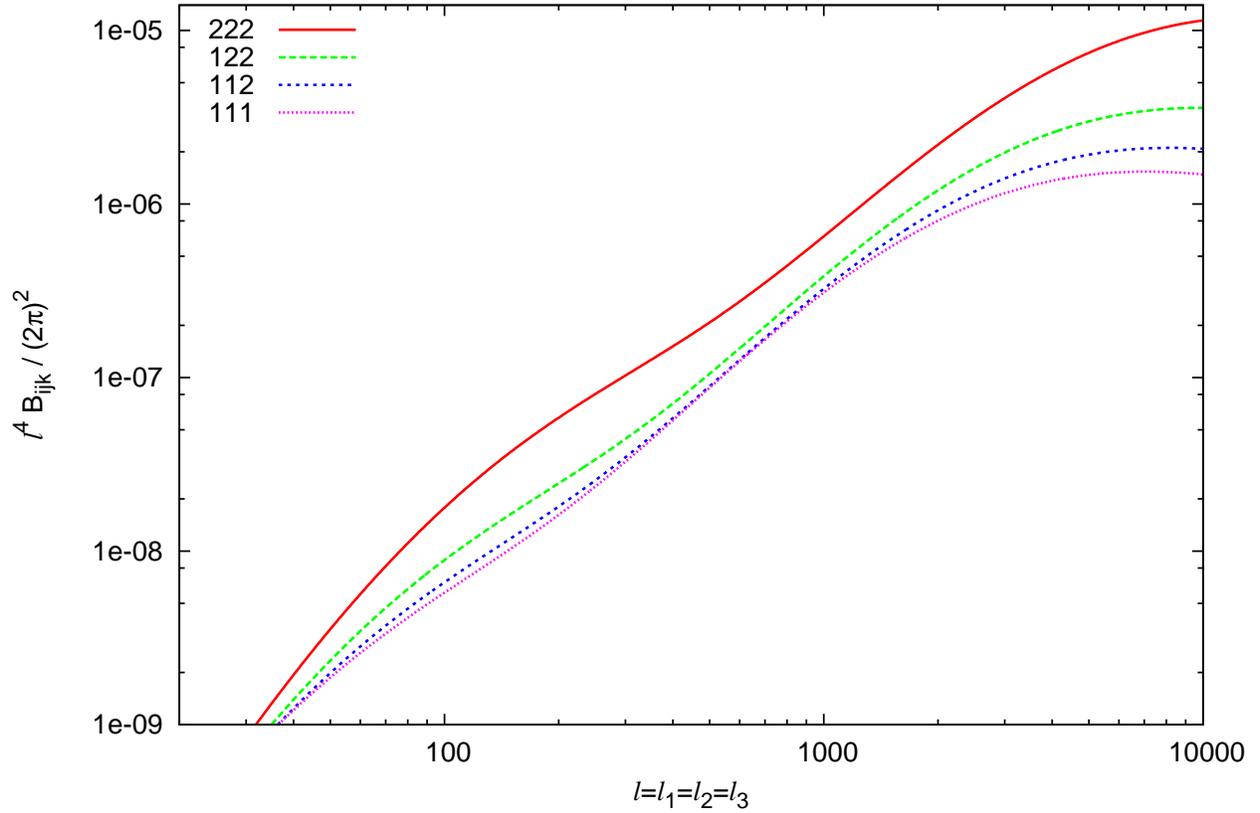}
\caption{The convergence auto- and cross-bispectra in Eq.~(\ref{eq:bspec2}) are shown for the base $\Lambda$CDM model with a cosmological constant, $\Omega_m=0.27$, $\Omega_{\Lambda}=0.73$, $n_s=0.96$, and $\sigma_8=0.84$. The redshift distribution given in Eq.~(\ref{eq:fz}) is split into two bins with boundary $z=0.8$, the median of the distribution. The auto-bispectra of the lower and higher redshift bins are labeled `111' and `222', respectively, while the cross-bispectra are labeled `112' and `122'.}\label{fig:tomography2}
\end{figure}

\subsection{Weak gravitational lensing tomography}

Weak lensing tomography takes advantage of additional information on the redshift of source galaxies in surveys, which allows the galaxy sample to be split into multiple redshift bin slices. By correlating separately the shapes of galaxies in each redshift bin (i.e. auto-power spectra), this provides additional depth or redshift dependent information to the cosmic shear signal, and can probe physical properties like the growth rate of large-scale structure, which is a function of redshift and directly influences the theoretical matter power spectrum (Eq.~(\ref{eq:mps})). In addition to the auto-power spectra for each bin, one can also consider cross-power spectra or bispectra between tomographic bins. This not only adds more constraining information, and has been recognized as the optimal method for extracting cosmological information (e.g., \cite{Hu1999,Huterer2002,SimonEtAl2004,TakadaJain2004}), but can also be used in methods to deal with systematics in weak lensing such as the intrinsic alignment of galaxies (see Sec.~\ref{mitigation}). A tomographic galaxy sample is chosen such that each redshift bin $i$ spans from some $\chi_i(z_i)$ to $\chi_{i+1}(z_{i+1})$, with a mean $\bar{\chi}_i(\bar{z}_i)$. The weighting function is then given by  
\begin{align}
W_i(\chi)&=\frac{3}{2}H_0^2\frac{\Omega_m}{a(\chi)}\int_{\chi_i}^{\chi_{i+1}}d\chi' f_i(\chi')\sin_k(\chi)\frac{\sin_k(\chi'-\chi)}{\sin_k(\chi')},\label{eq:weightbin}
\end{align}
for $\chi<\chi_{i+1}$ and zero otherwise, assuming that the true comoving distance of the source galaxy is known. The distribution $f_i(\chi)$ is the normalized comoving galaxy distribution in the $i$-th redshift bin. For spectroscopic redshift information, the true redshifts of galaxies in the $i$-th bin lie between $z_i$ and $z_{i+1}$, while for photometric redshifts (photo-z) with some photo-z probability distribution function (PDF) $p(z|z^p)$, the true redshift distribution can be smeared outside the assumed redshift bin boundaries, and the integral in Eq.~(\ref{eq:weightbin}) must be taken from zero to infinity for completeness. 

A typical photo-z PDF can be written \citep{MaBernstein2008}
\begin{align}
p(z|z^P)=\frac{1-p_{\textrm{cat}}}{\sqrt{2\pi}\sigma(z^P)}\exp\left[\frac{(z-z^P)^2}{2\sigma^2(z^P)}\right]+\frac{p_{\textrm{cat}}}{\sqrt{2\pi}\sigma(z^P)}\exp\left[\frac{(z-f_{\textrm{bias}}z^P)^2}{2\sigma^2(z^P)}\right],\label{eq:pdf}
\end{align}
with a photo-z uncertainty $\sigma(z^p)=\sigma_{ph}(1+z^p)$ and fraction of catastrophic outliers $p_{cat}$ which are biased by a factor $f_{bias}$. The galaxy distribution $f_i(\chi)$ is then modified according to $p(z|z^p)$. Equations (\ref{eq:spec}) \& (\ref{eq:bspec}) are then
\begin{align}
P_{ij}(\ell)&=\int_0^{\chi_1}d\chi\frac{W_i(\chi)W_j(\chi)}{\sin_k^2(\chi)}P_{\delta}(k=\frac{\ell}{\sin_k(\chi)};\chi),\label{eq:pstomo}\\
B_{ijk}(\ell_1,\ell_2,\ell_3)&=\int_0^{\chi_1}d\chi\frac{W_i(\chi)W_j(\chi)W_k(\chi)}{\sin_k^4(\chi)}B_{\delta}(k_1=\frac{\ell_1}{\sin_k(\chi)},k_2=\frac{\ell_2}{\sin_k(\chi)},k_3=\frac{\ell_3}{\sin_k(\chi)};\chi).\label{eq:bspec2}
\end{align}
We show tomographic power spectra and bispectra for two photo-z bins in Figs. \ref{fig:tomography} \& \ref{fig:tomography2}. The two photo-z bins use a galaxy redshift distribution given by (e.g., \cite{WittmanEtAl2000})
\begin{align}
n(z)=\frac{z^2}{2z_0^3}e^{-z/z_0},\label{eq:fz}
\end{align}
which has mean redshift $z_{mean}=3z_0$. We have chosen $z_0=0.3$ to match planned survey depths for Stage IV weak lensing surveys (e.g., LSST and Euclid), as classified by \cite{DETFReport2006}. The two bins are separated by the median redshift of the sample, $z_{med}=0.8$. We assume a photo-z uncertainty $\sigma_{ph}=0.05$, but without catastrophic outliers. These tomographic cross-power spectra and bispectra have similar power to the full power spectrum and bispectrum, and the inclusion of the full set of tomographic spectra significantly increases the available information and constraining power of weak lensing surveys. Beyond increased cosmological information, the use of tomographic bins is also essential for many methods used to distinguish between weak lensing and intrinsic alignment, which is discussed in detail in Sec.~\ref{mitigation}.

\subsection{Cosmic shear as a probe for cosmological study}\label{cosmicweak}

\begin{figure}
\center
\includegraphics[height=\textwidth,angle=270]{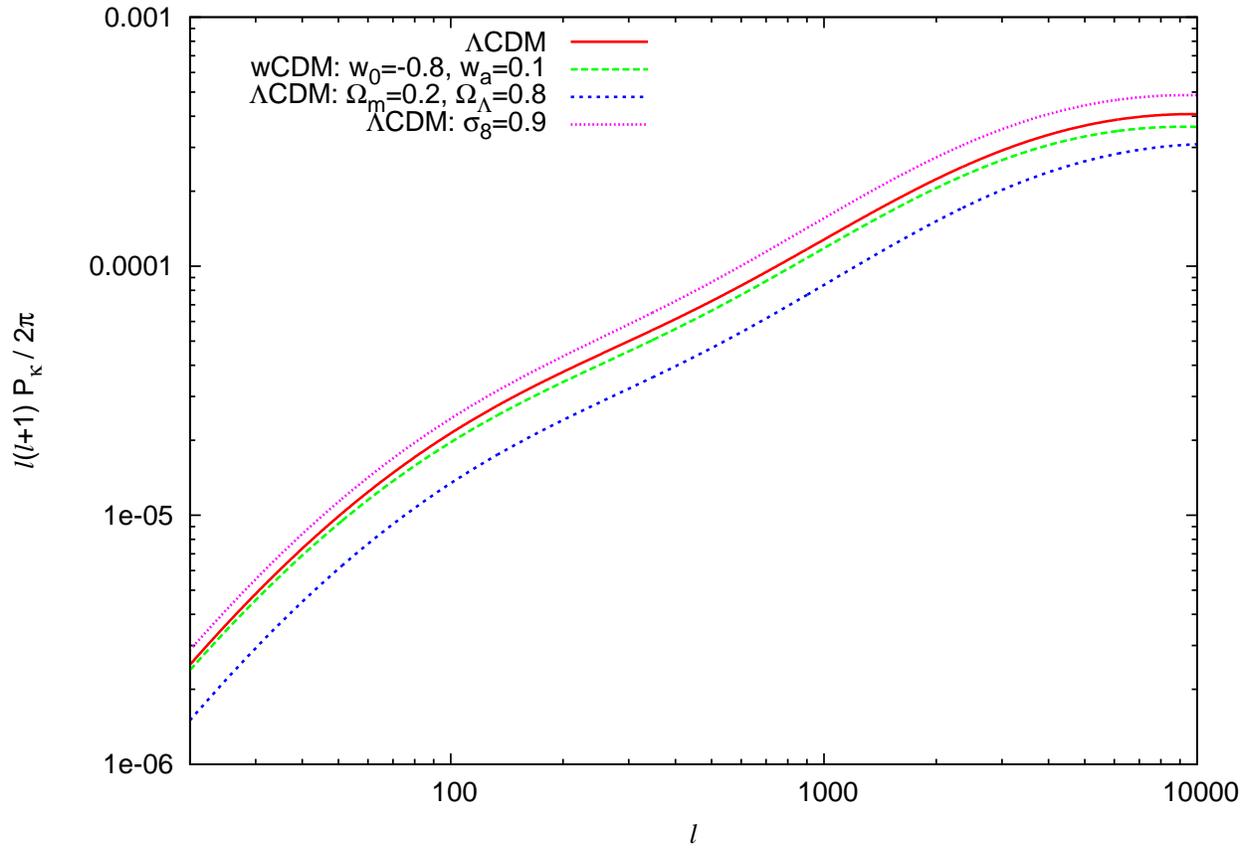}
\caption{The convergence power spectrum in Eq.~(\ref{eq:spec}) is shown for several CDM models. The base $\Lambda$CDM model is taken to have a cosmological constant $\Lambda$, $\Omega_m=0.27$, $\Omega_{\Lambda}=0.73$, $n_s=0.96$, and $\sigma_8=0.84$. The wCDM model has dark energy equation of state $w=w_0+w_a(1-a)$. The galaxy distribution is given by Eq.~(\ref{eq:fz}).}\label{fig:pscosmo}
\end{figure}

\begin{figure}
\center
\includegraphics[height=\textwidth,angle=270]{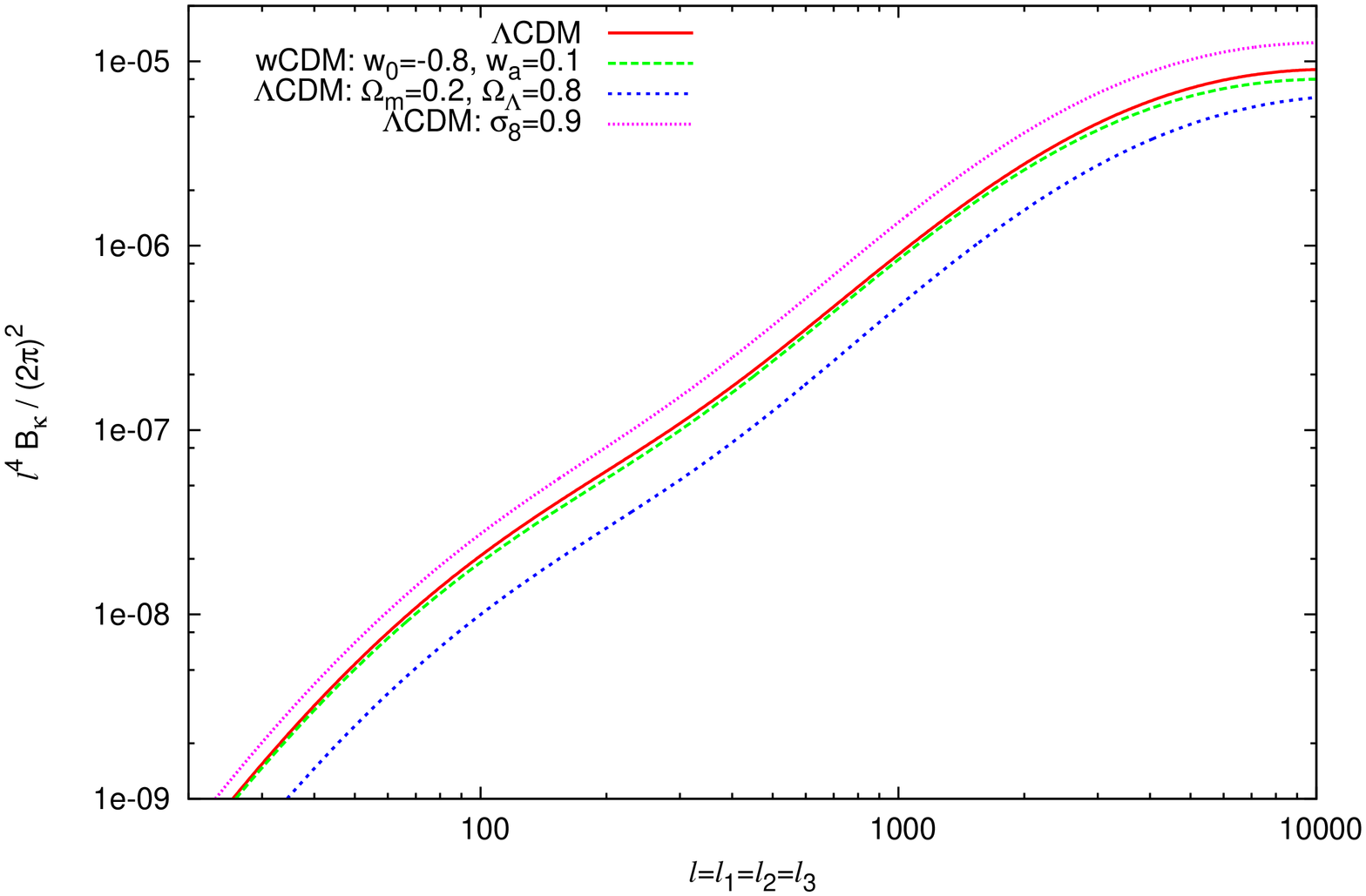}
\caption{The convergence bispectrum in Eq.~(\ref{eq:bspec}) is shown for several CDM models. The base $\Lambda$CDM model is taken to have a cosmological constant $\Lambda$, $\Omega_m=0.27$, $\Omega_{\Lambda}=0.73$, $n_s=0.96$, and $\sigma_8=0.84$. The wCDM model has dark energy equation of state $w=w_0+w_a(1-a)$. The galaxy distribution is given by Eq.~(\ref{eq:fz}).}\label{fig:bscosmo}
\end{figure}

Weak gravitational lensing has emerged as one of the most promising cosmological probes. It is an ideal probe to trace the distribution of the dark and baryonic matter in the universe and to determine the matter density parameter (e.g., \cite{Schneider1996,BaconRefregierEllis2000,VanwaerbekeEtAl2000,RhodesRefregierGroth2001,WilsonEtAl2001,HoekstraEtAl2002,Hu2002a,Mellier2002,VanwaerbekeEtAl2002,BrownEtAl2003,JarvisEtAl2003,MasseyEtAl2004,Schneider2004,TaylorEtAl2004,BaconEtAl2005,MasseyEtAl2005,BenjaminEtAl2007,MasseyEtAl2007,Heavens2009,JoudakiCoorayHolz2009,SimonTaylorHartlap2009,Huterer2010,MasseyEtAl2010,SchrabbackEtAl2010,JeeEtAl2013,KilbingerEtAl2013,heymans,KitchingEtAl2014,FuEtAl2014,HuffEtAl2014,ShanEtAl2014,vanwaerbekeEtAl2013}). Weak lensing is sensitive to the growth rate of large-scale structure in the Universe as well as its expansion history. This allows it, in combination with other cosmological data, to put stringent constraints on important cosmological parameters such as the Hubble constant, the matter density parameters, and the matter fluctuation amplitude. This also makes it a powerful probe of dark energy models and parameters (e.g., \cite{HuTegmark1999,Zhu2000,Hu2002,Hu2002a,PenLuVanwaerbekeMellier2003,Bernardeau2003,BenabedWaerbeke2004,BernsteinJain2004,HuJain2004,KuhlenKeetonMadau2004,Ishak,UpadhyeIshakSteinhardt2005,HannestadEtAl2006,7i,ShapiroDodelson2007,VaccaColombo2008,Heavens2009,HollensteinEtAl2009,JoudakiCoorayHolz2009,ChongchitnanKing2010,DebonoEtAl2010,Ellis2010,SchrabbackEtAl2010,ShapiroEtAl2010,WeinbergEtAl2013,CaoCovoneZhu2012,VanderveldMortonsonHuEifler2012,HeavensAlsingJaffe2013,KitchingEtAl2014}). Last but not least, weak lensing has also proved to be a powerful probe to test general relativity and modified gravity theories at cosmological scales (e.g., \cite{Song2005,2c,IshakUpadhyeSpergel2006,ZhaoBaconTaylorHorne2006,HutererLinder2007,2j,ZhangLiguoriBeanDodelson2007,AcquavivaEtAl2008,DanielCaldwellCorrayMelchiorri2008,2k,TsujikawaTatekawa2008,BeynonBaconKoyama2009,1d,HearinZentner2009,2i,SongDore2009,ThomasAbdallaWeller2009,ZhaoPogosianSilvestriZylberberg2009,BeanTangmatitham2010,DanielEtAl2010,JainKhoury2010,TerenoSemboloniSchrabback2010,DossetEtAl2011b,DossettMoldenhauerIshak2011,ThomasEtAl2011,AsabaEtAl2013,KirkEtAl2013,SimpsonEtAl2013}). 

Another useful feature of cosmic shear is that in addition to constraints derived from its power spectrum, the bispectrum is equally and particularly important for upcoming high precision surveys as listed in the introduction, many of which will be capable of making high confidence measurements of 3-point weak lensing statistics. The bispectrum has been shown to break degeneracies in the cosmological parameters and to probe additional physics like the non-Gaussianity of large-scale structure \citep{BernardeauEtAl1997,waerbekeEtAl1999,matarrese,waerbekeEtAl2001,verde,TakadaJain2003,TakadaJain2004,MunshiEtAl2008,28,VafaeiEtAl2010,huterer,SemboloniEtAl2010,munshi,FuEtAl2014}. \cite{TakadaJain2004} have demonstrated, for example, that we should expect improvements by a factor of 2-3 in constraining power of cosmological constraints for deep lensing surveys when the bispectrum is included. This has recently been used by the Canada France Hawaii Telescope Lensing Survey (CFHTLenS), who report improved constraints on $\Omega_m$, $\sigma_8$, and their combination when 3-point information is included, despite only a 2$\sigma$ detection of the third moment of the aperture mass \citep{FuEtAl2014}. 

Several surveys have already provided cosmic shear measurements and put complementary constraints on cosmological parameters \citep{BaconRefregierEllis2000,VanwaerbekeEtAl2000,RhodesRefregierGroth2001,
HoekstraEtAl2002,VanwaerbekeEtAl2002,BrownEtAl2003,JarvisEtAl2003,
Heymansetal2004c,MasseyEtAl2005,BenjaminEtAl2007,MasseyEtAl2007,SchrabbackEtAl2010,JeeEtAl2013,KilbingerEtAl2013,heymans,KitchingEtAl2014,FuEtAl2014,HuffEtAl2014}. Comparisons between theoretical predictions and measurements of the convergence power spectrum, bispectrum, or other equivalent statistical measures like the aperture mass can place direct constraints on the cosmological parameters $H_0$, $\Omega_m$, $\Omega_k$, $\Omega_{\Lambda}$, and the dark energy equation of state $w$ through the growth factor $D$ and distance calculations, and explicitly as they appear in Eq.~(\ref{eq:weighting}). The amplitude of the matter power spectrum depends directly on the value of the parameter $\sigma_8$ and its scale dependence on $n_s$, the spectral index. The matter power spectrum is also sensitive to modifications to gravity, which will manifest themselves via changes to the growth of structure and distances.

To briefly demonstrate the effects of varying the cosmological parameters on the predicted power spectrum and bispectrum, the convergence power spectrum is depicted in Fig.~\ref{fig:pscosmo} for various FLRW CDM cosmologies. These include: the concordance $\Lambda$CDM model with a cosmological constant, $\Omega_m=0.27$, $\Omega_{\Lambda}=0.73$, $n_s=0.96$, and $\sigma_8=0.84$; a wCDM model with equation of state $w=w_0+w_a(1-a)$ for $w_0=-0.8$ and $w_a=0.3$; a $\Lambda$CDM model with $\Omega_m=0.2$ and $\Omega_{\Lambda}=0.80$; and a $\Lambda$CDM model with $\sigma_8=0.9$. The equilateral ($\ell=\ell_1=\ell_2=\ell_3$) convergence bispectrum is also shown in Fig.~\ref{fig:bscosmo} for the same CDM cosmologies. We assume a galaxy redshift distribution given by Eq.~(\ref{eq:fz}).

      \subsection{Systematics and sources of bias in weak lensing}\label{systematics}

One of the primary challenges to successfully using weak lensing for cosmological constraints is the prevalence of observational and physical systematics and biases, which pose major challenges for large ongoing and planned lensing surveys that aim to place precision constraints on cosmological parameters. These systematic effects must be accounted for in order to make full use of the potential of ongoing and planned weak lensing surveys (see for example, \cite{CroftMetzler2000,HeavensRefregierHeymans2000,BaconRefregierCloweEllis2000,CatelanKamionkowskiBlandford2001,ErbenEtAl2001,BernsteinJarvis2002,BrownTaylorHamblyDye2002,KingSchneider2002,HirataSeljak2003a,Refregier2003,VanwaerbekeMellier2003,HeymansEtAl2004,IshakHirataMcdonaldSeljak2004,TakadaWhite2004} and the review \cite{MunshiEtAl2008} and references therein). Systematic effects in weak lensing can impact constraints either directly through the shape and redshift measurements of galaxies or through the statistical correlations of galaxy shapes. They can also enter into constraints through limitations in our theoretical modeling. These effects include challenges to measuring the shape of galaxies, which are smeared or distorted due to atmospheric, camera, or reduction effects, calibration biases, difficulties in accurately determining the redshift of such large ensembles of objects, and fundamental limits to our current understanding of the matter power spectrum and its nonlinear evolution for standard and nonstandard cosmologies. The primary physical systematic effect which biases weak lensing statistics is a correlation of the intrinsic alignment of galaxies (their ellipticities prior to lensing), which is the topic of this review. 

It is thus vital to understand and control these and other systematic effects of weak lensing in order to fully explore its potential as a tool for precision cosmology (e.g., \cite{BaconRefregierCloweEllis2000,CroftMetzler2000,HeavensRefregierHeymans2000,Kaiser2000,LeePen2000,PenLeeSeljak2000,CatelanKamionkowskiBlandford2001,CrittendenNatarajanPenTheuns2001,ErbenEtAl2001,LeePen2001,BernsteinJarvis2002,BrownTaylorHamblyDye2002,HuOkamoto2002,Jing2002,KingSchneider2002,LeePen2002,HeymansHeavens2003,HirataSeljak2003a,KingSchneider2003,VanwaerbekeMellier2003,HeymansEtAl2004,IshakHirataMcdonaldSeljak2004,TakadaJain2004,TakadaWhite2004,Ishak,IshakHirata2005}). We will briefly mention some of these here, but recommend more extensive reviews on gravitational lensing (e.g., \cite{SEF,Mellier1999,BSWLReview2001,Wittman2002,Refregier2003,VanwaerbekeMellier2003,Schneider2006,HoekstraJain2008,MunshiEtAl2008,Heavens2009,Bartelmann2010,Huterer2010,MasseyEtAl2010,WeinbergEtAl2013} and references therein) for a more thorough discussion of these and additional systematics. 

The most problematic source of systematics in shape measurement is in simply determining an estimate of the ellipticity, which, though not necessarily the true ellipticity of the galaxy, is unbiased when averaged over random source orientations. The primary challenge to this is smearing or distortion of the image due, for example, to atmospheric, camera, or reduction effects. This is characterized by a point spread function (PSF), mapping the true isophote to the measured isophote \citep{ValdesJarvisTyson1983,BonnetMellier1995,KaiserSquiresBroadhurst1995,LuppinoKaiser1997,Kuijken1999,BaconRefregierCloweEllis2000,Kaiser2000,ErbenEtAl2001,BernsteinJarvis2002,VanwaerbekeMellier2003,HirataSeljak2003a,Refregier2003b,JainJarvisBernstein2006,MasseyEtAl2007a}. The PSF can lead to a multiplicative bias in the ellipticity measurement, a systematic over- or under-detection of shear due to smearing out of the image, which is difficult to isolate. The PSF can also be anisotropic, which induces an additional shear. Recently, the use of gravitational lensing of the CMB to calibrate the multiplicative bias in galaxy lensing has been proposed \citep{Vallinotto2012,DasEtAl2013,KitchingEtAl2014}. Other calibration and bias effects have also been discussed, for example, by \cite{HirataSeljak2003a,HeymansHeavens2003,IshakHirataMcdonaldSeljak2004,HutererKeetonMA2005,IshakHirata2005,Ishak,HutererEtAl2006,IlbertEtAl2006,vanwaerbekeEtAl2006,ZhangJ2010}.

To use weak lensing tomography for precise cosmological constraints, and in particular to address other sources of systematics like intrinsic alignment, redshift information for each galaxy is necessary. Uncertainties in determining accurate redshifts for large samples of galaxies and their distribution can lead to systematic errors and biases in constraints from weak lensing measurements. It is infeasible to take true (spectroscopic) redshift measurements for all galaxies in a large survey, and so photometric redshift (photo-z) estimates are used. This has an associated probability distribution function (PDF), which depends on the process by which the photo-z estimate has been arrived at. There are several techniques for discriminating between galaxies at different redshifts (e.g., \cite{ConnollyEtAl1995,Benitez2000,BudavariEtAl2000,CollisterLahav2004,VanzellaEtAl2004,IlbertEtAl2006,SchneiderEtAl2006,Newman2008,BernsteinHuterer2009,ZhangEtAl2010,SheldonEtAl2012,WittmanDawson2012,McQuinnWhite2013,MenardEtAl2013,dePutterEtAl2014,GoreckiEtAl2014} and references therein), including utilizing photometric information for template fitting, using spectroscopic samples to calibrate the full photo-z sample, and using cross-correlations with the galaxy density or spectroscopic samples. A variety of methods for photo-z calibration were recently analyzed by \cite{SanchezEtAl2014}, and the accuracy of photo-z determinations is a significant driver of survey design in order to deal with systematics like intrinsic alignment (e.g., \cite{LaureijsEtAl2011}). 

One of the dominant limitations to precise theoretical modeling of the matter power spectrum and bispectrum at small, nonlinear scales is the impact of baryonic physics (e.g., \cite{White2004,ZhanKnox2004,JingEtAl2006,ZentnerRuddHu2008,MeadEtAl2010,SemboloniEtAl2011,SembaloniEtAl2013,YangEtAl2013,ZentnerEtAl2013,EiflerEtAl2014,VelanderEtAl2014}), which causes changes in small scale (e.g., single halo) clustering relative to a dark matter only model and thus the matter power spectrum and bispectrum. At these strongly nonlinear scales, the impact of baryonic physics on the spectrum is potentially degenerate with the effects of intrinsic alignment. The simultaneous calibration of both effects is thus of great importance to future weak lensing surveys that seek to use the power spectrum or bispectrum at nonlinear scales for cosmological constraints.

\begin{figure}
\center
\includegraphics[width=0.65\textwidth,natwidth=610,natheight=642]{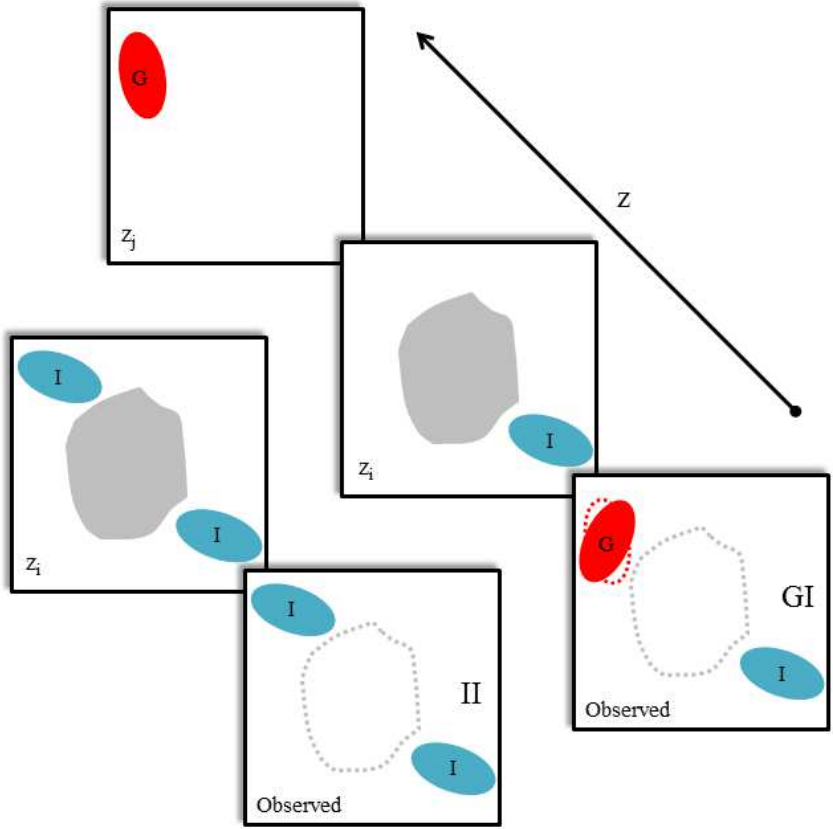}
\caption{A physical representation of the 2-point intrinsic alignment correlations $GI$ and $II$ is shown. Panels to the bottom right of each series represent the observed view of the galaxies on the sky, with unlensed galaxy shapes and the location of the lensing structure shown as dotted outlines. Panels to the upper left in each series represent the relevant physical components at different redshift slices with $z_i<z_j$. The left series of panels shows the $II$ correlation, where two galaxies (labeled $I$ and colored blue) are both intrinsically aligned with the tidal field of a structure (shown in grey) at $z_i$. This will tend to produce a correlation between the galaxy shapes. The right series of panels shows the $GI$ correlation, where a single galaxy is intrinsically aligned by a structure at $z_i$, while a background galaxy (labeled $G$ and colored red) at $z_j>z_i$ is lensed by the same structure. The direction of shearing tends to be orthogonal to the intrinsic alignment, and thus this produces an anti-correlation. Source: Reproduced from \protect\cite{TroxelIshak2012b}.}\label{fig:2ptia}
\end{figure}

              \section{The intrinsic alignment of galaxies}\label{backIA}

The shapes of galaxies, expressed in terms of their ellipticities, can be used to measure the shear $\gamma$ (or alternatively convergence $\kappa$). However, this cosmic shear signal, or extrinsic alignment, can be heavily contaminated by the intrinsic shape or ellipticity of galaxies, which is a much larger contributor to single galaxy shapes than the effects of gravitational shear. There is a dominant, (Gaussian) random component to this intrinsic ellipticity, which does not contribute to the correlation of shapes. There is a second component of the intrinsic ellipticity that is due to the correlated intrinsic alignment of galaxies with the gravitational tidal field of large-scale structure or in local environments with the intrinsic alignment of other galaxies. These correlated alignments can initially be driven by stretching or compression of initially spherically collapsing mass distributions in some gravitational gradient (e.g., \cite{CatelanKamionkowskiBlandford2001}), described in Sec.~\ref{la}, or by the mutual acquisition of angular momentum through tidal torquing \citep{Sciama1955,Peebles1969,Doroshkevich1970,White1984} of aspherical protogalactic mass distributions during galaxy formation, described in Sec.~\ref{qa}. This galaxy ellipticity or angular momentum alignment and the potential for finding correlations in the alignments of galaxies has been extensively studied; see for example \cite{Djorgovski1987} and references therein for an early review of the topic. Beyond these large-scale gravitational mechanisms, there is also evidence from both simulations and measurements of small-scale alignments of galaxies (e.g., Secs.~\ref{detections} \& \ref{sims}) to indicate that nonlinear or baryonic physics, merger history, and gas infall may play a significant role in late-time alignments (or mis-alignments) of galaxies.

To introduce these effects, we can label to first order the measured shear as 
\begin{equation} 
\gamma^{\textrm{obs}}=\gamma+\gamma^I,\label{eq:gammai}
\end{equation}
where $\gamma$ is the true gravitational shear and $\gamma^I$ represents only the correlated part of this intrinsic alignment of galaxies, unlike in Eq. \ref{eq:efrome} where $e^I$ denotes the actual intrinsic ellipticity. Since we are concerned only with the weak limit, we can work with the lensing convergence $\kappa$ instead. From the measured $\gamma^{\textrm{obs}}$, we instead obtain 
\begin{equation} 
\kappa^{\textrm{obs}}=\kappa+\kappa^I. \label{eq:kappai}
\end{equation}
This observed shear or convergence can be propagated through the 2- and 3-point correlation functions to construct several intrinsic alignment correlations, which will be discussed and given a physical description in the following sections.

             \subsection{The 2-point intrinsic alignment correlations} 

For the 2-point correlation function, we will assume that two galaxies $i$ and $j$ with redshifts $z_i$ and $z_j$, respectively, are observed such that $z_i<z_j$. This is demonstrated qualitatively using the simple representation of the correlated intrinsic alignment signal in Eq.~(\ref{eq:gammai}), such that the observed 2-point shear correlation function is actually composed of up to four parts: 
\begin{align}
\langle\gamma^{\textrm{obs}}_i\gamma^{\textrm{obs}}_j\rangle=&\langle\gamma_i\gamma_j\rangle+\langle\gamma^I_i\gamma_j\rangle+\langle\gamma_i\gamma^I_j\rangle+\langle\gamma^I_i\gamma^I_j\rangle.
\end{align}
\begin{itemize}
\item{The first term $\langle\gamma_i\gamma_j\rangle$ represents the true weak lensing component of the measurement, the gravitational shear--gravitational shear correlation, which is often labeled $GG$. This is the component of the observed shear-shear correlation that we seek to use, for example, to constrain a cosmological model or to test gravity on large scales.} 
\item{The second and third terms $\langle\gamma^I_i\gamma_j\rangle+\langle\gamma_i\gamma^I_j\rangle$ are the same physical correlation, but with differently oriented components. They represent the gravitational shear--intrinsic alignment correlation, labeled $GI$. This $GI$ correlation is shown in the right series of panels in Fig.~\ref{fig:2ptia}, where some galaxy $i$ at a redshift $z_i$ is between the observer and galaxy $j$ at redshift $z_j>z_i$.  The structure located at $z_i$ (shown in gray) produces a tidal field that leads to the intrinsic alignment of the nearby galaxy labeled $I$ (shown in blue), while also lensing the background galaxy labeled $G$ (shown in red) at $z_j$. The third term should be negligible in this case if $z_a$ are true redshifts, because the shear signal of galaxy $i$, which is caused by the matter between this galaxy and the observer, should be independent of any background object of sufficient separation.

As depicted in Fig.~\ref{fig:2ptia}, the gravitational shear and intrinsic alignment tend to act in orthogonal directions in a simple alignment model which is driven by interaction with the tidal field, which means that $GI$ is actually an anti-correlation. More complex models that include the impact, for example, due to the merger history of the galaxy or baryonic infall may result in an intrinsic alignment for some galaxies which causes a positive correlation with gravitational shear. This is commented on further in terms of merger rates of late-type galaxies in Secs. \ref{detections} \& \ref{sims}.} 
\item{Finally, the fourth term $\langle\gamma^I_i\gamma^I_j\rangle$ represents the intrinsic alignment -- intrinsic alignment correlation, labeled $II$, and is due to two physically close galaxies ($z_i\approx z_j$) being mutually aligned by the gravitational tidal field of the structures surrounding or near to them. This is shown in the left series of panels in Fig.~\ref{fig:2ptia}, where two galaxies at $z_i$ are both aligned with the tidal field of the structure (shown in gray).}
\end{itemize}
Altogether, the measured shear 2-point correlation can be represented as the sum of
\begin{equation} 
\langle\gamma^{\textrm{obs}}\gamma^{\textrm{obs}}\rangle=GG+IG+GI+II.
\end{equation}
Isolating the intrinsic alignment components $GI$, $IG$, and $II$ from the pure lensing signal $GG$ is a nontrivial exercise, which is discussed in detail in Sec.~\ref{mitigation}.

\begin{figure}
\center
\includegraphics[width=0.675\textwidth]{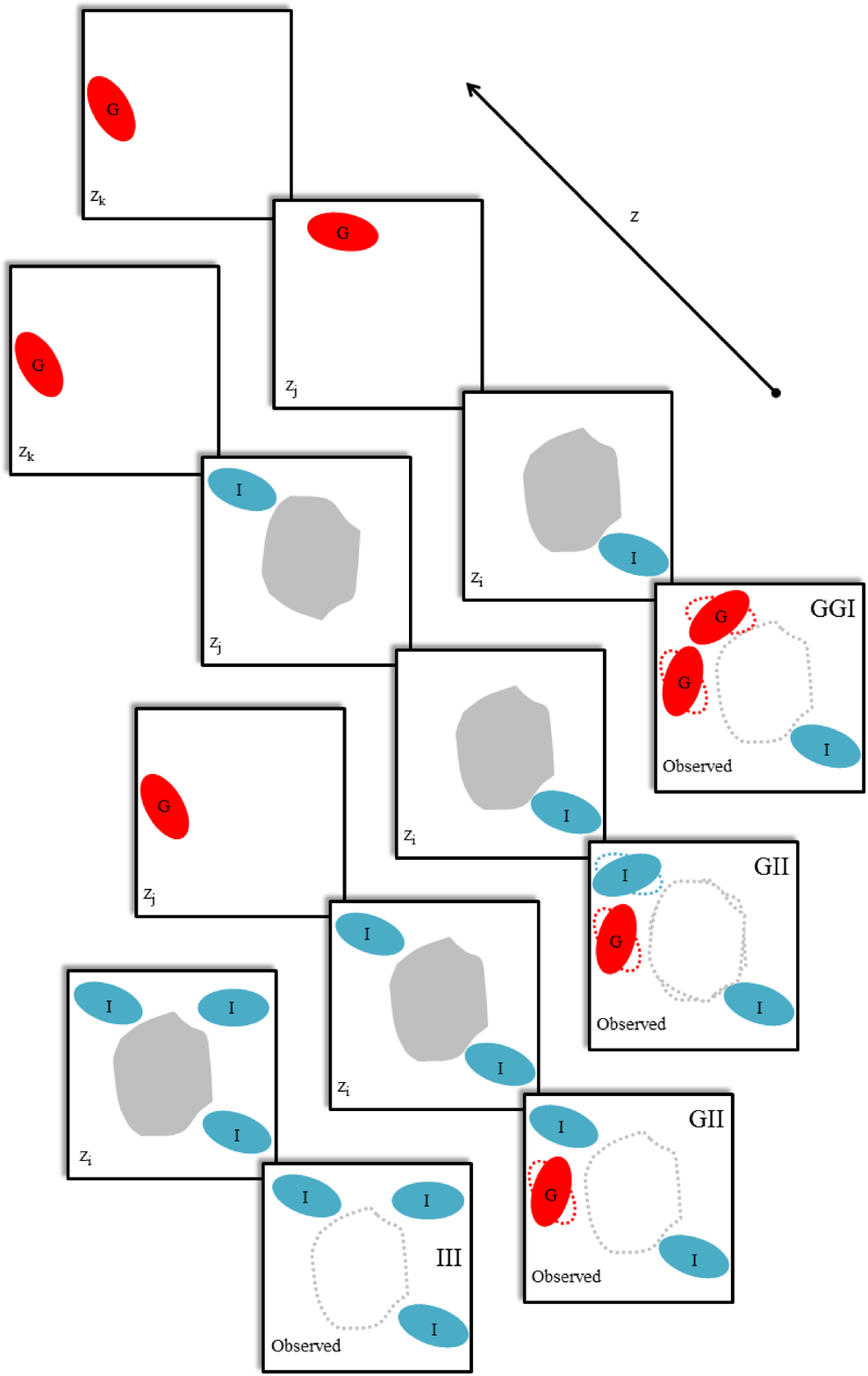}
\caption{A physical representation of the 3-point intrinsic alignment correlations $GGI$, $GII$, and $III$ is shown. Panels to the bottom right of each series represent the observed view of the galaxies on the sky, with unlensed galaxy shapes and the location of the lensing structure(s) shown as dotted outlines. Panels to the upper left in each series represent the relevant physical components at different redshift slices with $z_i<z_j<z_k$. The bottom series of panels shows the $III$ correlation, where three galaxies (labeled $I$ and colored blue) are intrinsically aligned with the tidal field of a structure (shown in gray) at $z_i$. This will tend to produce a correlation between the galaxy shapes. The top series of panels shows the $GGI$ correlation, where a single galaxy is intrinsically aligned by a structure at $z_i$, while two background galaxies (labeled $G$ and colored red) at $z_k,z_j>z_i$ are lensed by the same structure. Finally, the middle series of panels show the $GII$ correlation, where two foreground galaxies at the same or different redshifts are intrinsically aligned by local structures, which in turn lens a background galaxy. Both the $GGI$ and $GII$ correlations can change sign based on triangle shape and scale. Source: Reproduced from \protect\cite{TroxelIshak2012b}.}\label{fig:3ptia}
\end{figure}

\begin{figure}
\center
\includegraphics[height=\textwidth,angle=270]{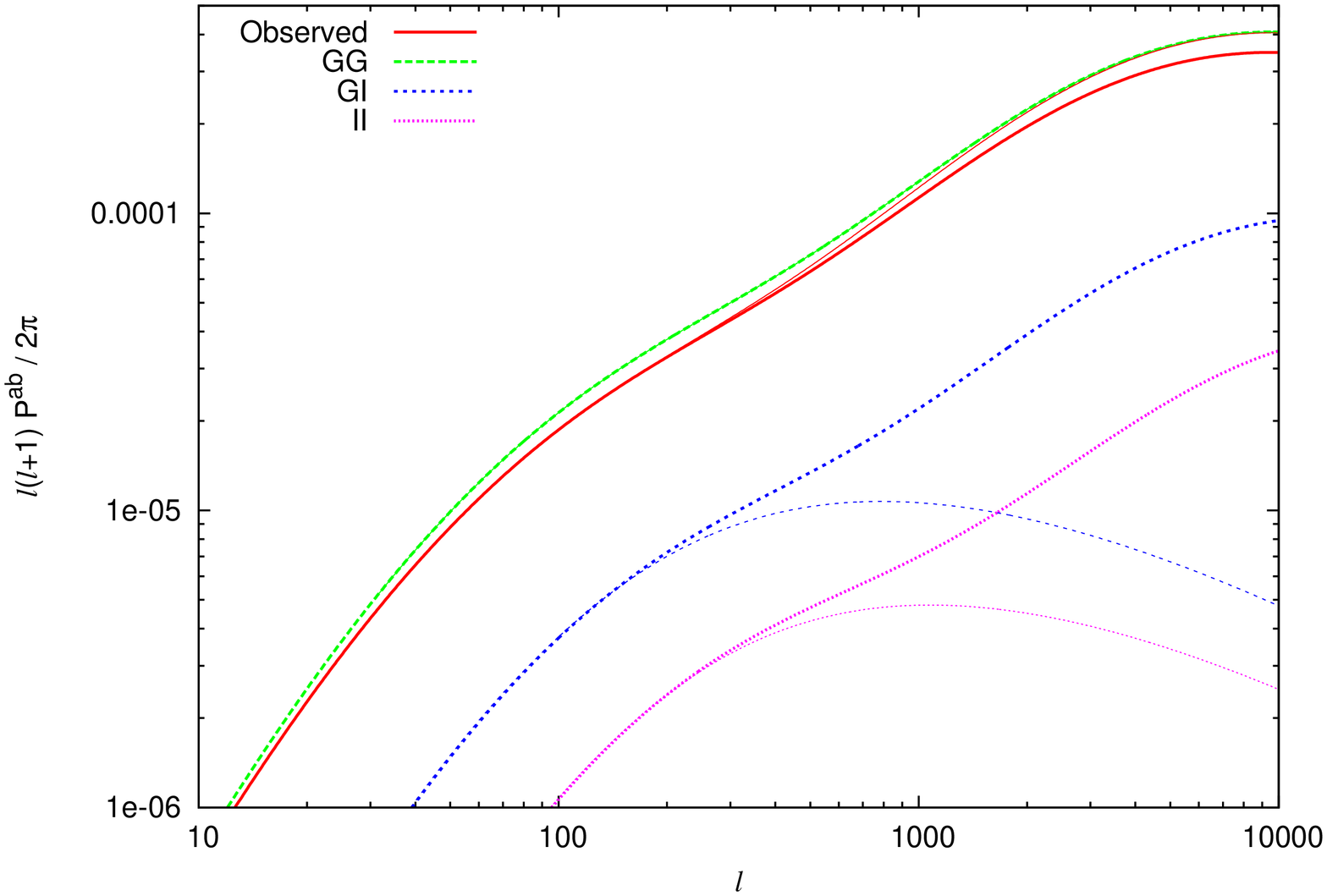}%
\caption{The intrinsic alignment power spectra are shown with their impact on the observed lensing spectrum for the base cosmology discussed in Sec.~\ref{formalisms}. Both the linear alignment model (thin lines) and the ad hoc nonlinear alignment model (thick lines) are shown, with $C_1$ chosen to match the normalization of \protect\cite{HirataSeljak2004}. The models represent the fiducial models discussed in Secs. \ref{la} \& \ref{nla} that agree with low redshift observations of early-type galaxy alignment. The linear alignment model under-predicts the intrinsic alignment signal on small scales (large $\ell$). The magnitude of the impact of intrinsic alignment on the observed lensing spectrum is comparable to the changes in cosmology shown in Fig.~\ref{fig:pscosmo} for even a deep, stage IV survey.}\label{fig:psia}
\end{figure}

           \subsection{The 3-point intrinsic alignment correlations} 

Using the simple representation above, we can also decompose the measured 3-point shear correlation into four parts that represent the associated 3-point intrinsic alignment correlations with $z_i<z_j<z_k$, where permuted terms have been combined:
\begin{align}
\langle\gamma^{\textrm{obs}}_i\gamma^{\textrm{obs}}_j\gamma^{\textrm{obs}}_k\rangle&=\langle\gamma_i\gamma_j\gamma_k\rangle+\langle\gamma^I_i\gamma_j\gamma_k\rangle+2\textrm{ perm.}+\langle\gamma^I_i\gamma^I_j\gamma_k\rangle+2\textrm{ perm.}+\langle\gamma^I_i\gamma^I_j\gamma^I_k\rangle.
\end{align}
\begin{itemize}
  \item {The first term $\langle\gamma_i\gamma_j\gamma_k\rangle$ represents the true weak lensing component of the measurement, labeled $GGG$. This can be combined with the $GG$ correlation above to better constrain cosmological parameters by breaking degeneracies between $\Omega_m$ and $\sigma_8$. It is also able to place limits on the levels of primordial non-Gaussianity, which is not possible at the 2-point level.}
  \item {The second term $\langle\gamma^I_i\gamma_j\gamma_k\rangle$ and permutations represent the $GGI$ correlation, where a galaxy is intrinsically aligned by a nearby matter structure, which in turn contributes to the lensing of two background galaxies. For true redshifts, where $z_i<z_j<z_k$, the permutations are zero or negligible compared to the $\langle\gamma^I_i\gamma_j\gamma_k\rangle$. The $GGI$ correlation is shown in the top series of panels in Fig.~\ref{fig:3ptia}, where a galaxy labeled $I$ (shown in blue) at $z_i$ is intrinsically aligned with the tidal field of a structure (shown in gray). This structure then lenses two galaxies labeled $G$ (shown in red) at redshifts $z_j,z_k>z_i$. Like $GI$, this correlation is typically an anti-correlation, but can change sign based on the triangle shape, scale, and redshift of the galaxy triplet.} 
  \item {The fifth term $\langle\gamma^I_i\gamma^I_j\gamma_k\rangle$ and permutations is labeled $GII$. In this case, two foreground galaxies are intrinsically aligned by structure(s) which in turn lens a background galaxy. This is represented in the two middle series of panels in Fig.~\ref{fig:3ptia}, the upper-most of which is actually a cross-over between the GGI and GII correlations. As shown in these two series of panels, the $GII$ correlation can either be a positive or negative correlation, and the case where $z_i\approx z_j$ generally has a larger magnitude. In simple representations (see Sec.~\ref{modellingbs}), the sign of $GII$ typically changes as a function of $\ell$ between scales where the gravitational shear--intrinsic alignment and the intrinsic alignment--intrinsic alignment contributions dominate. This behavior is dependent, though, on the triangle shape, scale, and redshift of the galaxy triplet.} 
  \item {The last term $\langle\gamma^I_i\gamma^I_j\gamma^I_k\rangle$ is the $III$ correlation between the intrinsic ellipticities of three spatially close galaxies which are intrinsically aligned by a nearby or surrounding structure.}
  \end{itemize}
Altogether, the measured shear 3-point correlation can be represented by the sum 
\begin{equation} 
\langle\gamma^{\textrm{obs}}\gamma^{\textrm{obs}}\gamma^{\textrm{obs}}\rangle=GGG+GGI+2\textrm{ perm.}+GII+2\textrm{ perm.}+III.
\end{equation}
Methods to disentangle these intrinsic alignment ($GGI$, $GII$, and $III$) and gravitational shear ($GGG$) components are also discussed in detail in Sec.~\ref{mitigation}.

\subsection{Analytic description of the intrinsic alignment power spectra and bispectra}

Assuming that one has knowledge of the underlying intrinsic shear field, Limber's approximation \citep{Limber1954,Kaiser1992} can be employed to write the intrinsic alignment power and bispectra described above. Instead of the weighting function in Eq.~(\ref{eq:weightbin}), which is dependent on the lensing efficiency, the contribution of the intrinsic alignment of galaxies is simply proportional to the normalized galaxy density. We will refer to $W^G_i(\chi)$ as the lensing weight from Eq.~(\ref{eq:weightbin}), while $W^I_i(\chi)=W^g_i(\chi)=f_i(\chi)$ is the intrinsic alignment weighting function. This is the same weighting function used in the Limber approximation for the galaxy density, and will be used in Sec.~\ref{sc2} to constrain the intrinsic alignment signal.

The two intrinsic alignment power spectra are then written
\begin{align}
P^{IG}_{ij}(\ell)&=\int_0^{\chi_1}d\chi\frac{W^I_i(\chi)W^G_j(\chi)}{\sin_k^2(\chi)}P_{\delta\bar{\gamma}^I}(k=\frac{\ell}{\sin_k(\chi)};\chi)\\
P^{II}_{ij}(\ell)&=\int_0^{\chi_1}d\chi\frac{W^I_i(\chi)W^I_j(\chi)}{\sin_k^2(\chi)}P_{\bar{\gamma}^I}(k=\frac{\ell}{\sin_k(\chi)};\chi),
\end{align}
where $P_{\bar{\gamma}^I}$ and $P_{\delta\bar{\gamma}^I}$ are some 3D spectra that can be related to the 3D matter power spectrum on large enough scales. Some proposed methods for modeling these spectra are reviewed in Sec.~\ref{models}. We have plotted the $GI$ and $II$ spectra relative to $GG$ in Fig.~\ref{fig:psia} for the intrinsic alignment models discussed in Secs. \ref{la} \& \ref{nla}. For the deep survey described in Sec.~\ref{formalisms}, $II$ is negligible compared to $GI$, which is of order $10\%$ of the lensing signal. The magnitude of the impact of intrinsic alignment on the observed lensing spectrum is comparable to the changes in cosmology shown in Fig.~\ref{fig:pscosmo} for even a deep, stage IV survey. In fact, the given wCDM model in Fig.~\ref{fig:pscosmo} looks very similar to the effects of $GI$ on the spectrum. A deviation from a cosmological constant of 20\% in $w_0$ for the wCDM model changes the magnitude of the spectrum by about $8\%$ at $\ell=1000$, while the reduction due to intrinsic alignment is about $10\%$

The three intrinsic alignment bispectra are similarly written
\begin{align}
B^{IGG}_{ijk}(\ell_1,\ell_2,\ell_3)&=\int_0^{\chi_1}d\chi\frac{W^I_i(\chi)W^G_j(\chi)W^G_k(\chi)}{\sin_k^4(\chi)}B_{\delta\delta\bar{\gamma}^I}(k_1=\frac{\ell_1}{\sin_k(\chi)},k_2=\frac{\ell_2}{\sin_k(\chi)},k_3=\frac{\ell_3}{\sin_k(\chi)};\chi)\\
B^{IIG}_{ijk}(\ell_1,\ell_2,\ell_3)&=\int_0^{\chi_1}d\chi\frac{W^I_i(\chi)W^I_j(\chi)W^G_k(\chi)}{\sin_k^4(\chi)}B_{\delta\bar{\gamma}^I\bar{\gamma}^I}(k_1=\frac{\ell_1}{\sin_k(\chi)},k_2=\frac{\ell_2}{\sin_k(\chi)},k_3=\frac{\ell_3}{\sin_k(\chi)};\chi)\\
B^{III}_{ijk}(\ell_1,\ell_2,\ell_3)&=\int_0^{\chi_1}d\chi\frac{W^I_i(\chi)W^I_j(\chi)W^I_k(\chi)}{\sin_k^4(\chi)}B_{\bar{\gamma}^I}(k_1=\frac{\ell_1}{\sin_k(\chi)},k_2=\frac{\ell_2}{\sin_k(\chi)},k_3=\frac{\ell_3}{\sin_k(\chi)};\chi),
\end{align}
with associated 3D bispectra $B_{\delta\delta\bar{\gamma}^I}$, $B_{\delta\bar{\gamma}^I\bar{\gamma}^I}$, and $B_{\bar{\gamma}^I}$. Unlike the power spectrum, little work has been done to properly characterize these 3D bispectra, though the relationships between the mean intrinsic shear and the underlying density field in Secs. \ref{la} \& \ref{qa} remain applicable. However, \cite{SemboloniHeymansVanwaerbekeSchneider2008} have showed that lensing bispectrum measurements are typically more strongly contaminated by intrinsic alignment compared to the lensing spectrum measurements, with $GGI$ being as large as $15-20\%$ compared to the $GGG$ lensing signal. This is discussed further in Secs. \ref{modellingbs} \& \ref{sims}.

\subsection{Modeling and characterizing the intrinsic alignment correlations}\label{models}

To evaluate the impact of the intrinsic alignment of galaxies on the cosmic shear signal, it is useful to have an analytic description of the associated intrinsic alignment power spectra and bispectra. This requires some physical model to describe the impact of the tidal field due to the underlying density distribution in the universe on the observed shapes of galaxies, similarly to the description of the lensing spectrum or bispectrum in Eq.~(\ref{eq:bspec}) in terms of the matter power spectrum. Simple models of this effect on large scales can be generally broken down into two categories, linear and quadratic alignment models, based on the order at which the alignment responds to the gravitational potential due to the linear density field $\delta_{m}$. These two models are associated with separate physical causes of the intrinsic galaxy alignment. On smaller scales, several attempts have been made to more accurately model the intrinsic alignment, including ad hoc inclusions of the nonlinear matter power spectrum and attempts to build a `halo' or semi-analytic intrinsic alignment model. A brief discussion of fitting functions for the intrinsic alignment signal from simulation measurements is also given in Sec.~\ref{sims}.

\subsubsection{Linear alignment models}\label{la}

The most commonly used model for intrinsic alignment in cosmic shear studies on large scales is the linear alignment model of \cite{HirataSeljak2004,HirataSeljak2010}, which we will refer to generally as the 'linear alignment model', and which is based on the intrinsic alignment prescription of \cite{CatelanKamionkowskiBlandford2001}. The linear alignment model follows a similar argument as linear galaxy bias theory; that is, large-scale correlations or fluctuations in the mean intrinsic ellipticity field of triaxial elliptical galaxies should be due to large-scale fluctuations in the primordial potential in which the galaxy formed during the matter dominated epoch. The relationship between gravitational shear and the linear density perturbation field is known to be
\begin{equation}
\bm{\gamma_i}=(\gamma_{i+},\gamma_{i\times})=\partial^{-2}\int_0^{\infty}d\chi W(\chi,\chi_i)(\partial_x^2-\partial_y^2,2\partial_x\partial_y)\delta_m(\chi\hat{n}_i),
\end{equation}
for partial derivatives with respect to angular position, and density perturbation $\delta_m$ at comoving distance $\chi_i$ and in angular direction $\hat{n}_i$. $W(\chi,\chi_i)$ is the integrand of Eq.~(\ref{eq:weighting}), such that $W(\chi)=\int d\chi_i W(\chi,\chi_i)$.

Following \cite{CatelanKamionkowskiBlandford2001}, one can similarly write a linear relationship between the mean intrinsic shear and the primordial Newtonian potential (or equivalently, the density perturbation field),
\begin{equation}
\bm{\gamma^I}=-\frac{C_1}{4\pi G}(\nabla_x^2-\nabla_y^2,2\nabla_x\nabla_y)S[\Psi_p].\label{eq:gammapsi}
\end{equation}
$S$ is a smoothing filter which removes galaxy-scale fluctuations and $\nabla$ is a comoving derivative. The right hand side of Eq.~(\ref{eq:gammapsi}) is a unique representation of the possible linear and local functions of $\Psi_p$, since higher order derivative terms should be negligible on large scales. The particular choice of smoothing for the potential $\Psi_p$ can also be considered to be a free component of the linear alignment model, but there is as yet no clear best choice in smoothing scale or method. The idea of a linear alignment model is also not specifically an early-time phenomenon, but could refer to linear dependence on the potential at any time. Most discussions of the linear alignment model, however, assume an early-time 'freeze-in' of alignments based on the primordial potential.

$C_1$ is a normalization constant, which has traditionally been determined from observation. A positive $C_1$ corresponds to alignment with the tidal field, in contrast with the tangential shearing due to gravitational lensing. The value of $C_1$ could in principle also be determined by a suitable analytic model of galaxy alignment or through hydrodynamical simulations of sufficient size and resolution to allow the unbiased measurement of individual galaxy shapes.

In Fourier space, the primordial potential can be related to the linear density perturbation field by
\begin{equation}
\Psi_p(\bm{k})=-4\pi G\frac{\bar{\rho}_m(z)}{\bar{D}(z)}a^2k^{-2}\delta_{m}(\bm{k}),
\end{equation}
where $\bar{D}(z)\propto D/a$ is the normalized growth factor. The density-weighted mean intrinsic shear $\bar{\gamma}^I=(1+\delta_g)\gamma^I$ at some redshift can then be written as \citep{HirataSeljak2004,HirataSeljak2010}
\begin{equation}
\bm{\bar{\gamma}^I}(\bm{k})=C_1 a^2\frac{\bar{\rho}}{\bar{D}}\int d^3\bm{k_1}\frac{(k^2_{2x}-k^2_{2y},2k_{2x}k_{2y})}{k_2^2}\delta_{m}(\bm{k_2})\left[\delta^{(3)}(\bm{k_1})+\frac{b_1}{(2\pi)^3}\delta_{m}(\bm{k_1})\right],\label{eq:weightint1}
\end{equation}
where $b_1$ is the linear galaxy bias ($\delta_g=b_1 \delta_m$) and $\bm{k}=\bm{k_1}+\bm{k_2}$ is assumed to lie perpendicular to the line of sight on the x-axis. The density weighting of $\gamma^I$ cannot be ignored as intrinsic alignment typically occurs in environments with galaxies near to each other, where $\delta_g\ge 1$.

The associated E- and B-mode power spectra for $\bar{\gamma}^I$ are then related to the linear matter power spectrum $P_{\delta}$ at some redshift through
\begin{align}
P^{EE}_{\bar{\gamma}^I}(k)=&C_1^2a^4\frac{\bar{\rho}^2}{\bar{D}^2}\left\{P_{\delta}(k)+\frac{b^2_g}{(2\pi)^3}\int d^3\bm{k_1}\left[f_E(\bm{k_1})+f_E(\bm{k_2})\right]f_E(\bm{k_2})P_{\delta}(k_1)P_{\delta}(k_2)\right\},\label{eq:iie}\\
P^{BB}_{\bar{\gamma}^I}(k)=&C_1^2a^4\frac{\bar{\rho}^2}{\bar{D}^2}\frac{b^2_g}{(2\pi)^3}\int d^3\bm{k_1}\left[f_B(\bm{k_1})+f_B(\bm{k_2})\right]f_B(\bm{k_2})P_{\delta}(k_1)P_{\delta}(k_2),\label{eq:iib}
\end{align}
where $f_{(E,B)}(k)=([k_x^2-k_y^2]/k^2,2k_xk_y/k^2)$. To first order in $P_{\delta}$, $P_{\bar{\gamma}^I}$ is purely E-mode, and the cross-power between the mean weighted shear and matter density at some redshift is given by
\begin{equation}
P^{EE}_{\delta\bar{\gamma}^I}(k)=-C_1a^2\frac{\bar{\rho}}{\bar{D}}P_{\delta}(k).\label{eq:ig}
\end{equation}
The B-mode spectrum for $\bar{\gamma}^I$ and second order term in Eq.~(\ref{eq:iib}) are both similar in magnitude, typically an order of magnitude smaller than the leading term in the E-mode spectrum, and are thus often neglected when utilizing the linear alignment model.

\subsubsection{Quadratic alignment models}\label{qa}

For spiral galaxies, the observed ellipticity is due to an inclination of the disk with respect to the line of sight, and thus the orientation of its angular momentum vector. This produces a quadratic relationship between the mean ellipticity and the primordial potential, since a tidal field both causes an anisotropic moment of inertia, leading to the spin up of angular momentum in the galaxy, while also applying a torque. While this theory of tidal torquing is widely used in studies of spiral galaxy alignment and evolution, its validity has not been well demonstrated, which may be an important caveat given the lack of confirmation of such quadratic models through large-scale measurements of intrinsic alignment. The second order contribution to the mean intrinsic shear due to tidal torquing is given by \citep{CatelanKamionkowskiBlandford2001,CrittendenNatarajanPenTheuns2001,MackeyWhiteKamionkowski2002} 
\begin{equation}
\bm{\gamma^I}=C_2(T^2_{x\mu}-T^2_{y\mu},2T_{x\mu}T_{y\mu}),
\end{equation}
with the tidal tensor
\begin{equation}
T_{\mu\nu}=\frac{1}{4\pi G}\left(\nabla_{\mu}\nabla_{\nu}-\frac{1}{3}\delta_{\mu\nu}\nabla^2\right)S[\Psi_p].
\end{equation}
The density weighted mean intrinsic shear is then \cite{HirataSeljak2004}
\begin{equation}
\bm{\bar{\gamma}^I}_X(\bm{k})=C_2a^4\frac{\bar{\rho}^2}{(2\pi)^3\bar{D}^2}\int d^3\bm{k}_1'd^3\bm{k}_2'h_X(\hat{\bm{k}}_1',\hat{\bm{k}}_2')\delta_{m}(\bm{k}_1')\delta_{m}(\bm{k}_2')\left[\delta^{(3)}(\bm{k}_3')+\frac{b_1}{(2\pi)^3}\delta_{m}(\bm{k}_3')\right],
\end{equation}
for $\hat{\bm{k}}_a=\bm{k}_a/|k_a|$ and $\bm{k_3'}=\bm{k}-\bm{k_1'}-\bm{k}_2'$. For $X\in{E,B}$, $h_E=h_{xx}-h_{yy}$ and $h_B=2h_{xy}$, where
\begin{align}
h_{\lambda\mu}(\hat{\bm{u}},\hat{\bm{v}})=\left(\hat{u}_{\mu}\hat{u}_{\nu}-\frac{1}{3}\delta_{\mu\nu}\right)\left(\hat{v}_{\lambda}\hat{v}_{\nu}-\frac{1}{3}\delta_{\lambda\nu}\right).
\end{align}

The E- and B-mode power spectra can then be written as, replacing $h_E$ with $h_B$ for $P^{BB}_{\bar{\gamma}^I}$,
\begin{align}
P^{EE}_{\bar{\gamma}^I}(k)=&C_2^2 a^8 \frac{\bar{\rho}^4}{\bar{D}^4}\Bigg[\frac{2}{(2\pi)^3}\int d^3\bm{k}_1 h^2_E(\hat{\bm{k}}_1',\hat{\bm{k}}_2')P_{\delta}(k_1)P_{\delta}(k_2)\\
&+\frac{2b_1^2}{3(2\pi)^6}\int d^3\bm{k}_1'd^3\bm{k}_2' \left(h_E(\hat{\bm{k}}_1',\hat{\bm{k}}_2')+h_E(\hat{\bm{k}}_2',\hat{\bm{k}}_3')+h_E(\hat{\bm{k}}_3',\hat{\bm{k}}_1')\right)^2P_{\delta}(k_1')P_{\delta}(k_2')P_{\delta}(k_3')\Bigg].\nonumber
\end{align}
We will refer to this prescription for the intrinsic alignment power spectrum as the 'quadratic alignment model'. The quadratic alignment model predicts no $GI$ spectrum for a Gaussian $\delta_m$ and linear galaxy biasing and evolution of the density field, which are frequently made assumptions. The lack of a $GI$ spectrum for the quadratic model is consistent, though, with a null detection of $GI$ in various survey samples of late-type or blue galaxies on large scales, which are discussed in Sec.~\ref{detections}. Due to the relative success of the linear alignment model in measurements using strongly biased early-type or red galaxies and its ease of use in making predictions, the quadratic model has typically received less attention in the literature. Other investigations related to the tidal alignment of spin have also been performed by \cite{Lee2004,SchaeferMerkel2012,GiahiSaravaniSchaefer2014}. For a more thorough discussion of galactic angular momentum and its impact on galaxy alignment, we refer the reader to the review by \cite{Schaefer2009}.

\subsubsection{Modifications to the linear alignment model}\label{nla}

The linear alignment model described above is designed to capture large-scale features of the intrinsic alignment signal for elliptical galaxies, and its applicability at small scales is unclear, where nonlinear physics may enhance or reduce the mean intrinsic shear. This was addressed in an ad hoc way, for example, by modifying the implementation of Eqs. (\ref{eq:iie})-(\ref{eq:ig}) to use the nonlinear matter power spectrum in order to enhance the small scale magnitude of the intrinsic alignment spectra, following a suggestion by \cite{HirataEtAl2007}. This was shown by \cite{BridleKing2007} to better match previous models of intrinsic shear correlations on small scales by \cite{HeavensRefregierHeymans2000,HeymansEtAl2004} (the HRH$^{*}$ model, modified from the original HRH model of \cite{HeavensRefregierHeymans2000}) and recent observations of both $II$ and $GI$ by \cite{MandelbaumEtAl2006}. The HRH$^{*}$ model predicts, for example, the correlated mean intrinsic shear, $w_{++}(r_p)$, along the axis between pairs of galaxies at separation $r_p$ in the plane of the sky. The correlation function $w_{++}$ can be related to $P_{\bar{\gamma}^I}$ by
\begin{equation}
P_{\bar{\gamma}^I}(k)=2\pi\int dr_p w_{++}(r_p)J_0(kr_p)r_p,
\end{equation}
where $J_0$ is a Bessel function of the first kind. In the HRH$^{*}$ model, $w_{++}$ is
\begin{equation}
w_{++}(r_p)=\frac{A}{8\mathcal{R}^2}\int dr_{||}\left[1+\left(\frac{r}{r_0}\right)^{-\gamma_{gg}}\right]\frac{1}{1+(r/B)^2},
\end{equation}
with amplitude $A$, 3D separation $r^2=r_{||}^2+r_p^2$, galaxy clustering parameters $r_0=5.25 h^{-1}$ Mpc and $\gamma_{gg}=1.8$ \citep{HeymansEtAl2004}, and free parameter $B=1 h^{-1}$ Mpc. $\mathcal{R}\approx0.87$ converts the measured ellipticity into a shear. 

The inclusion of the nonlinear matter power spectrum by \cite{BridleKing2007}, often referred to as the 'nonlinear alignment model' (NLA) in the literature, produces much better agreement between the linear alignment model and the HRH$^{*}$ model on small scales ($r_p<2 h^{-1}$ Mpc), as constrained by comparison to observations in \cite{HeymansEtAl2004,MandelbaumEtAl2006}. The name 'nonlinear alignment model' is misleading, however, as the model is not truly nonlinear. To avoid confusion with future attempts to include true nonlinear corrections to the linear alignment models, we will instead refer to the modification to the linear alignment model by \cite{BridleKing2007} as the ad hoc nonlinear alignment model ('ad hoc' NLA), as the nonlinear matter power spectrum includes the late-time nonlinear evolution of the density field in determining the intrinsic alignment redshift and scale dependence. While the inclusion of the nonlinear matter power spectrum produces the desired improvement in fit for the linear alignment model to small scale predictions and measurements, it has no consistent basis in physical theory. See also \cite{BlazekMcquinnSeljak2011} for a discussion of some inconsistencies in the original nonlinear linear alignment model.

In more recent years, further modifications to this approach have been made to attempt to reconcile this (e.g., \cite{KirkRassatHostBridle2011,KirkLaszloBridleBean2012}), late-time clustering linear alignment models (LC-LA), where the intrinsic alignments of galaxies are assumed to be seeded at early times and thus the $II$ term is related to the linear matter power spectrum, while the $GI$ term is related to the geometric mean of the linear and nonlinear matter power spectra, which allows for late-time non-linear evolution of the density field. \cite{HirataEtAl2007,JoachimiMandelbaumAbdallaBridle2011} also modified the linear alignment models to include an additional parameterized luminosity and redshift dependence to the galaxy--intrinsic alignment spectrum, such that (e.g., \cite{JoachimiMandelbaumAbdallaBridle2011})
\begin{equation}
P_{gI}(k,z,L)\equiv A b_1 P_{\delta\bar{\gamma}^I}(k,z)\left(\frac{1+z}{1+z_0}\right)^{\alpha}\left(\frac{L}{L_0}\right)^{\beta}.\label{eq:NLA}
\end{equation}
The reference parameters are chosen to be $z_0=0.3$ and a luminosity $L_0$ corresponding to $M_r=-22$. Best-fit values for various galaxy samples of $\{A,\alpha,\beta\}$ are given in Table 4\footnote{$\alpha$ is denoted $\eta_{other}$ in \protect\cite{JoachimiMandelbaumAbdallaBridle2011}} of \cite{JoachimiMandelbaumAbdallaBridle2011}. Additional redshift scaling was shown to be unnecessary to fit measurements, while there is strong evidence for the inclusion of a luminosity dependence, which is not captured by the base linear alignment models. Recent hydrodynamical simulation measurements by \cite{TennetiEtAl2014b} in the MassiveBlack-II simulation \cite{mbii} qualitatively agree with measurements by \cite{JoachimiMandelbaumAbdallaBridle2011}, with $\alpha$ consistent with zero, but a weaker luminosity dependence characterized by $\beta$. This may be due simply to difference in the galaxy samples used to measure $\beta$ in the two works. Work has also progressed to attempt to describe the intrinsic alignment signal on small scales through alternative methods, like the halo and semi-analytical approaches discussed below. 

\subsubsection{Halo alignment models}\label{haloia}

To provide a more physically motivated basis for expanding the linear alignment model to smaller scales, \cite{SchneiderBridle2010} developed a framework for modeling the impact of galaxy intrinsic alignment through a halo model approach, following earlier work by \cite{SmithWatts2005} to include the effects of triaxial halos and intrinsic alignment into the halo model. The halo model, which views the universe as filled with structure represented by dark matter halos of varying mass, has been very successful for predictions of galaxy clustering \citep{SherrerBertschinger1991,ScoccimarroEtAl2001,CooraySheth2002}. The positions of galaxies are then dependent on the resulting distribution of dark matter. This leads to multiple contributions to the matter power spectrum from correlations between galaxies within a single dark matter halo (`1h' terms) and between different halos (`2h' terms). 

In order to use the halo model approach to calculate the theoretical intrinsic alignment signal, \cite{SchneiderBridle2010} assigned a central galaxy to each halo, which is surrounded by (nearly) radially aligned satellite galaxies. The one- or two-halo terms are then potentially central--central ('cc'), central--satellite ('cs'), or satellite--satellite ('ss') correlations. The two-halo central--central correlation of halo pairs is assumed to follow the linear alignment model for both $P^{2h,cc}_{\bar{\gamma}^I}$ and $P^{2h,cc}_{\delta\bar{\gamma}^I}$ (i.e. Sec.~\ref{la}), where the central galaxy is assumed to be precisely centered in and to share the ellipticity of the parent halo. 

The remaining non-zero one-halo E-mode spectra at some redshift were given by \cite{SchneiderBridle2010} as 
\begin{align}
P^{1h,ss}_{\bar{\gamma}^I}(k)=&\int dm n(m) \frac{\langle N^s_g(N^s_g-1)|m\rangle}{\bar{n}_g^2}|w(k,\theta_k|m)|^2\bar{\gamma}^2(m)\label{eq:1hss}\\
P^{1h,ss}_{\delta\bar{\gamma}^I}(k)=&\int dm n(m) \frac{m}{\bar{\rho}} \frac{\langle N^s_g|m\rangle}{\bar{n}_g}|w(k,\theta_k|m)|\bar{\gamma}(m)u(k|m),
\end{align}
while the non-zero 2-halo terms were given as
\begin{align}
P^{2h,ss}_{\bar{\gamma}^I}(k)=&\int dm_1 n(m_1) \frac{\langle N^s_g|m_1\rangle}{\bar{n}_g}|w(k,\theta_k|m_1)|\bar{\gamma}(m_1)\nonumber\\
&\times\int dm_2 n(m_2) \frac{\langle N^s_g|m_2\rangle}{\bar{n}_g}|w(k,\theta_k|m_2)|^2\bar{\gamma}(m_2) P_{2h}(k|m_1,m_2)\\
P^{2h,ss}_{\delta\bar{\gamma}^I}(k)=&\int dm_1 n(m_1) \frac{\langle N^s_g|m_1\rangle}{\bar{n}_g}|w(k,\theta_k|m_1)|\bar{\gamma}(m_1)\\
&\int dm_2 n(m_2) \frac{m_2}{\bar{rho}}u(k|m_2)  P_{2h}(k|m_1,m_2)P_{2h}(k|m_1,m_2)\nonumber\\
P^{2h,cs}_{\bar{\gamma}^I}(k)=&C_1\frac{\bar{\rho}}{\bar{D}}P_{\delta}^{lin}\int dm n(m) \frac{\langle N^s_g|m\rangle}{\bar{n}_g}b_h(m)|w(k,\theta_k|m)|\bar{\gamma}(m)u(k|m).\label{eq:2hcs}
\end{align}
These depend on the halo-halo power spectrum $P_{2h}$, the linear matter power spectrum $P_{\delta}$, and the 3D density weighted, projected ellipticity of galaxies $\bar{\gamma}^I(r,m,c)=\bar{\gamma}(r,m,c)e^{2i\phi}\sin\theta N_g u(r|m,c)$, which generally depends on position in the halo, halo mass, and concentration. $\bar{\gamma}$ is the magnitude of the projected ellipticity field, $N_g$ is the number of galaxies in the halo, $\bar{n}_g$ is the mean galaxy number density, $u(\bm{r}|m,c)\equiv \rho_{NFW}(\bm{r},m)/m$, $w(k|m)\equiv \bar{\gamma}^I(k|m)/\bar{\gamma}(m)$, and $\langle N^s_g|m\rangle$ and $\langle N^s_g(N^s_g-1)|m\rangle$ are the first and second moments of the distribution of galaxies within a halo of mass $m$.

This basic model assumes that satellite galaxies are exactly radially oriented with respect to the parent halo. \cite{SchneiderBridle2010} demonstrated that the inclusion of a probability distribution of satellite galaxy alignments with respect to the halo radial direction, based in part on the measured probability distribution of alignments by \cite{KnebeEtAl2008} from numerical simulations, causes a systematic multiplicative reduction in the amplitude of the intrinsic correlation by some factor $\bar{\gamma}_{scale}^2=0.21^2$, which is independent of the halo mass. The factor $\bar{\gamma}_{scale}$ captures the dominant, multiplicative impact on the power spectrum of having some fraction of galaxies with a non-radial orientation. Fitting functions were provided for the one-halo terms at some redshift,
\begin{align}
P^{1h,ss}_{\bar{\gamma}^I,fit}(k)=&\bar{\gamma}^2_{scale}\frac{(k/p_1)^4}{1+(k/p_2)^{p_3}}\\
P^{1h,ss}_{\delta\bar{\gamma}^I,fit}(k)=&-\bar{\gamma}_{scale}\frac{(k/p_1)^2}{1+(k/p_2)^{p_3}},
\end{align}
where $p_i(z)=q_{i1}\exp{q_{i2}z^{q_{i3}}}$. Best-fit values for $q_{ij}$ are given in \cite{SchneiderBridle2010}. The component spectra in Eqs. (\ref{eq:1hss})-(\ref{eq:2hcs}), when taken in sum, reduce naturally to the linear alignment model on large scales, while providing a more physically motivated boost to the intrinsic alignment signal on small scales to match requirements from small-scale models based on numerical simulations, as well as recent direct measurements. 

The framework presented by \cite{SchneiderBridle2010} is a necessary first step toward producing more accurate, physically motivated models of intrinsic alignment on smaller scales for use in evaluating its impact in studies of weak gravitational lensing. As mentioned by the authors, improvements are likely possible by relaxing the initial assumptions made in developing the model, including: incorporating appropriately mixed spiral and elliptical populations of central and satellite galaxies, which have been shown to have different contributions to the intrinsic alignment signal (e.g., Secs. \ref{la}-\ref{qa}, \ref{detections}, \ref{sims}); allowing for misalignment between central galaxies and the parent halo, as indicated by numerical simulations (e.g., Sec.~\ref{sims}); developing a more realistic description of the mean projected intrinsic ellipticity function, which includes variation as a function of radius, mass of the halo, and galaxy type; and including an appropriate anisotropic description of satellite distribution and relative alignment with the major/minor axes in a non-spherical halo (see for example, \cite{FaltenbacherEtAl2007} and references therein). Recent work (e.g., \cite{SchneiderFrenkCole2012,TennetiEtAl2014,SifonEtAl2014}) has provided some insight into how these effects can be properly taken into account. Despite significant room for improvement, much of which is dependent on advances in our understanding of galaxy formation and evolution (e.g., the transition between one-halo and two-halo regimes), as well as the capability of providing large hydrodynamical simulations of structure formation, the halo approach for intrinsic alignment modeling disentangles to some degree the impact due to small scale correlations (one-halo) with expectations from linear large-scale physics (two-halo). This allows, for example, the fine resolution study of halos with baryonic and other effects included, to constrain the morphology, scale, mass, and redshift dependence of the input functions for the model.

\subsubsection{Semi-analytic alignment models}\label{sam}

Beyond the halo model approach to building an analytical description of the intrinsic alignment of galaxies, one might also consider a semi-analytical approach that folds in analytical models of intrinsic alignment on large scales, information from observations and dark matter simulations of cosmological scale, and small-scale galaxy and alignment properties from hydrodynamical simulations, which are as yet limited in size. \cite{JoachimiEtAl2013a,JoachimiEtAl2013b} presented such an approach to build a semi-analytical model to describe intrinsic galaxy alignment across a wide range of galaxy properties. We will summarize the process used to design the models, but refer the reader to \cite{JoachimiEtAl2013b} for a full discussion of the models' predictions for intrinsic alignment correlations across galaxy properties and redshifts, as they are too numerous to discuss fully here.

\begin{figure}
\center
\includegraphics[height=.53\textwidth,angle=270]{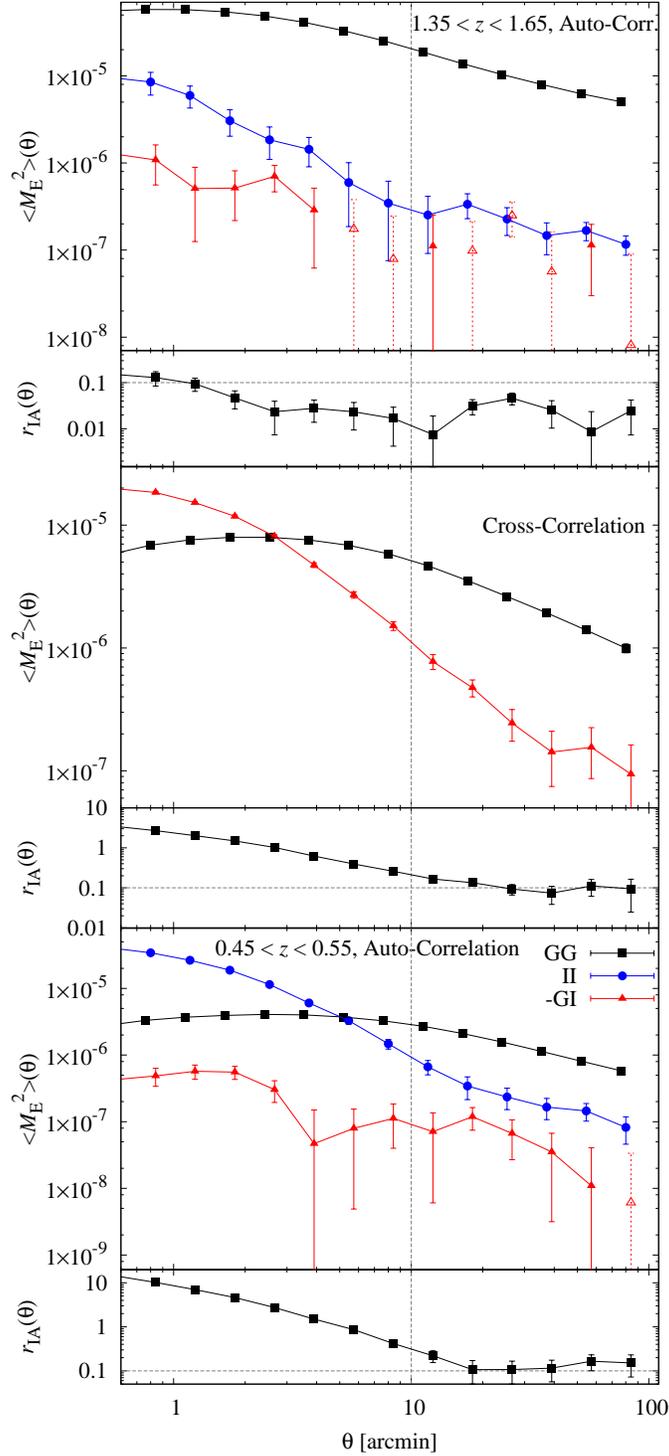}
\caption{The aperture-mass dispersion $\langle M^2_{E}\rangle(\theta)$ for the true weak lensing signal ($GG$ - black squares) and the intrinsic alignment signals ($|GI|$ - red triangles; and $II$ - blue circles) from the semi-analytical model of \protect\cite{JoachimiEtAl2013b} discussed in Sec.~\ref{sam}. Predictions for a future survey with $RIZ<24.5$ and $z_{med}\approx 0.9$ are shown for two redshift bins $0.45 < z < 0.55$ and $1.35 < z < 1.65$. The main panels show the aperture mass dispersion, while small panels show the fractional intrinsic alignment contamination $r_{IA}(\theta)$ relative to the lensing signal ($GG$). Top panels: The auto-correlation in the higher redshift bin. Middle panels: The cross-correlation between the redshift bins. Bottom panels: The auto-correlation in the lower redshift bin. Source: Reproduced with permission from \protect\cite{JoachimiEtAl2013b}, Oxford University Press on behalf of the Royal Astronomical Society.}\label{fig:sam}
\end{figure}

The semi-analytical models are approached from the perspective that we must merge large-scale correlations of galaxy ellipticities, the statistical distribution of galaxy alignments, and small-scale halo satellite alignments to achieve a fully self-consistent description of intrinsic alignment on all scales. The models take into account both early and late-type galaxies, as well as central and satellite galaxies. \cite{JoachimiEtAl2013a,JoachimiEtAl2013b} accomplish this by using dark matter halo properties determined from the Millennium Simulation \citep{SpringelEtAl2005}, but combined with (semi-)analytical models which describe the link between baryonic galaxy shapes and halo properties, including shape and alignment of satellite galaxies, based on hydrodynamical simulations (see Sec.~\ref{hydro}). These potential models are then vetted against constraints on intrinsic alignments from observations (see Sec.~\ref{detections}).

The types of galaxy alignment models used for central galaxies follow previous modeling techniques for early and late-type galaxies. Early type central galaxies are assumed to follow either the halo shape, determined using either the simple (overall halo shape) or reduced (inner halo region given increased weight) inertia tensor, with a galaxy misalignment with the halo shape (e.g., \cite{OkumuraJingLi2009}) included in some cases. Late-type central galaxies are assumed to be aligned perpendicularly to the halo angular momentum axis, with varying disk thickness-to-length ratios, and in some cases a galaxy misalignment (e.g., \cite{Bett2012}). Satellite galaxies, both late and early type, are assumed to have their primary alignment in the direction of the halo center (radial alignment). Late-type satellite galaxies have shape variations similar to the central late-type galaxies, while satellite early type galaxies have a halo distribution following from Millennium Simulation measurements, either from the simple or reduced inertia tensor, with alternate shape modifications following \cite{KnebeEtAl2008}.

An example of these models is shown in Fig.~\ref{fig:sam}, where predictions for a future survey of depth $z_{med}\approx 0.9$ are given in terms of the aperture mass dispersion $\langle M^2_{E}\rangle(\theta)$,
\begin{align}
\langle M^2_{E}\rangle(\theta)=\frac{1}{2}\int_{0}^{2\theta}d\vartheta\frac{\vartheta}{\theta^2}\left[\xi_{+}(\vartheta)T_{+}\left(\frac{\vartheta}{\theta}\right)+\xi_{-}(\vartheta)T_{-}\left(\frac{\vartheta}{\theta}\right)\right],
\end{align}
which is related to the angular correlation functions $\xi_{\pm}$ by the appropriately chosen weight functions $T_{\pm}$ \citep{SchneiderEtAl1998,SchneiderEtAl2002}. Also shown is the fractional intrinsic alignment contamination $r_{IA}(\theta)$
\begin{align}
r_{IA}(\theta)\equiv \frac{|\langle M^2_{E,GI}\rangle(\theta)+\langle M^2_{E,II}\rangle(\theta)|}{\langle M^2_{E,GG}\rangle(\theta)}.
\end{align}
For the lower redshift bin and the cross-correlation, $r_{IA}(\theta)\ge 10\%$ for all angular scales considered, while for the higher redshift bin, $1\%<r_{IA}(\theta)<10\%$, indicating that the impact of intrinsic alignment is still significant in planned surveys when computed using the results of the semi-analytical models derived by \cite{JoachimiEtAl2013a,JoachimiEtAl2013b}.

\subsubsection{Modeling of the intrinsic alignment bispectrum}\label{modellingbs}

The principles discussed above also hold for the analytical modeling of the impact of intrinsic alignment at the level of the bispectrum. However, little work has been done to rigorously explore the intrinsic alignment effects in the bispectrum, since there has until now been no survey capable of making precise enough measurements of the lensing bispectrum to warrant significant concern about the effects of intrinsic alignment in higher order correlations. Several approaches for mitigating the intrinsic alignment have been extended to the bispectrum in preparation for strong detections of the bispectrum predicted in ongoing and future surveys, however, and we describe these in Sec.~\ref{mitigation}.

Measurements of the 3-point intrinsic alignment correlations have been made by \cite{SemboloniHeymansVanwaerbekeSchneider2008} in numerical simulations, following the process developed by \cite{HeymansEtAl22006}. This is discussed in detail in Sec.~\ref{sims}, but it was shown that for low redshift samples ($z_{med}\approx 0.4$) the $III$ component dominates the lensing signal by an order of magnitude, while for deeper surveys ($z_{med}\approx 0.7$), the intrinsic alignment component comprises about $15\%$ of the $GGG$ lensing signal. In both cases, these are significantly stronger contaminations than at the 2-point level, and indicate that work to model, constrain, and mitigate the influence of intrinsic alignment on cosmological constraints using the bispectrum is very important. \cite{SemboloniHeymansVanwaerbekeSchneider2008} also provides fitting formulae for the dominant $GGI$ and $GII$ correlations in deeper surveys, which are discussed in Sec.~\ref{sims}.

In addition to simulation fitting formulae, some initial analytic estimates have also been made for the 3-point intrinsic alignment bispectra. In order to evaluate the performance of the intrinsic alignment self-calibration techniques for the bispectrum, \cite{TroxelIshak2012b,TroxelIshak2012c,TroxelIshak2012a} extended the linear alignment and halo models of Secs. \ref{nla} \& \ref{haloia} for the intrinsic alignment spectrum in an ad hoc way to the bispectrum, propagating the 2-point intrinsic alignment model through the perturbation theory result for the bispectrum given in Eq.~(\ref{eq:bfit}). This process produces reasonable magnitudes for the various 3-point intrinsic alignment correlations compared to \cite{SemboloniHeymansVanwaerbekeSchneider2008}, including a scale-dependent change of sign for the $GII$ correlation for deep surveys. The use of the given effective kernel for the intrinsic alignment bispectra is not physically motivated, however, and significant work is left to be done to appropriately express theoretical intrinsic alignment bispectra. 

\cite{ShiJoachimiSchneider2010} alternately constructed a parameterized toy model for the 3D $GGI$ bispectrum to evaluate the performance of the 3-point nulling technique, which is related to the 3D matter bispectrum such that
\begin{align}
B_{\delta\delta\bar{\gamma}^I}(k_1,k_2,k_3;\chi)\equiv& -A B_{\delta}(k_{ref},k_{ref},k_{ref};\chi{z_{med}})\left(\frac{1+z}{1+z_{med}}\right)^{r-2}\\
&\times\left[\left(\frac{k_1}{k_{ref}}\right)^{2(s-2)}+\left(\frac{k_2}{k_{ref}}\right)^{2(s-2)}+\left(\frac{k_3}{k_{ref}}\right)^{2(s-2)}\right].\nonumber
\end{align}
The matter bispectrum is evaluated at the median redshift $z_{med}$ of the survey and at some scale $k_{ref}$, which was chosen to be weakly nonlinear. $A$ sets the magnitude of the bispectrum relative to simulation results, where $GGI/GGG\approx 10\%$, and $r$ and $s$ are free parameters, with default values of $r=0$ and $s=1$. 

\subsection{Impacts of intrinsic alignment on cosmological constraints} \label{impacts}

The impact of intrinsic alignment on our ability to constrain cosmological models is an evolving question, which will ultimately depend on our understanding of the intrinsic alignment signal and the success of mitigation techniques employed in ongoing and future lensing surveys. The combined development of such techniques and quantifying their success and the total impact of intrinsic alignment is expected to take a central place in work leading up to first science results from these surveys. Work thus far in modeling, measurement, and mitigation of the intrinsic alignment signal, however, indicates that the presence of intrinsic alignment in the lensing signal may be a significant (e.g., Fig.~\ref{fig:sam}), but manageable obstacle in the pursuit of weak lensing as a precision cosmological probe. 

The intrinsic alignment components to the lensing spectrum are generally dependent on both redshift and scale (see Secs. \ref{la}-\ref{modellingbs}), and thus while they tend to introduce an overall bias in the magnitude of the spectrum or bispectrum, this bias is more complex than a simple scaling parameter. It is clear from a comparison of Figs. \ref{fig:pscosmo} \& \ref{fig:psia} that the intrinsic alignment signal can mimic changes in cosmological parameters like $\Omega_m$ or $\sigma_8$ and modifications to the standard $\Lambda$CDM model. For example, a wCDM cosmology with significant departure from a cosmological constant alters the lensing spectrum in ways very similar to the $GI$ intrinsic alignment component. In the deep survey described above, a 20\% change in $w_0$ leads to a reduction in the magnitude of the spectrum by about 8\% at $\ell=1000$, while the presence of intrinsic alignment leads to a similar reduction of about $10\%$.

\begin{figure}
\includegraphics[width=\columnwidth]{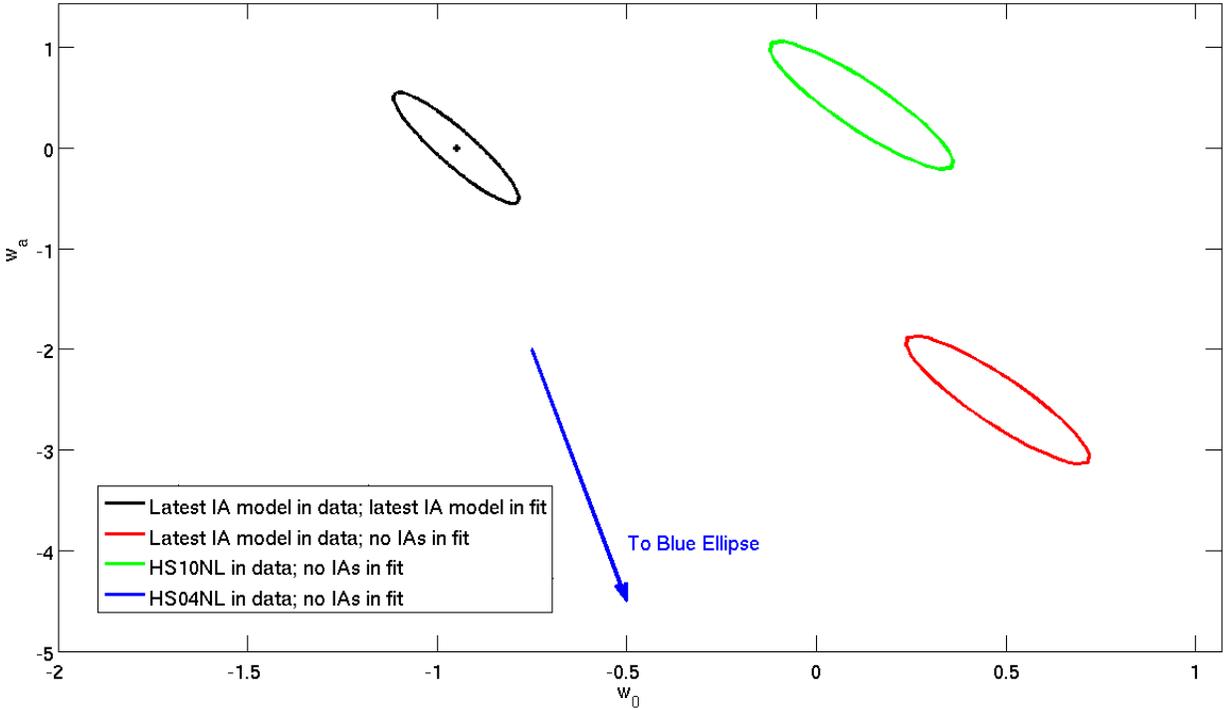}%
\caption{The effects of ignoring the presence of intrinsic alignment on determining the dark energy equation of state for the shear-shear correlation. Shown are 95\% confidence contours for three intrinsic alignment models: the nonlinear linear alignment model described in Sec.~\ref{nla} ('HS04NL'), its corrected version with modified redshift scaling according to \protect\cite{HirataSeljak2010} ('HS10NL'), and the model described in \protect\cite{KirkRassatHostBridle2011} ('latest IA model'). These models differ primarily in their redshift scaling and incorporation of nonlinearities, but share the same magnitude scaling that matches low redshift measurements of intrinsic alignment in bright, early-type galaxies. Not including the effects of intrinsic alignment based on these models in a parameter fit leads to catastrophic biasing of the equation of state parameters, though these results must be considered in the context that the low redshift galaxy sample used to confirm the magnitude of the models is not representative of fainter, high redshift galaxies that make up the majority of galaxies in future surveys. Source: Reproduced with permission from \protect\cite{KirkRassatHostBridle2011}, Oxford University Press on behalf of the Royal Astronomical Society.}\label{fig:iaimpact}
\end{figure}

This impact of intrinsic alignment on dark energy constraints has been quantified, for example, by \cite{BridleKing2007,JoachimiBridle2010,KirkRassatHostBridle2011}. \cite{KirkRassatHostBridle2011} introduced various linear alignment-based models to the data but ignored its effect on constraints, producing significant biases in the determination of $w_0$ and $w_a$ in the dark energy equation of state relative to the case where the model was assumed to be known. This is shown in Fig.~\ref{fig:iaimpact} with 95\% confidence contours for the nonlinear linear alignment model ('HS04NL') \citep{HirataSeljak2004,HirataEtAl2007,BridleKing2007} and its correction with modified redshift scaling ('HS10NL') \citep{HirataSeljak2010}, and the modification discussed in Sec.~\ref{nla} by \cite{KirkRassatHostBridle2011} ('latest IA model'). These models differ primarily in their redshift scaling and incorporation of nonlinearities, but share the same magnitude scaling that matches low redshift measurements of intrinsic alignment in bright, early-type galaxies. This is compared to the fiducial model with a cosmological constant and a proper treatment of the intrinsic alignment signal. Even for the model used by \cite{KirkRassatHostBridle2011}, which predicts a small amplitude of intrinsic alignment contamination, the determination of the dark energy equation of state parameters are catastrophically biased for a Stage IV survey when the intrinsic alignment is ignored. Ignoring the effect of intrinsic alignment, of course, also strongly limits any attempts to test gravity on cosmological scales with cosmic shear \citep{KirkLaszloBridleBean2012}, which is particularly suited for sampling the growth rate of structure across a wide range of redshifts. These conclusions depend, though, on whether the low redshift normalization is sufficient to characterize the actual signal in the fainter, high redshift samples that will form the bulk of galaxies in future surveys.

Impacts on constraints of $\sigma_8$ tend to be less catastrophically biased, on the order of several percent. \cite{HirataEtAl2007} found a constraint of $0.004<\Delta \sigma_8<0.1$, while \cite{MandelbaumEtAl2010} used redshift information from the WiggleZ survey to better constrain this to be near $\Delta \sigma_8=\pm 0.03$ for a Canada France Hawaii Telescope Legacy Survey (CFHTLS)-like survey with other cosmological parameters fixed. Similarly, \cite{JoachimiMandelbaumAbdallaBridle2011} placed estimates of the bias on the dark energy equation of state, $\Omega_m$ and $\sigma_8$ for a CFHTLS-like survey, where for their best case, $\Delta \sigma_8\approx 0.03$, $\Delta \Omega_m\approx-0.03$, and $\Delta w_0\approx 30\%$. 

The above estimates are typically done using variants on the linear alignment model in Secs. \ref{la}\&\ref{nla}, which assumes a linear relationship with large-scale density perturbations. Models like the quadratic model in Sec.~\ref{qa}, instead are produced due to coupled angular momenta, and are constrained from current observations of late-type galaxies to produce an intrinsic alignment correlation that is smaller in magnitude than that predicted by the linear alignment model and measured for early-type galaxies. \cite{CapranicoMerkelSchaefer2012} investigated the effects of these quadratic models for intrinsic alignment on cosmological parameters for a Euclid-like survey, finding negligible bias in the dark energy question of state, and bias at the level of $\approx 2\sigma$ for $\sigma_8$ and $\Omega_m$.  Beyond the use of the shear or convergence spectrum, the impact of intrinsic alignment on the number of false peaks in studies which use convergence maps has also been discussed by \cite{Fan2007}, and in 3D weak lensing studies by \cite{SimonTaylorHartlap2009,MerkelSchaefer2013}.

Finally, the presence of correlations of intrinsic alignment, while they degrade cosmological constraints produced using gravitational lensing, also hold the potential to provide additional or complementary cosmological information. This has begun to be explored in recent years. For example, \cite{SchmidtJeong2012} showed that tensor mode contributions from intrinsic alignment in the linear alignment model is much stronger than that due to gravitational lensing, and could boost the potential of galaxy surveys to constrain a stochastic gravitational wave background. \cite{ChisariEtAl2014a} also explored the impact of tensor modes due to intrinsic alignments. Similarly, \cite{ChisariDvorkin2013} discuss the potential of the cross-correlation between intrinsic alignment and galaxy density to constrain local primordial non-Gaussianity and Baryon Acoustic Oscillations (BAO). The influence of non-Gaussianity on the intrinsic spin alignments of halos is also discussed by \cite{LeePen2008,HuiZhang2008}, while that of gravitational waves on shear measurements including intrinsic alignment is discussed by \cite{SchmidtPajerZaldarriaga2014}.

The intrinsic alignment correlations themselves also are direct measurements of the structure formation history which produces the shape and alignments of the galaxies being sampled (e.g., \cite{LeePen2001}). We have already discovered (e.g., Sec.~\ref{detections}) that the intrinsic alignment signal has a strong dependence on galaxy type. This indicates the strong potential of the intrinsic alignment correlations themselves, when measured over a range of galaxy samples, to provide information about large-scale structure and galaxy formation as a function of galaxy properties, environment, and redshift. While theoretical modeling of how the galaxy formation history impacts the intrinsic alignment of galaxies is not sufficiently advanced to make accurate predictions of the intrinsic alignment signal on smaller scales, direct measurements of these correlations in future surveys will provide needed insight to improve these intrinsic alignment models.

\section{Measurements of the large-scale correlated intrinsic alignment signal}\label{detections}

In recent years there have been many attempts to directly measure the intrinsic alignment of galaxies in correlations within weak lensing surveys. These measurements include detections of both the correlation between the intrinsic ellipticity of galaxies ($II$) and the cross-correlation between intrinsic ellipticity and gravitational shear ($GI$). In general, these detections rely on either limiting measurements to situations where the intrinsic alignment signal is dominant (e.g., using large, low-redshift surveys or galaxies physically close to one another in redshift space) or to developing more complex estimators or algorithms to disentangle the lensing and intrinsic alignment information. We review in this section a number of relevant large-scale measurements or measurement-based predictions of the large-scale intrinsic alignment signal, which are most relevant to developing models or methods to measure or mitigate the intrinsic alignment signal in future large galaxy lensing surveys. There is an additional, large set of literature on measuring the intrinsic alignment properties of galaxies on smaller scales that we have not discussed in detail here, but to which we kindly refer the reader. These related studies are particularly suited for studying the detailed evolution and formation properties of galaxy alignment and its relation to galaxy evolution. 

\subsection{Methods for measuring intrinsic alignment correlations in weak lensing surveys}

\subsubsection{Projected auto- and cross-correlation functions}\label{sec:projected_corr}

Projected correlation functions have been used by many authors to measure the intrinsic alignment signal in various surveys. These correlation functions are defined in real space and can be measured directly in the survey. For galaxy pairs, the  auto- and cross-correlation functions can be written in terms of the ensemble average as 
\begin{align}
\xi_{XX}(\textbf{r})&=\langle \bar{\gamma}_{X}^{I}(\textbf{x})\bar{\gamma}_{X}^{I}(\textbf{x+r})\rangle\nonumber \\ 
\xi_{gX}(\textbf{r})&=\langle \delta_g(\textbf{x})\bar{\gamma}_{X}^{I}(\textbf{x+r})\rangle,
\label{2points_corr}
\end{align}
where $\textbf{r}$ is the separation vector and $X\in\{+,\times\}$. These 2-point correlations and their 3-point equivalents are related to the power spectra and bispectra and the aperture mass statistics by relations similar to Eqs. (\ref{eq:corr2ps}) \& (\ref{eq:apmstat}). The separation vector can be divided in redshift space into a line-of-sight-separation component, $\Pi$, and a transverse separation component, $r_p$. The projected correlations are then defined as the integral along the line of sight
\begin{equation}
w_{g+}(r_p) = \int_{-\Pi}^{+\Pi}\xi_{g+}(r_p,\Pi)\,\rm d \Pi,
\label{projected_corr}
\end{equation}
and similarly for $w_{++}(r_p)$ and  $w_{\times\times}(r_p)$.

One can then measure directly these correlation functions by means of estimators that can be calculated directly from the galaxy position and ellipticity (shape) measurements of a given survey. For example, \cite{MandelbaumHirataIshakSeljakBrinkmann2006,HirataEtAl2007} developed an estimator to measure the correlation functions $\xi$ in a survey by generalizing the usual LS \citep{LandySzalay1993} estimator for the galaxy correlation function
\begin{equation}
\hat\xi(r_p,\Pi) = \frac{(D-R)^2}{RR} = \frac{DD-2DR+RR}{RR},
\label{eq:lsxi}
\end{equation}
where for pairs of galaxies with separation $r_p$ and $\Pi$, $DD$ is the number of galaxy pairs from the survey catalog, $RR$ is the number of galaxy pairs in a random catalog, and $DR$ is the number of pairs between the survey catalog and the random catalog. 

In analogy to Eq.~(\ref{eq:lsxi}), \cite{MandelbaumHirataIshakSeljakBrinkmann2006} defined for the galaxy-intrinsic shear correlation function the estimator 
\begin{equation}
\hat\xi_{g+}(r_p,\Pi) = \frac{S_+(D-R)}{RR} = \frac{S_+D-S_+R}{RR},
\label{eq:lsxids}
\end{equation}
where $S_+D$ is the sum over all pairs with separations $r_p$ and $\Pi$ of the $+$ component of shear, i.e. 
\begin{equation}
S_+D = \sum_{i\neq j| r_p,\Pi} \frac{e_+(j|i)}{2\cal R}, 
\label{S+D}
\end{equation}
with $e_+(j|i)$ being the $+$ component of the ellipticity of galaxy $j$ ($S_+$) measured relative to galaxy $i$ ($D$), and ${\cal R}$ is the shear responsivity. A similar expression to Eq.~(\ref{S+D}) defines $S_{+}R$ with respect to a random catalog galaxy position.

In a similar way, \cite{MandelbaumHirataIshakSeljakBrinkmann2006} used for the intrinsic shear-intrinsic shear correlation functions the estimators
\begin{align}
\hat\xi_{++} &= \frac{S_+S_+}{RR} \nonumber \\
\hat\xi_{\times\times} &= \frac{S_\times S_\times}{RR},
\label{eq:lsxiss}
\end{align}
where
\begin{align}
S_+S_+ &= \sum_{i\neq j| r_p,\Pi} \frac{e_+(j|i)e_+(i|j)}{(2{\cal R})^2} \nonumber \\
S_\times S_\times &= \sum_{i\neq j| r_p,\Pi} \frac{e_\times(j|i)e_\times(i|j)}{(2{\cal R})^2}. 
\end{align}

\subsubsection{Auto- and cross-correlation functions with galaxy orientation angle dependency}\label{statmeas}

In \cite{FaltenbacherLiWhiteJingWang2009}, two new statistical measures of the intrinsic alignment correlations were developed for use in the Sloan Digital Sky Survey \citep{SDSSDR6} and the Millennium Simulation \citep{SpringelEtAl2005}. An alignment correlation function was defined such that pairs of galaxies are summed as a function not only of separation, but also of the projected angle $\theta_p$ between the major axis of the primary galaxy relative to some reference galaxy. The alignment correlation function is then written as $\xi(\theta_p,r_p,\Pi)$. The projected correlation function given by the line-of-sight integration is then 
\begin{align}
w_{p}(\theta_p,r_p) &= \int_{-\Pi}^{+\Pi}\xi_{}(\theta_p,r_p,\Pi)\,\rm d \Pi.
\label{projected_corr_theta}
\end{align}
\cite{FaltenbacherLiWhiteJingWang2009} defined the estimator 
\begin{equation}
\xi(\theta_p,r_p,\Pi)=\frac{Q\tilde{R}/N_{\tilde{R}}}{QR/N_{R}}-1,
\label{projected_corr_theta_est}
\end{equation}
where $Q$ represents the primary galaxy shape sample considered, $\tilde{R}$ is the reference sample, and $R$ represents the random sample. $N_{\tilde{R}}$ and $N_{R}$ are the number of galaxies in the reference and random samples, respectively. $QR$ and $Q\tilde{R}$ are the counts of cross-pairs between the samples as indicated.

The $w_{p}(\theta_p,r_p)$ statistic has been motivated both by use at small scales (sub-Mpc), in order to measure the alignment between central and satellite galaxies in clusters, and also on large scales, where it can measure the alignment of a galaxy sample with respect to the large-scale structure of the universe, which is represented by large-scale distributions of some reference sample of galaxies. This statistic was further discussed and adapted to the linear alignment model (Sec.~\ref{la}) in \cite{BlazekMcquinnSeljak2011}.

\cite{FaltenbacherLiWhiteJingWang2009} also defined a $\cos(2\theta)$-statistic, which describes the average of $\cos(2\theta_p)$ for all galaxy pairs that are considered at a given projected separation 
\begin{equation}
\langle \cos(2\theta_p)\rangle(r_p)=\frac{\int_{0}^{\pi/2} \cos(2\theta_p)w_{p}(\theta_p,r_p)d\theta_p}{\int_{0}^{\pi/2}w_{p}(\theta_p,r_p)d\theta_p}.  
\label{cos_2_theta}
\end{equation}
The values taken by this statistic are well-defined and informative, ranging between -1 and +1. This statistic is related to Eq.~(\ref{projected_corr}) by 
\begin{equation} 
\tilde{w}_{g+}(r_p)=w_p(r_p)\langle \cos(2 \theta_p) \rangle(r_p),
\end{equation}
where $\tilde{w}_{g+}$ is the unweighted $w_{g+}$. $\tilde{w}_{g+}(r_p)$ is related to $w_g(r_p,\theta)$ by
\begin{equation} 
\tilde{w}_{g+}(r_p)=\frac{2}{\pi}\int^{\pi/2}_{0}d\theta \cos(2\theta)w_g(r_p,\theta).
\end{equation}

\subsubsection{E- and B-mode decompositions of intrinsic alignment autocorrelations}

The projected distortion field of galaxy shapes and images as expressed in terms of their ellipticities can be uniquely decomposed in a gradient type component (E-mode) and curl type component (B-mode). Gravitational weak lensing produces only E-modes to first-order in the deflection angle while the intrinsic alignment of galaxies ($II$) produces both E- and B-modes. This decomposition thus allows one to discriminate between the two signals \citep{CrittendenNatarajanPenTheuns2002} (see also Sec.~\ref{ebmode}). As described in \cite{CrittendenNatarajanPenTheuns2002}, the E- and B-mode components of the auto-correlation functions can be written in real space as 
\begin{eqnarray}
\xi_E(r)  &= &\frac{1}{2} [\xi_{+}(r) + \xi_{\times}(r)] 
+ \frac{1}{2}\nabla^4 \chi^{-1} [\xi_{+}(r) - \xi_{\times}(r)] \label{eq:invE} \nonumber\\ 
\xi_B(r) &= & \frac{1}{2}[\xi_{+}(r) + \xi_{\times}(r)] 
- \frac{1}{2}\nabla^4 \chi^{-1} [\xi_{+}(r) - \xi_{\times}(r)]\label{eq:invB},
\end{eqnarray}
where $\nabla^4 = 8D^2 + 8r^2 D^3 + r^4 D^4$, the operator $\chi =
r^4 D^4$ and $D \equiv \frac{1}{r}\frac{\partial}{\partial r}$.

It then follows that the projected E- and B-mode components of the correlation function can be written in terms of a linear combinations of the projected correlations, $w_{\pm}(r_p) \equiv w_{++}(r_p) \pm w_{\times \times}(r_p)$, as \citep{BlazekMcquinnSeljak2011}
\begin{align}
\label{l}
w_{(E,B)}(r_p) = \frac{w_+(r_p) \pm w'(r_p)}{2},
\end{align}
where $w'(r_p)$ is the non-local function of $w_-(r_p)$ given by
\begin{align}
\label{m}
w'(r_p) \equiv  w_-(r_p) + 4 \int_{r_p}^{\infty}dr' \frac{w_-(r')}{r'} -12r_p^2 \int_{r_p}^{\infty}dr' \frac{w_-(r')}{r'^3}.
\end{align}
\cite{HirataSeljak2004,BlazekMcquinnSeljak2011} decomposed the linear alignment model in Sec.~\ref{la} into expressions for the E- and B-mode components in $P(\ell)$ and $w(r_p)$, which predict a zero B-mode to first order in $\delta_m$.

\subsection{Measurements of the large-scale intrinsic alignment signal}\label{det2}

Successful attempts to measure observationally an intrinsic ellipticity signal initially focused on detections of intrinsic ellipticity-intrinsic ellipticity correlations ($II$) in the form of spin-spin correlations (first weakly detected by \cite{PenLeeSeljak2000} in the Tully Catalog) and in the magnitude of the variance of the mean intrinsic galaxy ellipticity of low-redshift galaxies \citep{BrownTaylorHamblyDye2002}, which was found to be non-zero on scales between 1-100 arcmin and consistent with analytical predictions of intrinsic alignment by \cite{CrittendenNatarajanPenTheuns2001}, with an ellipticity variance of
\begin{align}
\label{eq:ellvariancecrit}
\sigma^2(\theta)\approx A z^{-2n}(1+(\theta/\theta_0)2)^{-n}.
\end{align}
The measurement by \cite{BrownTaylorHamblyDye2002} was later extended by \cite{HeymansEtAl2004}. A variety of other early studies of galaxy spin correlations have previously been reviewed by \cite{Schaefer2009} and include, for example, \cite{LeePen2001,LeePen2002,NavarroEtAl2004}. Such studies focused exclusively on correlations between intrinsic alignments of galaxies, either in the form of spin or ellipticity alignment, until the prediction of the long-range correlation between intrinsic ellipticity and the gravitational tidal field ($GI$) by \cite{HirataSeljak2004}. Measurements of $GI$ have generally had more consistent success than those of $II$ in large-scale surveys, where $II$ is often found to be consistent with zero (e.g., \cite{MandelbaumHirataIshakSeljakBrinkmann2006,MandelbaumEtAl2010}) despite generally strong detections of $GI$ in many studies.

\begin{figure}
\begin{center}
\includegraphics[height=.6\textwidth,angle=270]{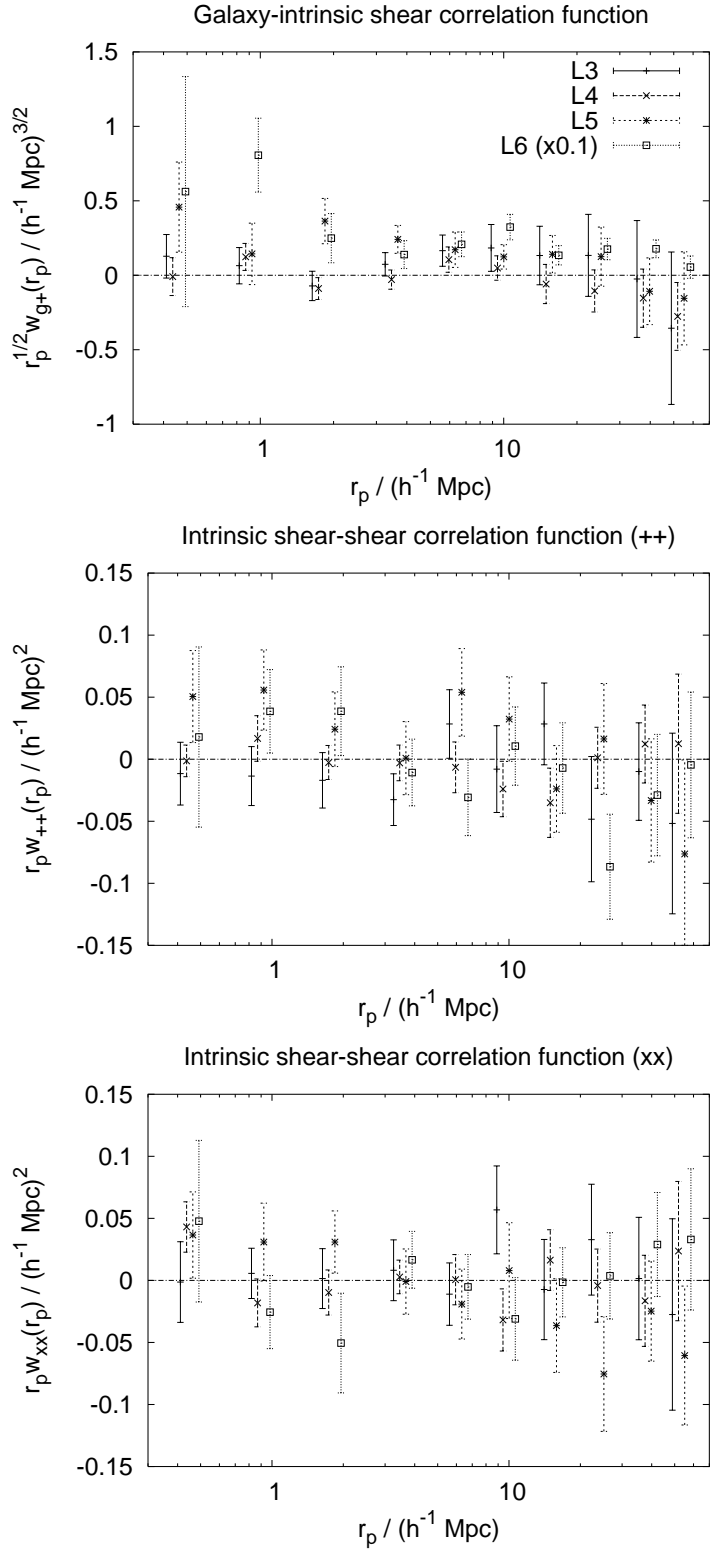}
\caption{\label{fig:IA_SDSS}The projected correlation functions $w_{g+}(r_p)$, $w_{++}(r_p)$, and $w_{\times\times}(r_p)$ obtained from the SDSS L3, L4, L5, and L6 galaxy samples using Pipeline II in \protect\cite{MandelbaumHirataIshakSeljakBrinkmann2006}. Each of the 10 bins contains the same range in $r_p$ for all samples, but some of the error bars have been slightly displaced horizontally for readability (except for L5). The L6 data have been multiplied by 0.1 to fit on the same scale. Errors shown are the 68\% confidence level. Source: Reproduced from \protect\cite{MandelbaumHirataIshakSeljakBrinkmann2006}.}
\end{center}
\end{figure}

This $GI$ correlation was first detected by \cite{MandelbaumHirataIshakSeljakBrinkmann2006}. They used various spectroscopic galaxy samples in the Sloan Digital Sky Survey Data Release 6 (SDSS DR6, \cite{SDSSDR6}) to search for the $II$ and $GI$ correlations. The authors used the projected correlation functions described in Sec.~\ref{sec:projected_corr} to measure $w_{g+}(r_p)$, $w_{++}(r_p)$, and $w_{\times\times}(r_p)$ over a range of transverse pair separations $0.3 < r_p < 60$ $h^{-1}$Mpc. They fit these measured correlation functions to a power-law model of the intrinsic alignment correlations given, for example, by 
\begin{equation}
w_{g+}(r_p) = A\left(\frac{r_p}{1h^{-1}\,{\rm 
Mpc}}\right)^{\alpha},
\label{eq:w-power}
\end{equation}
where $A$ represents the amplitude and $\alpha$ is a power-law exponent. A similar expression for $w_{++}$ and $w_{\times\times}$ can be written. Their results indicated a first detection of the large-scale intrinsic ellipticity-density correlation ($GI$) with confidence level greater than 99\% for galaxies with $L > L_{*}$ in the L5 and L6 galaxy samples (as well as in the overall sample of 265,908 spectroscopic galaxies), with a non-zero amplitude of the correlation function $w_{g+}(r_p)$. The analysis made no significant detection of the $II$ correlation, but was able to place upper limits on its amplitude. These results are summarized for reference in Table \ref{tab:IA_SDSS} and Fig.~\ref{fig:IA_SDSS} \citep{MandelbaumHirataIshakSeljakBrinkmann2006}. The $w_{g+}$ correlation was found to persist to the largest scales probed (i.e., 60 $h^{-1}$Mpc) and to have a sign consistent with theoretical models. This $GI$ signal was found to be dominated by the brightest galaxies, and it was suggested that weak lensing surveys should consider rejection of the brightest cluster galaxies from catalogs to limit contamination by the $GI$ correlation. 

\begin{table}
\begin{center}
\caption{\label{tab:IA_SDSS}Best-fit parameters for the power-law models $A\,\,[r_p/(1 \mbox{Mpc}/h)]^{\alpha}$ to the intrinsic alignment signal \protect\citep{MandelbaumHirataIshakSeljakBrinkmann2006}; the 95 per cent confidence intervals shown here may include no constraint on $\alpha$ if the amplitude is consistent with zero at this level.}
\vspace{5 mm}
\begin{tabular}{c c c c}
\hline\hline
SDSS Sample & function & $A$ ($h^{-1}$Mpc) & $\alpha$ \\
\hline
    & $w_{g+}(r_p)$ & $0.082^{+0.106}_{-0.079}$ & $-0.18^{+\infty}_{-\infty}$ \\
L3  & $w_{++}(r_p)$ & $-0.018^{+0.027}_{-0.025}$ & $-1.13^{+\infty}_{-\infty}$ \\
    & $w_{\times\times}(r_p)$ & $0.005^{+0.030}_{-0.022}$  & $-0.68^{+\infty}_{-\infty}$ \\
\hline
    & $w_{g+}(r_p)$ & $0.020^{+0.115}_{-0.085}$ & $0.013^{+\infty}_{-\infty}$ \\
L4  & $w_{++}(r_p)$ & $(-5.7^{+1972}_{-1314})\times 10^{-5}$ & $-5.5^{+\infty}_{-\infty}$ \\
    & $w_{\times\times}(r_p)$ & $(3.8^{+259}_{-3.8})\times 10^{-4}$ & $-7.1^{+5.8}_{-\infty}$ \\
\hline
    & $w_{g+}(r_p)$ & $0.30^{+0.28}_{-0.22}$ & $-0.66^{+0.54}_{-0.46}$ \\
L5  & $w_{++}(r_p)$ & $0.031^{+0.035}_{-0.031}$ & $-1.9^{+1.2}_{-\infty}$ \\
    & $w_{\times\times}(r_p)$ & $0.011^{+0.030}_{-0.029}$ & $-2.4^{+\infty}_{-\infty}$ \\
\hline
    & $w_{g+}(r_p)$ & $3.8^{+3.5}_{-2.2}$ & $-0.77^{+0.29}_{-0.30}$\\
L6  & $w_{++}(r_p)$ & $0.04^{+0.45}_{-0.48}$ & $-1.8^{+\infty}_{-\infty}$\\
    & $w_{\times\times}(r_p)$ & $-0.25^{+1.05}_{-0.49}$ & $-2.1^{+\infty}_{-\infty}$\\
\hline
    & $w_{g+}(r_p)$ & $0.098^{+0.067}_{-0.069}$ & $-0.59^{+0.65}_{-2.30}$ \\
All & $w_{++}(r_p)$ & $(4.3^{+9.3}_{-4.3})\times 10^{-3}$ & $-2.8^{+\infty}_{-\infty}$ \\
    & $w_{\times\times}(r_p)$ & $(7.2^{+13.0}_{-7.2})\times 10^{-3}$ & $-2.1^{+\infty}_{-\infty}$ \\
\hline\hline
\end{tabular}
\end{center}
\end{table}

\begin{table*}
\begin{center}
\caption{\label{tab:IA_SDSS2} The best-fit model parameters to Eq.~(\ref{eq:wdp}) using SDSS and 2SLAQ LRGs \protect\citep{HirataEtAl2007}. Errors are the 95\% confidence limits. The amplitude $A_0$ is in units of 0.01$h^{-1}\,$Mpc.}
\vspace{5 mm}
\begin{tabular}{c c c c c c}
\hline\hline
Fit region & $A_0/(0.01h^{-1}\,$Mpc$)$ & $\alpha$ & $\beta$ & $\gamma$ &
$\chi^2$/dof\\
\hline
\multicolumn{6}{c}{Fits to SDSS+2SLAQ}\\
\hline
$r_p>11.9h^{-1}\,$Mpc &
$+6.0^{+2.6}_{-2.2}$ &
$-0.88^{+0.31}_{-0.34}$ &
$+1.51^{+0.73}_{-0.69}$ &
$-1.00^{+2.40}_{-3.19}$ &
33.3/28 
\\
$r_p>7.5h^{-1}\,$Mpc &
$+6.4^{+2.5}_{-2.1}$ &
$-0.85^{+0.24}_{-0.25}$ &
$+1.41^{+0.66}_{-0.63}$ &
$-0.27^{+1.88}_{-2.46}$ &
42.8/36\\
$r_p>4.7h^{-1}\,$Mpc &
$+5.9^{+2.3}_{-2.0}$ &
$-0.73^{+0.19}_{-0.19}$ &
$+1.48^{+0.64}_{-0.63}$ &
$-0.56^{+2.02}_{-2.74}$&
54.9/44\\
\hline
\multicolumn{6}{c}{Fits to SDSS only}\\
\hline
$r_p>11.9h^{-1}\,$Mpc &
$+7.1^{+3.4}_{-2.7}$ &
$-0.95^{+0.32}_{-0.35}$ &
$+1.43^{+0.73}_{-0.71}$ &
$+1.94^{+4.75}_{-4.52}$&
21.3/20 \\
$r_p>7.5h^{-1}\,$Mpc &
$+7.4^{+2.9}_{-2.4}$ &
$-0.88^{+0.24}_{-0.25}$ &
$+1.31^{+0.67}_{-0.66}$ &
$+2.39^{+4.52}_{-4.30}$ &
27.9/26 \\
$r_p>4.7h^{-1}\,$Mpc &
$+6.6^{+2.7}_{-2.2}$ &
$-0.74^{+0.19}_{-0.18}$ &
$+1.44^{+0.63}_{-0.62}$ &
$+1.81^{+4.52}_{-4.40}$&
34.0/32\\
\hline\hline
\end{tabular}
\end{center}
\end{table*}

In a follow up analysis to \cite{MandelbaumHirataIshakSeljakBrinkmann2006}, \cite{HirataEtAl2007} explored a more detailed characterization of the $GI$ correlation. They used a combination of samples including 36,278 Luminous Red Galaxies (LRGs) from the SDSS spectroscopic sample with redshift range $0.15 < z < 0.35$, 7,758 LRGs from the 2dF-SDSS LRG and QSQ (2SLAQ) survey with $0.4 < z < 0.8$, and other SDSS subsamples. The formalism of Sec.~\ref{sec:projected_corr} and the pipelines of \cite{MandelbaumHirataIshakSeljakBrinkmann2006} were expanded in this study to explore the correlations as a function of redshift. The results included over 3$\sigma$ detections of the $GI$ correlation 
on large scales up to 60 $h^{-1}$Mpc for all LRG subsamples within the SDSS and a 2-$\sigma$ detection for the bright subsample of the 2SLAQ. They also introduced an empirical parameterization for the large-scale $GI$ correlation of LRGs, with power-law dependence on the galaxy luminosity, redshift, and transverse separation such that 
\begin{equation}
w_{\delta +}(r_p) = A_0 \left(\frac{r_p}{r_{\rm pivot}}\right)^\alpha
\left( \frac L{L_0}\right)^\beta \left(\frac{1+z}{1+z_{\rm 
pivot}}\right)^\gamma.
\label{eq:wdp}
\end{equation}
Here there are 4 parameters $\{A_0,\alpha,\beta,\gamma\}$ and the galaxy luminosity $L$, which do not have the same meaning as the parameterization of the intrinsic alignment signal given in Eq. \ref{eq:NLA}. The normalization $L_0$ corresponds to absolute magnitude $M_r=-22$. The resulting best-fit model parameters $\{A_0,\alpha,\beta,\gamma\}$ are given in Table \ref{tab:IA_SDSS2} \citep{HirataEtAl2007}. 

\cite{LeePen2007} also used the SDSS-DR6 spectroscopic galaxy sample to perform a comparative study of intrinsic alignment measurements between blue and red galaxies. In order to measure the intrinsic alignment signal, the authors used the 2D projection of the intrinsic spin correlation function defined as \citep{LeePen2001}
\begin{equation}
\label{eqn:2dcorr}
\eta_{2D}(r) \approx 
\frac{25}{96}a^2_{\rm l}\frac{\xi^{2}(r;R)}{\xi^{2}(0;R)} + 
\frac{5}{4}\varepsilon_{\rm nl}\frac{\xi(r;R)}{\xi(0;R)},
\end{equation}
where $r$ is the three dimensional separation of a galaxy pair and $\xi(r;R)$ is the 2-point correlation function of the linear density field smoothed on the Lagrangian galactic scale $R$. The parameters $a_{\rm l}$ and $\varepsilon_{\rm nl}$ represent the magnitude of the small- and large-scale correlations, respectively.

The authors selected 434,849 galaxies with axis ratio $b/a \le 0.8$, and found a 3-$\sigma$ detection of the correlation signal in the redshift range $0\le z \le 0.4$ for both blue and red galaxy samples.  
For the bright blue galaxies, the signal followed a quadratic scaling (i.e., $\eta_{2D}(r)\propto \xi^2(r)$), consistent with models of tidal torquing, but had a strong amplitude only up to separations of $r \le 3h^{-1}$Mpc. 
For the bright red galaxies, the scaling was found to be linear, again consistent with theoretical models of large-scale intrinsic alignment, and the signal remained detectable up to larger separations of $r\sim 6h^{-1}$Mpc. Following this work, \cite{Lee2011} used the spectroscopic galaxy sample of SDSS-DR7 \citep{SDSSDR7} to measure the intrinsic spin correlations \citep{PenLeeSeljak2000} using late-type spiral galaxies in the redshift range $0 \le z \le 0.02$. The author found an intrinsic alignment correlation at the 3.4$\sigma$ and 2.4$\sigma$ significance levels at separations of 1$h^{-1}$Mpc and 2$h^{-1}$Mpc, respectively. Measurements were again consistent with analytic models based on tidal torquing for late-type spiral galaxies, and the intrinsic correlations of the galaxy spin axes were found to be stronger than those of the underlying dark halos, consistent with recent findings from simulations (Sec.~\ref{sims}).

In the work of \cite{OkumuraJingLi2009}, the ellipticity correlation function, $c_{ab}(r) = \langle e_a(x) e_b(x + r)\rangle$, was measured using 83,773 spectroscopic LRGs from SDSS-DR6 with redshift range $0.16 \le z \le 0.47$. The authors found a detection of positive alignment between pairs of LRGs at separations of up to 30$h^{-1}$Mpc. They found marginal dependence on luminosity and no significant evidence for redshift dependence. They also considered an N-body simulation to study the effect of misalignment between central LRGs and their host dark matter halos, putting tight constraints on such a misalignment (see Sec.~\ref{misalignment} for further discussion of the misalignment angle). They found that the simulation predicts the same profile for the ellipticity correlations but with an amplitude about 4 times stronger than their measurements if the central LRGs are assumed to be perfectly aligned with their host halos. However, when misalignment is allowed, the authors were able to place a constraint on the misalignment angle dispersion of $\sigma_{\theta}=35.4^{+4.0}_{-3.3} \deg$. The authors stress that this misalignment must be taken into consideration to accurately account for intrinsic alignments in weak lensing surveys. In a follow-up paper, \cite{OkumuraJing2009}, the authors examined whether the $GI$ correlation function of the LRGs in the SDSS-DR6 can be modeled while taking into account the distribution function of the misalignment angle. They made accurate measurements of the $GI$ correlation, confirming previous results which they used in turn to put constraints on the misalignment angle. By fitting the projected correlation function $w_{g+}(r_p)$ to the data they derive the constraints $\sigma_{\theta}=34.9^{+1.9}_{-2.1} \deg$. This is in agreement with the value above from the $II$ correlations but tighter due to better constraints on the $GI$ signal. Additionally, the authors found a correlation between the axis ratios of the LRGs and their intrinsic alignments, an effect that they suggest should be taken into account in modeling intrinsic alignment for future weak lensing surveys.     

After developing the alignment correlation function and the $\cos(2\theta)$ statistics described in Sec.~\ref{statmeas}, \cite{FaltenbacherLiWhiteJingWang2009} investigated the alignment between galaxies and large-scale structure from the SDSS-DR6 \cite{SDSSDR6} and the Millennium Simulation \citep{SpringelEtAl2005}. The authors found a positive detection of the alignment signal for LRGs up to projected separations of 60 $h^{-1}$Mpc, but no large-scale alignment for blue galaxies consistent with the results above in the SDSS. In the Millennium Simulation, they found a mean projected misalignment between a halo and its central region of amplitude $\sim 20^{\textrm{o}}$ that decreases slightly with the luminosity of the central region. When the central region alignments are assigned to the luminous red central galaxies, the simulation results were in agreement with the SDSS results. They found that this misalignment can cause an overestimation of the observed alignment by more than a factor of two. 

\cite{LiJingFaltenbacherWang2013} then measured the intrinsic alignment of galaxies in the CMASS galaxy sample from the Baryon Oscillation Spectroscopy Survey (BOSS-DR9) \citep{AhnEtAl2012,DawsonEtAl2013} at redshift $z\approx 0.6$. The intrinsic alignment 2-point correlation function and the $\cos(2\theta_p)$ statistic (Sec.~\ref{statmeas}) were measured, and they found a significant alignment signal out to approximately 70$h^{-1}$Mpc. Using large-scale simulations, they found alignments of dark halos with masses greater than $10^{12}h^{-1}\rm M_{Solar}$ that have the same scale dependence as the observed galaxies, but with stronger amplitudes. They attribute part of this discrepancy to a misalignment between the central galaxies and their host halos. They also found that more massive galaxies show stronger alignments.

\cite{MandelbaumEtAl2010} combined galaxy shape measurements from SDSS and spectroscopic redshifts from the WiggleZ Dark Energy Survey \citep{DrinkwaterEtAl2010} to search for intrinsic alignment correlations. They used the methodology of \cite{MandelbaumEtAl2006} to attempt to measure the $GI$ correlation at intermediate-redshift ($z\approx 0.6$) for blue galaxies. The result was a null measurement for the full WiggleZ sample as well as for two redshift subsamples. The result allowed them to put upper limits on the contamination of weak lensing measurements by the intrinsic alignment of blue galaxies on large scales. They found that for a CFHTLS-like cosmic shear survey dominated by WiggleZ-like galaxies, there should be a small enough contamination by intrinsic alignment so that any bias on the value of $\sigma_8$ should be smaller than the statistical errors of the survey.

Explicitly taking into account photometric redshift uncertainties in the measurement of intrinsic alignment correlations, \cite{JoachimiMandelbaumAbdallaBridle2011} used the MegaZ-LRG photometric galaxy sample \citep{CollisterEtAl2007,AbdallaEtAl2011} plus various spectroscopic galaxy samples from the SDSS in order to measure galaxy position--shape correlations. The authors combined the SDSS shape measurements with photometric information from the MegaZ-LRG catalog allowing them to derive constraints for early-type galaxies with a redshift range extending to $\sim 0.7$. The authors developed and used a formalism that uses photometric redshifts and includes their scatter in measuring and modeling the galaxy position--shape correlations. The formalism takes into consideration the effect of the photometric redshift scatter on the spread of the number density-intrinsic shear correlations along the line of sight. It also accounts for the effects of other signals such as galaxy-galaxy lensing. The derivation of this photometric formalism is described in Section 4 and Appendix A of \cite{JoachimiMandelbaumAbdallaBridle2011}. The authors used a variant of the correlation estimator in Eq.~(\ref{eq:lsxids}), and the wide ranges in redshift and luminosity of galaxies in the survey allowed the authors to better constrain the redshift and luminosity evolution of the galaxy number density-intrinsic shear correlations, and to extrapolate their results to estimate the contamination from these correlations in future weak lensing surveys. 

For separations larger than $6h^{-1}$Mpc, \cite{JoachimiMandelbaumAbdallaBridle2011} found that these correlations are consistent with the separation and redshift dependencies of a modified nonlinear version of the linear alignment model (see Secs. \ref{la} \& \ref{nla}). In order to better fit observations, an extra redshift and luminosity dependence was introduced, such that the signal is proportional to $(1+z)^{\alpha}$ with $\alpha=-0.3\pm0.8$ and to $L^{\beta}$ with $\beta=1.1^{+0.3}_{-0.2}$ (see Eq.~(\ref{eq:NLA})). The intrinsic alignment power spectrum normalization was found to be $C_1=(0.077\pm 0.008)/\rho_{cr}$ for galaxies at redshift $z=0.3$ and $M_r-22$. The specific values obtained for the various luminosity and redshift sub-samples from MegaZ and SDSS that the authors considered are given in Table 4 of \cite{JoachimiMandelbaumAbdallaBridle2011}. Finally, based on the constraints they derived on the intrinsic alignment correlations, and assuming no intrinsic alignment contribution from blue galaxies, the authors estimated the bias on cosmological parameters as determined from a CFHTLS-like tomographic cosmic shear survey. They found that the bias is smaller than the 1-$\sigma$ statistical errors when all the sub-samples are combined. However, for future weak lensing surveys with significantly higher statistical power, the same intrinsic alignment signal will constitute a serious systematic causing significant degradation in the constraints of cosmological parameters. 

Galaxy position--shape correlation measurements have also been attempted at relatively smaller scales than above within the context of galaxy-galaxy lensing, which has a slightly modified formalism from typical cosmic shear correlations and thus is not reproduced here, where one considers lensing of background galaxies (sources) by foreground galaxies (lenses) (e.g., \cite{BrainerdEtAl1996,FischerEtAl2000,BernsteinNorberg2002,HirataEtAl2004}). Galaxy-galaxy lensing typically provides a stronger signal than cosmic shear, and can suffer from an intrinsic alignment correlation due to satellite objects associated with the lens galaxy being misidentified as source galaxies \citep{HirataEtAl2004,BlazekEtAl2012}. Measurements of galaxy-galaxy lensing can also be used to isolate information on the $GI$ correlation, as described by \cite{BlazekEtAl2012}. For example, \cite{HirataEtAl2004} has put stringent constraints on the intrinsic alignment signal in galaxy-galaxy lensing to be $-0.0062 < \Delta\gamma< +0.0066$ (99.9 per cent confidence) at transverse separations of 30--446 $h^{-1}$ kpc, where $\Delta\gamma$ is the mean intrinsic shear due to intrinsic alignment (see appendix A in \cite{HirataEtAl2004}). In this context of galaxy-galaxy lensing, \cite{BlazekEtAl2012} recently developed a methodology for isolating the effect of $GI$ from a photometric galaxy-galaxy lensing measurement by splitting the sample into two sets of lens-source pairs, allowing for the simultaneous isolation of the intrinsic alignment and lensing contributions to the correlated shear of galaxies. Applying this methodology to a lens sample of SDSS LRGs from DR7 selected by \cite{KazinEtAl2010} and a source sample with shape measurements from the SDSS DR8 photometric data selected by \cite{ReyesEtAl2011}, they find an intrinsic alignment signal consistent with zero. An intrinsic alignment measurement consistent with zero around stacked clusters in photo-z samples was also found by \cite{Chisari2014b}.

\begin{figure}
\centering
\begin{tabular}{c}
\includegraphics[width=.5\columnwidth]{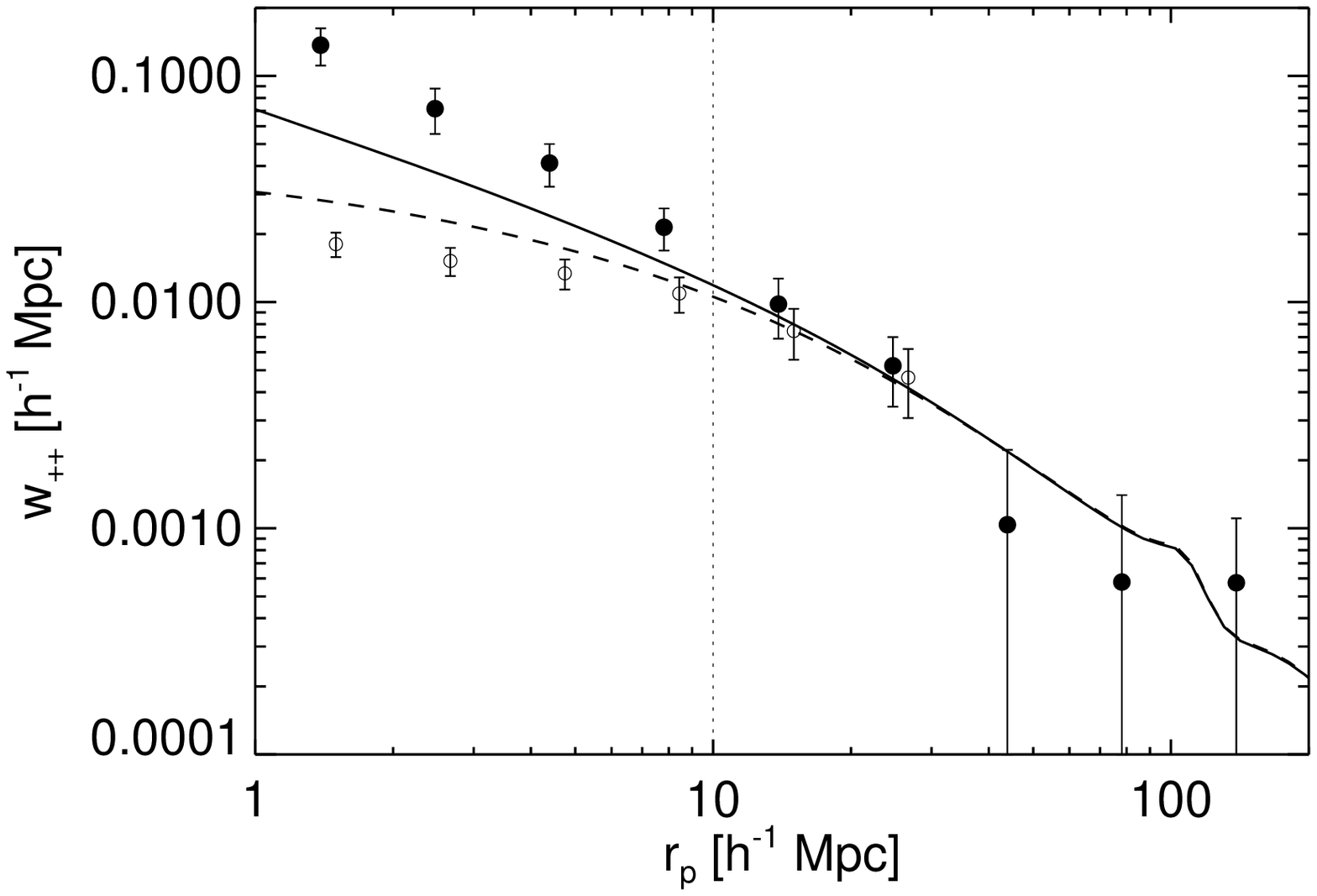}\\
\includegraphics[width=.5\columnwidth]{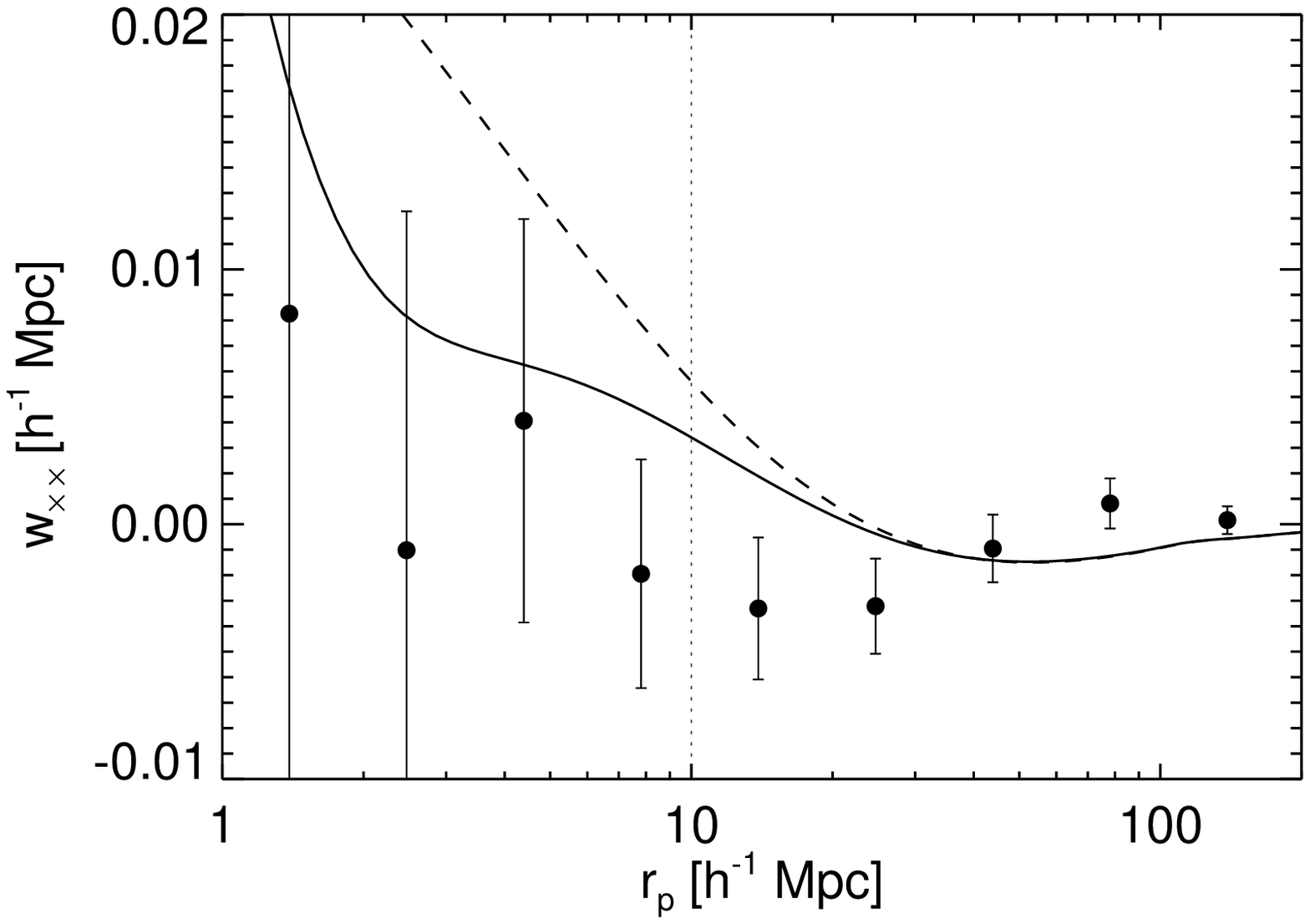}\\
\includegraphics[width=.5\columnwidth]{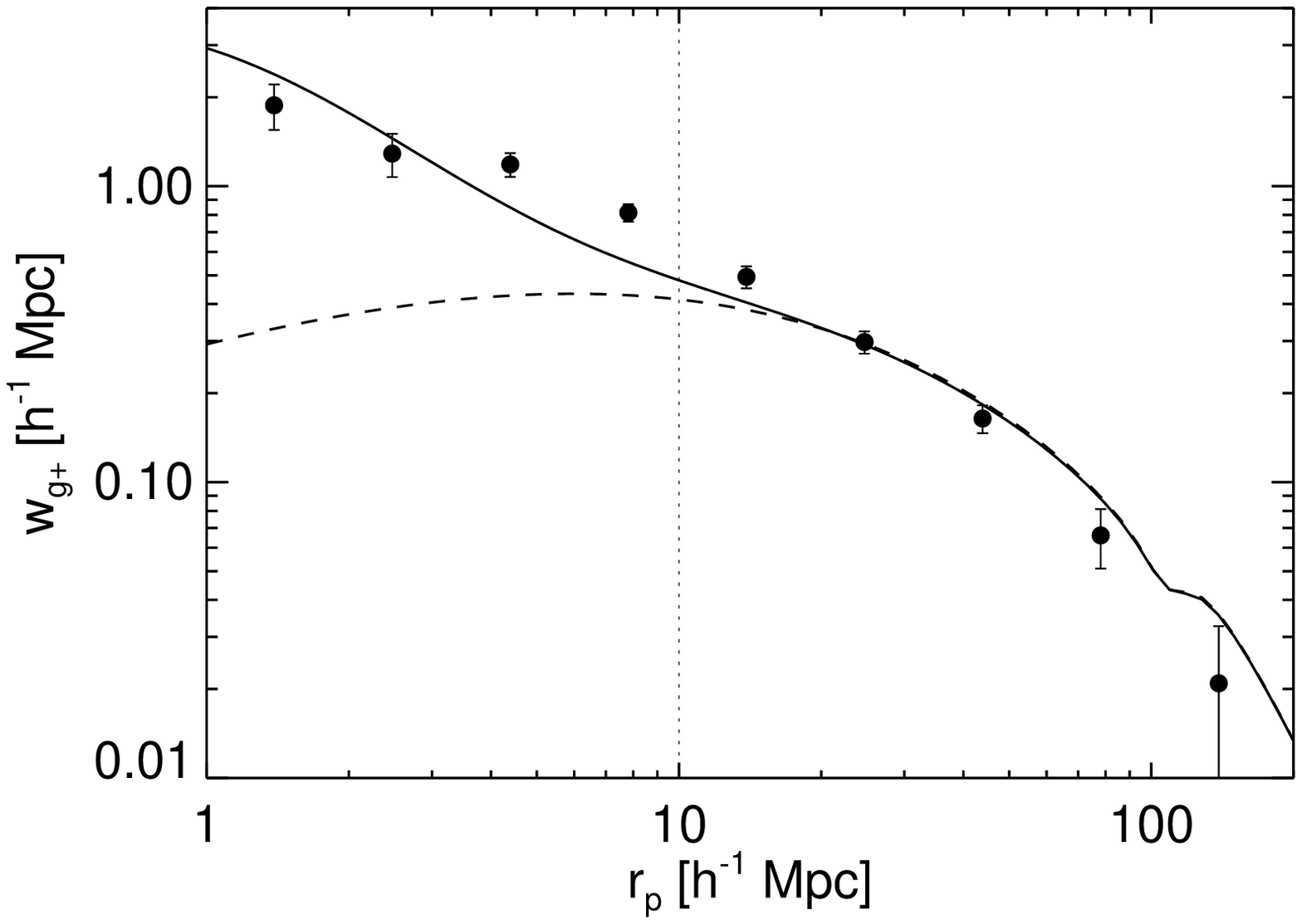}\\
\end{tabular}
\caption{The linear alignment model (dashed lines, Sec.~\ref{la}) and the nonlinear linear alignment model (solid lines, Sec.~\ref{nla}) are compared to observations. The models are normalized to projected measurements of $w_{++}$ in the top two panels and to measurements of $w_{g+}$ in the bottom panel. Top panel: Projected measurements by \protect\cite{OkumuraJingLi2009} of $w_{++}(r)$ are compared to model predictions. Open circles indicate measurements prior to correction by $(1+\xi_g(r))$. Middle panel: Projected measurements by \protect\cite{OkumuraJingLi2009} of $w_{\times\times}(r)$ are compared to model predictions. Bottom panel: Measurements by \protect\cite{OkumuraJing2009} of $w_{g+}(r)$ are compared to model predictions. Source: Reproduced with permission from \protect\cite{BlazekMcquinnSeljak2011}. \copyright 2011 SISSA Medialab Srl. and IOP Publishing. All rights reserved.}\label{fig:compIAmeas}
\end{figure}

These measurements contain a great deal of information related to the large-scale correlations of the intrinsic alignment of galaxies, which has been used to compare to and constrain various models of the intrinsic alignment signal on large scales. These comparisons have been combined to some extent in \cite{BlazekMcquinnSeljak2011}, for example, who recently performed a systematic analysis to test the linear alignment model using several statistics and methods (i.e., the projected correlation $w_{g+}(r_p)$, the alignment correlation function $w_p(\theta_p,r_p)$, the $\cos(2 \theta)$-statistic, and E- and B-mode decomposition). The authors found that the measurements used are generally consistent with the linear alignment model and its predictions for each statistic, and that the linear tidal alignment theory explains well a significant fraction of the observed ellipticity of LRGs on scales greater than or equal to 10 $h^{-1}$Mpc. This is shown in Fig.~\ref{fig:compIAmeas}, where the linear alignment model described in Sec.~\ref{la} (dashed lines) and nonlinear linear alignment model described in Sec.~\ref{nla} (solid lines) are compared to measurements. The top and middle panels of Fig.~\ref{fig:compIAmeas} compare projected measurements from \cite{OkumuraJingLi2009} of $w_{++}(r)$ and $w_{\times\times}(r)$, respectively, to model predictions, while measurements of $w_{g+}(r)$ from \cite{OkumuraJing2009} are compared to model predictions in the bottom panel of Fig.~\ref{fig:compIAmeas}.

Even when not attempting to make a focused measurement of intrinsic alignment in a galaxy shape catalog, surveys with enough statistical power can also take advantage of intrinsic alignment model assumptions to place limits on the level of intrinsic alignment contamination. Using a parameterized intrinsic alignment model (e.g., Sec \ref{margin}), one can simultaneously constrain the model's parameters along with cosmological parameters to obtain a simultaneous fit for the intrinsic alignment along with the cosmological model. CFHTLS \citep{FuEtAl2008} and CFHTLenS \citep{heymans} have both attempted to do this for intrinsic alignment models with a single scaling parameter $A$. \cite{FuEtAl2008} found a value of $A$ consistent with zero, and could place only weak upper limits on the intrinsic alignment contamination. \cite{heymans} were able to detect a nonzero amplitude for early-type galaxies. These approaches necessarily degrade cosmological constraints, as the number of parameters in the intrinsic alignment model increases.

Future measurements of intrinsic alignment in larger and deeper photometric and spectroscopic galaxy samples promise to place even better constraints on models of large-scale intrinsic alignment correlations, which pose a large challenge to the use of weak lensing in planned surveys to place precise and accurate constraints on cosmological parameters and models. The measurements discussed here represent only a selection of works most relevant to the study of intrinsic alignment as it impacts weak lensing, and are not meant to be a complete literature review of galaxy alignment measurements. For example, other works that focus on measurements of galaxy alignment on small or nonlinear scales, such as satellite alignments in clusters of galaxies, are numerous, but not presented here in any detail, as their connection to large-scale correlations of intrinsic alignment are not always clear, and significant work remains before a comprehensive intrinsic alignment modeling scheme (e.g., Secs. \ref{haloia} \& \ref{sam}) can effectively take advantage of such varied small scale measurements. It is likely that they will become important components in the calibration of future intrinsic alignment models tied to structure evolution.

\subsection{Measurements of the small-scale intrinsic alignment signal}

While the main focus of the measurement section above was on large-scale correlations of intrinsic alignment, it is worth mentioning a few comments and references on small-scale measurements. This brief sub-section is meant to provide a starting point for the interested reader in these other, potentially complementary fields of measurement. Each of these varying scales of intrinsic alignment measurement will likely provide input into models of intrinsic alignment in the future, particularly in the 1-halo regime, as measurements become ever more precise.

One interesting example of galaxy alignment measurements that may become impactful as survey size increases is the emerging consensus (following a long history of conflicting claims) in measurements of both spiral and elliptical galaxy alignments relative to large-scale filaments and walls of structure (e.g., \cite{JonesEtAl2010,VarelaEtAl2012,TempelEtAl2013,TempelLibeskind2013} and references therein), which are now becoming consistent with numerical predictions (see Sec.~\ref{sims}). Like these galaxy alignments relative to the large-scale filamentary structure in the universe, measurements of galaxy alignments within clusters have also had conflicting results indicating both random (e.g., \cite{HawleyPeebles1975,Thompson1976,Dekel1985,KampenRhee1990,TreveseEtAl1992,BernsteinNorberg2002,PankoEtAl2009,HaoEtAl2011,HungEbeling2012,SchneiderEtAl2013b,SifonEtAl2014,Chisari2014b}) and non-random (e.g., \cite{RoodSastry1972,Djorgovski1983,GodlowskiEtAl1998,GodlowskiEtAl2010,BaierEtAl2003,PlionisEtAl2003,PereiraKuhn2005,AugustssonBrainderd2006,FaltenbacherEtAl2007}) alignments of galaxies in clusters, despite being the early focus of intrinsic alignment measurements, but without a firm resolution as to the ultimate impact of such alignments on larger scale measurements. Some authors have also warned of the dependence on the methodology for shape measurement in identifying an alignment (e.g., \cite{HaoEtAl2011,SchneiderEtAl2013b}). 

Various alignments between clusters themselves and with their brightest central galaxies or the density field, however, do have some consistent confirmations in various studies (e.g., \cite{FullerEtAl1999,ChambersEtAl2000,ChambersEtAl2002,KimEtAl2002,HopkinsBahcallBode2005,AltayEtAl2006,HashimotoEtAl2008,WangEtAl2009,GodlowskiFlin2010,NiedersteEtAl2010,HaoEtAl2011,PazEtAl2011,SmargonEtAl2012}) after some early conflicting results (e.g., \cite{Binggeli1982,StrublePeebles1985,Flin1987,Lambas1988,UlmerEtAl1989,West1989,Plionis1994}). 

The statistical methods of measuring alignments in or near clusters and other components of large-scale structure are similar to some of those employed to measure the large-scale 2-point intrinsic alignment correlations. These estimators typically are related to the position angles of galaxies, rather than the ellipticity. Two of these in cluster studies are the correlation angle ($\theta_c$), the angle between the projected major axes of two clusters, and the pointing angle ($\theta_p$), the angle between the projected major axis of a cluster and the line connecting it to another cluster on the sky \cite{HopkinsBahcallBode2005}. Quantities like $\cos^2(\theta_c)$ are then measured as a function of separation, as in \cite{SmargonEtAl2012}, for example, to constrain the amount of intrinsic alignment between the structures. Similarly, for galaxies in halos, the radial alignment angle $\theta_r$, the angle between the galaxy major axis and a line connecting it to the Brightest Central Galaxy (BCG), the position angle $\theta_{pos}$, the angle between the BCG major axis and the line connecting the satellite galaxy to the BCG are often employed, and the direct alignment angle $\theta_d$, the angle between the major axes of the BCG and satellite galaxy (e.g., \cite{FaltenbacherEtAl2007}).

\section{Intrinsic alignment correlations from simulations}\label{sims}

With the first reported detections of cosmic shear that were consistent with predictions from large-scale structure (e.g., \cite{BaconRefregierEllis2000,VanwaerbekeEtAl2000,WittmanEtAl2000}), an initial investigation of the potential impact of a correlation between the previously proposed \citep{Coutts1996,LeePen2000,CatelanKamionkowskiBlandford2001} alignments of spatially close galaxies within a common tidal field ($II$) was performed by \cite{HeavensRefregierHeymans2000,CroftMetzler2000}. \cite{CroftMetzler2000}, for example, used an N-body numerical simulation of dark matter halo formation, using the ellipticity of the dark matter halo measured through the second order moment of the projected mass as a tracer for the visible ellipticity of the galaxy, and demonstrated that there was a 10-20\% contribution to the observed ellipticity correlation function due to correlated intrinsic ellipticities of galaxies on scales of up to 20 $h^{-1}$ Mpc (the simulation box size). 

Since this initial detection of correlations between halo ellipticities, the size and scope of numerical simulations have improved dramatically with computational capabilities, but some basic questions regarding intrinsic alignment remain challenging to address in simulations. As commented on by \cite{CroftMetzler2000}, it has since been shown that dark matter halos and their visible galaxies are misaligned to some degree. Indeed, one of the fundamental and still open questions regarding the use of simulations to predict or constrain the large-scale correlated intrinsic alignment signal is the degree to which we can accurately predict or include in a self-consistent way, the subhalo galaxy shapes on large enough scales so as to allow the statistical calculation of shear measurements across the volume spanned by currently planned surveys. This is particularly true for estimates of the $GI$ signal, which has contributions from across the full volume of the simulation, and thus require fine enough resolution in the simulation to accurately account for individual galaxy shapes and the baryonic physics involved, while simultaneously having a large enough volume to produce shear predictions for upcoming surveys. We focus again in this section on work that informs predictions of large-scale intrinsic alignment correlations, discussing small scale measurements in simulations primarily as it relates to our ability to predict the intrinsic shapes of galaxies in large volume simulations.

\begin{figure}
\center
\includegraphics[width=\columnwidth]{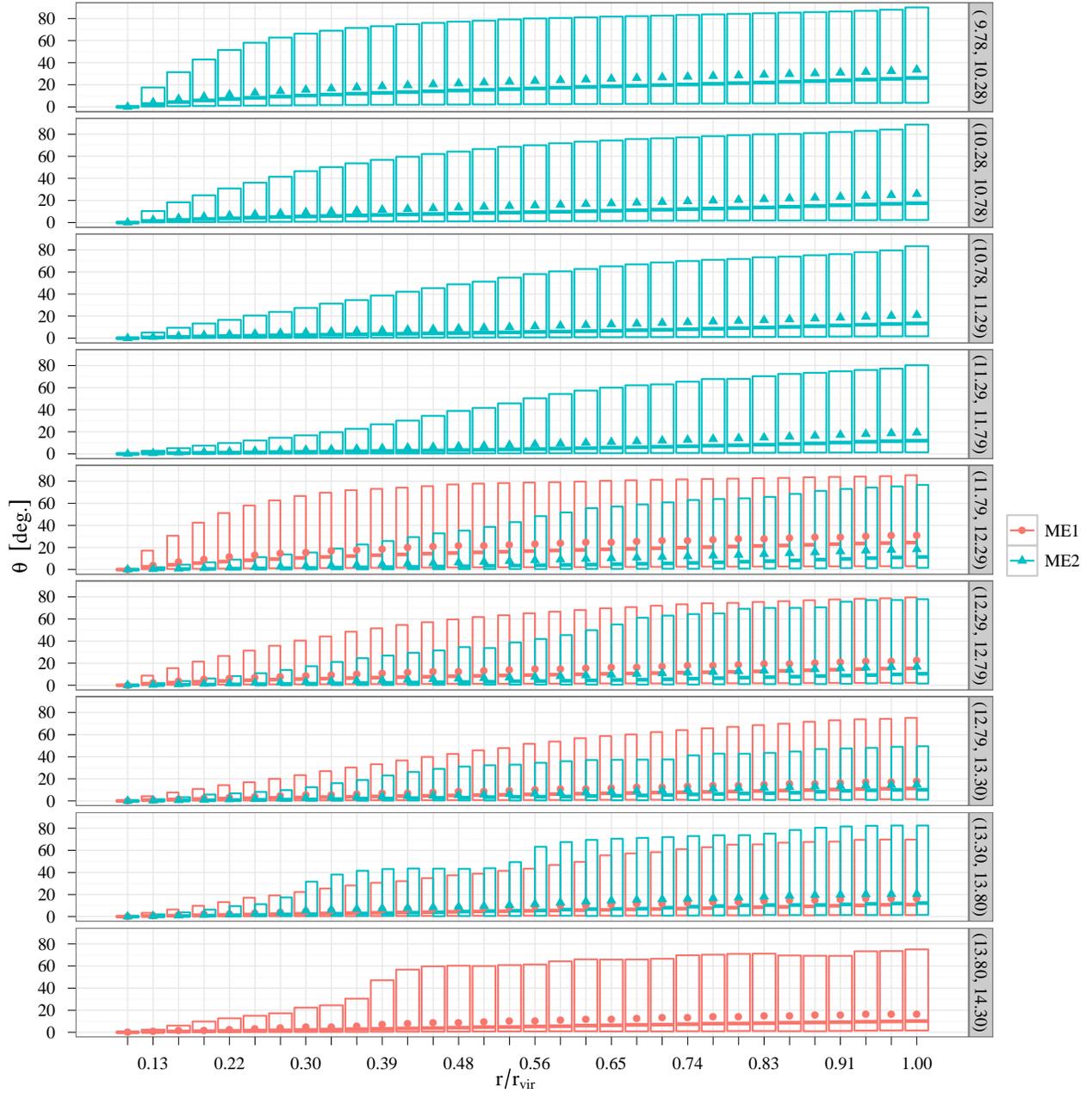}%
\caption{The angular separation between the major axes of a dark matter halo sampled in different radial bins at $z=0.5$ for various halo mass ranges (units of $\log(M_{200}/h^{-1}M_{\odot}))$ in the Millennium \& Millennium-2 Simulations (ME1 \& ME2, respectively). Shown are the mean (points) and median (lines) for each bin, with a box representing the central 50\% quantile range. There is a clear misalignment between inner and outer portions of the halo, with almost 25\% of halos having perpendicularly aligned inner and outer major axes. This misalignment has significant consequences when attempting to draw conclusions about intrinsic alignment in dark matter-only simulations. Source: Reproduced with permission from \protect\cite{SchneiderFrenkCole2012}. \copyright 2012 SISSA Medialab Srl. and IOP Publishing. All rights reserved.}\label{fig:misalign}
\end{figure}

\subsection{Measurements of intrinsic halo alignment correlations in dark matter simulations}\label{dmonly}

Following \cite{HeavensRefregierHeymans2000,CroftMetzler2000}, many large-scale intrinsic alignment measurements have been made in dark matter only simulations. Initial models of the $II$ contribution to intrinsic alignment were developed based partly on simulation results, which included the first identifications of the correlation due to intrinsic spiral and elliptical galaxy shapes by \cite{HeavensRefregierHeymans2000}. Elliptical galaxies were assumed to have the same ellipticity as their dark matter halos, while the disks of spiral galaxies were assumed to be perpendicular to the angular momentum axis of the halo. The spatial $II$ correlation function $\eta^{II}(r)=\langle e(\bm{x})e(\bm{x}+\bm{r})\rangle$, for ellipticity $e$, was fit to the simulation measurement with
\begin{equation}
\eta^{II}_{HRH}(r)=0.012 e^{-r/1.5 h^{-1}\textrm{Mpc}}.\label{eq:hrh}
\end{equation}
These results indicated shear correlations in shallow surveys like SuperCOSMOS and SDSS would be dominated by the intrinsic alignment signal and intrinsic alignments would be nonnegligible in deeper surveys. This was modified by \cite{HeymansEtAl2004} to have the form 
\begin{equation}
\eta^{II}_{HRH*}(r)=\frac{A}{1+(r/B)^2},\label{eq:hrhstar}
\end{equation}
for $B=1 h^{-1}$ Mpc and free parameter $A$, with best fit $A=0.0011$. This measurement forecasted a contamination in the lowest tomographic redshift bin of planned surveys of about ~7\%. \cite{Jing2002} then presented a specific power-law fitting formula for the $II$ correlation as a function of halo mass based on simulation results from \cite{Jing1998,JingSuto1998,JingSuto2002}
\begin{equation}
\eta^{II}_{Jing}(r)=2(3.6\times 10^{-2})\frac{[M_h/(10^{10}h^{-1} \textrm{Mpc})]^{0.5}}{r^{0.4}(7.5^{1.7}+r^{1.7})}.\label{eq:jing}
\end{equation}
These results confirmed that intrinsic alignment could contribute significantly to planned weak lensing surveys. 

Using the dark matter simulations by \cite{ValeWhite2003}, \cite{HeymansEtAl22006} included the potential impact from the $GI$ correlation for the first time, following previous assumptions that galaxy ellipticity follows either the parent halo ellipticity (elliptical galaxies) or angular momentum axis (spiral galaxies) (e.g. \cite{CroftMetzler2000,HeavensRefregierHeymans2000}). They compared results from these two models for galaxy ellipticity to observations by \cite{MandelbaumEtAl2006}, and provided parameterized fitting formulae for the correlation results.

Following \cite{HeymansEtAl2004}, \cite{HeymansEtAl22006} parameterized the $II$ correlation by Eq.~(\ref{eq:hrhstar}), with $B$ left as an additional free parameter for elliptical galaxies. The $GI$ correlation was parameterized to be redshift and scale dependent such that
\begin{equation}
\eta^{GI}(r)=\langle\gamma(\chi_s)e(\chi_l)\rangle=\mathcal{E}\frac{A}{\theta+\theta_0},\label{eq:h06gi}
\end{equation}
where $\chi_s$ is the comoving distance of the source galaxies, $\chi_l$ is the median comoving distance of the lens galaxy bins, $\mathcal{E}=D_l D_{ls}/D_s$ is the lensing efficiency, and there are free parameters for the amplitude $A$ and scale dependence $\theta_0$. For a fully elliptical galaxy sample, \cite{HeymansEtAl22006} found best-fit values of $A=-5.60\times 10^{-7}$ $h^{-1}$ Mpc and $\theta_0=1.83$ arcmin. For a mixed galaxy sample, $A=-1.29\times 10^{-7}$ $h^{-1}$ Mpc and $\theta_0=0.93$ arcmin. This predicts an intrinsic alignment contamination of up to 10\% for surveys with median redshift $z_m\approx 1$ on scales up to 20 arcminutes.

\cite{SemboloniHeymansVanwaerbekeSchneider2008} expanded the investigation of \cite{HeymansEtAl22006} to the 3-point correlations $GGI$, $GII$, and $III$, providing similar fitting formulae for the correlations and demonstrating a stronger intrinsic alignment contamination to the 3-point shear correlation. They assumed that the $GGI$ and $GII$ correlations can be decomposed into functions which depend only on comoving distance $\chi$ and only on angular scale $\theta$, such that
\begin{align}
\eta^{GGI}(\chi_{G_1},\chi_{G_2},\chi_L,\theta)=&E_{GGI}(\chi_{G_1},\chi_{G_2},\chi_L)F(\theta)\\
\eta^{GII}(\chi_{G},\chi_L,\theta)=&E_{GII}(\chi_{G},\chi_L)F(\theta),
\end{align}
where $\chi_{G_a}$ is the comoving distance of the source galaxies and $\chi_L<\min(\chi_{G_a})$ is the maximum lens distance. They then define
\begin{align}
E_{GGI}(\chi_{G_1},\chi_{G_2},\chi_L)=&\int_0^{\chi_L}d\chi'\frac{\sin_k(\chi_{G_1}-\chi')}{\sin_k(\chi_{G_1})}\frac{\sin_k(\chi_{G_2}-\chi')}{\sin_k(\chi_{G_2})}f(\chi')\\
E_{GII}(\chi_{G},\chi_L)=&\int_0^{\chi_L}d\chi'\frac{\sin_k(\chi_{G}-\chi')}{\sin_k(\chi_{G})}f^2(\chi')\\
F(\theta)=&A e^{-\theta/\theta_0},
\end{align}
for some comoving distribution of lenses $f(\chi)$. The best-fit values of $A$ and $\theta_0$ are given in Table 1 of \cite{SemboloniHeymansVanwaerbekeSchneider2008} for several choices of lens distribution and source redshifts. They found generally that intrinsic alignment more strongly contaminates the bispectrum relative to the power spectrum. Other investigations of intrinsic alignment in dark matter only simulations have also been carried out, for example, by \cite{DekelEtAl2001,FaltenbacherEtAl2002,PorcianiEtAl2002a,PorcianiEtAl2002b,AubertPichonColombi2004,HopkinsBahcallBode2005,LeeKangJing2005,AltayColbergCroft2006,AragonEtAl2007,KuhlenEtAl2007,FaltenbacherEtAl2008,LeeEtAl2008,PazStasyszynPadilla2008,PereiraEtAl2008,SousbieEtAl2008,PereiraBryan2010,CodisEtAl2012,AragonEtAl2013,LaigleEtAl2013}.

\subsection{Misalignment of galaxy and halo orientations in simulations}\label{misalignment}

A key question to the validity of many simulation results thus far for galaxy intrinsic alignment is whether dark matter halo properties can reasonably be used as tracers for galaxy alignment. Several attempts have been made to better identify alignments in simulations, and to explore more accurate ways to characterize or measure galaxy alignment. One approach includes attempts to characterize misalignments between dark matter halo and galaxy alignments, which may be stochastic or biased with a non-zero mean. Another is simply to push hydrodynamical simulations to larger volumes in order to measure correlations of galaxy alignments directly, without the need for relating the halo and galaxy alignment axes. The primary challenge to this is the expense of producing large enough volume hydrodynamical simulations to allow for a statistical characterization of the intrinsic alignment signal on large scales. 

Using higher resolution simulations and studying 3D alignments of halos, \cite{BailinSteinmetz2005} reported that angular momentum (disk orientation) and halo misalignment has a nonzero mean of about 25$^{\textrm{o}}$. \cite{FaltenbacherLiWhiteJingWang2009} compared results from the SDSS DR6 \citep{SDSSDR6} and Millennium Simulation \citep{SpringelEtAl2005,DeLuciaBlaizot2007}, finding a better agreement between simulation and survey data when the inner part of the halo is used to determine the projected galaxy alignment, as opposed to the total halo shape, which has a misalignment of about 25$^{\textrm{o}}$ relative to the inner region. By assuming a Gaussian distribution of misalignment angles with zero mean for LRGs and their host halos, \cite{OkumuraJing2009,OkumuraJingLi2009} measured a width of about 35$^{\textrm{o}}$ in the distribution, demonstrating that the assumption of a perfect projected alignment between LRG and parent halo leads to an over-estimation of the ellipticity correlation, in agreement with \cite{FaltenbacherLiWhiteJingWang2009}. Measuring the correlations of 3D dark matter halo shapes in Millennium and Millennium-2 Simulations \citep{SpringelEtAl2005,BoylanEtAl2009}, \cite{SchneiderFrenkCole2012} confirmed the misalignment of inner and outer halo regions with a mean of about 20$^{\textrm{o}}$, which is shown in Fig.~\ref{fig:misalign}. They also provided constraints for the $II$ correlation in terms of their halo-halo alignment correlations based on inner halo shapes, and for the $GI$ correlation in terms of their halo-mass alignment correlations. 

Confirming studies in dark matter simulations to find misalignments in inner and outer regions of a halo, \cite{HahnTeyssierCarollo2010} found similar results with resolved disk galaxies while studying a cosmic filament in a hydrodynamic adaptive mesh refinement (AMR) simulation. The spin vectors of gaseous and stellar disks were well aligned with the inner region of the host halo (median separation angle of 18$^{\textrm{o}}$), while poorly aligned with the total halo spin axis (median separation angle of 46-50$^{\textrm{o}}$). The alignment was also environment-dependent, with low density environments being more consistent with alignment in linear tidal torque theory, and higher density environments having less alignment, potentially due to nonlinear effects in high-density regions. 

\begin{table}
\center
\caption{Central and satellite group mean 3D misalignments for three halo mass bins at various redshifts by \protect\cite{TennetiEtAl2014} in the MassiveBlack-II hydrodynamical simulation \protect\citep{mbii}.\label{mbiialign1}}
\vskip .5cm\begin{tabular}{ l c c c c c c }
\hline\hline
 & \multicolumn{3}{c}{Central Galaxies} & \multicolumn{3}{c}{Satellite Galaxies} \\
 Halo Mass $(h^{-1} M_{\odot})$ & $z=0.06$ & 0.3 & 1.0 & $z=0.06$ & 0.3 & 1.0 \\
\hline
$10^{10}-10^{11.5}$ & $38.12^{\textrm{o}}$ & $37.39^{\textrm{o}}$ & $33.88^{\textrm{o}}$ & $35.60^{\textrm{o}}$ & $32.71^{\textrm{o}}$ & $32.88^{\textrm{o}}$ \\
$10^{11.5}-10^{13}$ & $29.10^{\textrm{o}}$ & $26.61^{\textrm{o}}$ & $21.98^{\textrm{o}}$ & $29.32^{\textrm{o}}$ & $28.52^{\textrm{o}}$ & $27.76^{\textrm{o}}$  \\
$>10^{13}$ & $14.76^{\textrm{o}}$ & $13.47^{\textrm{o}}$ & $10.33^{\textrm{o}}$ & $27.36^{\textrm{o}}$ & $26.48^{\textrm{o}}$ & $26.10^{\textrm{o}}$ \\
\hline\hline
\end{tabular}
\end{table}

\begin{table}
\center
\caption{Central and satellite group mean 3D misalignments for three subhalo mass bins at various redshifts by \protect\cite{TennetiEtAl2014} in the MassiveBlack-II hydrodynamical simulation \protect\citep{mbii}.\label{mbiialign2}}
\vskip .5cm\begin{tabular}{ l c c c c c c }
\hline\hline
 & \multicolumn{3}{c}{Central Galaxies} & \multicolumn{3}{c}{Satellite Galaxies} \\
 Subhalo Mass $(h^{-1} M_{\odot})$ & z=0.06 & 0.3 & 1.0 & z=0.06 & 0.3 & 1.0 \\
\hline
$10^{10}-10^{11.5}$ & $37.83^{\textrm{o}}$ & $37.07^{\textrm{o}}$ & $33.42^{\textrm{o}}$ & $29.00^{\textrm{o}}$ & $28.22^{\textrm{o}}$ & $28.21^{\textrm{o}}$ \\
$10^{11.5}-10^{13}$ & $28.68^{\textrm{o}}$ & $25.85^{\textrm{o}}$ & $21.30^{\textrm{o}}$ & $21.54^{\textrm{o}}$ & $20.43^{\textrm{o}}$ & $18.03^{\textrm{o}}$  \\
$>10^{13}$ & $14.00^{\textrm{o}}$ & $13.11^{\textrm{o}}$ & $9.61^{\textrm{o}}$ & $12.03^{\textrm{o}}$ & $11.73^{\textrm{o}}$ & $17.17^{\textrm{o}}$ \\
\hline\hline
\end{tabular}
\end{table}

More recently, \cite{TennetiEtAl2014} used the MassiveBlack-II hydrodynamical simulation \citep{mbii} to explore 2D and 3D alignments of stellar and dark matter components of halos and subhalos. They define a misalignment angle between the major axes of the ellipsoid defining both stellar and dark matter components of the halo. Specific results for the misalignment angle of central and satellite halos as a function parent and subhalo masses are given in Tables \ref{mbiialign1} \& \ref{mbiialign2}, which indicate a typical misalignment angle of about $10^{\textrm{o}}-30^{\textrm{o}}$ for halos of mass $10^{10}-10^{14}h^{-1}M_{\odot}$. These results confirm earlier hydrodynamical simulation measurements of halo shape and spin orientations, (mis-)alignments, and subhalo alignments by \cite{vandenBoschEtAl2002,BailinEtAl2005,CroftEtAl2009,BettEtAl2010,KnebeEtAl2010,PichonEtAl2011,DanovichEtAl2012,StewartEtAl2013}.

\subsection{Correlated intrinsic alignments of galaxies in hydrodynamical simulations}\label{hydro}

Hydrodynamical simulations are incredibly valuable in studies of galaxy intrinsic alignment as they provide an indication of what the observable matter is doing in galaxies. Given the evidence that dark matter halo ellipticity or angular momentum is a poor tracer for the actual or projected galaxy shape, it will be necessary in the coming years to push hydrodynamical simulations to sufficient volumes and resolutions to adequately measure the unbiased shapes of a large number of galaxies to truly constrain and calibrate models of intrinsic alignment with realistic samples of galaxy shapes. 

\cite{DuboisEtAl2014} specifically explored the alignment of blue, spin dominated galaxies in the Horizon-AGN simulation. They found an alignment of low-mass galaxy spins (alignments) with local filaments of large-scale structure, while high-mass galaxy spins tend to be aligned perpendicularly to the filaments. A transition mass of $M=3\times 10^{10}M_{\odot}$ is identified, consistent with \cite{CodisEtAl2012}. The high-mass galaxy misalignment is due to the merger of galaxies and was suggested by \cite{AubertPichonColombi2004,BailinSteinmetz2005}. This is opposed to low-mass galaxies, which form due to gas accretion, leading to spins parallel to the filament \citep{KimmEtAl2011,LaigleEtAl2013}. These recent high resolution results appear to support a growing consensus that galaxy spin alignment can be strongly influenced by large-scale filaments and sheets (e.g., \cite{AubertPichonColombi2004,BailinSteinmetz2005,HahnEtAl2007,AragonEtAl2007,PazStasyszynPadilla2008,SousbieEtAl2008}), which was predicted by \cite{LeePen2000,SugermanSummersKamionkowski2000}, and unify sometimes contradictory previous measurements (e.g., \cite{HattonNinin2001,FaltenbacherEtAl2002,HahnEtAl2007,ZhangEtAl2009}) with the identification of the transition mass. \cite{Cen2014} has also shown that mergers play a significant role in the reorientation of spin axes in galaxies.

\cite{CodisEtAl2014} performed a similar study of the Horizon-AGN simulation, using the angular momentum axis of a galaxy as a proxy for alignment. They found a null correlated alignment for redder galaxies, while measuring a potentially observable angular correlation in projected ellipticities in blue galaxies on the order of $\xi_{+}^{II}\approx 10^{-4}$. The measurable intrinsic alignment correlation among blue galaxies, coupled with a null result for red galaxies, is potentially conflicting with previous studies of intrinsic alignment (e.g., Secs. \ref{models} \& \ref{detections}), which find results for bluer or late-type galaxies consistent with zero on large scales and a stronger alignment for luminous red galaxies (LRGs). The authors discuss this discrepancy, and suggest, for example, the choice of angular momentum as a proxy for alignment being one reason for a null detection in red galaxies, as studies of red galaxies have typically used halo or stellar shape determination of the major axis, rather than angular momentum alignment which should be a subdominant factor in alignment of early-type galaxies. Angular momentum is instead most often associated with blue disk galaxies and the tidal torque picture of intrinsic alignment (Sec.~\ref{qa}). 

The detection of blue galaxy alignment by \cite{CodisEtAl2014} is significantly diminished when assuming a thicker galaxy disk, but the stronger detection relative to \cite{JoachimiEtAl2013b} may be motivated by galaxy spin being correlated with large-scale filament structure in the universe at competitive levels to alignment within the local dark matter halo (see for example, \cite{KimmEtAl2011,DuboisEtAl2014}), which is not captured by typical halo or semi-analytical models, as used in \cite{JoachimiEtAl2013b}. The decoherence of large-scale structure at later times is also suggested to explain the measurement by \cite{CodisEtAl2014} for blue galaxies, since the correlations are measured at relatively high redshift ($z>1$), and intrinsic correlations between spins would thus decrease at later times \citep{LeeEtAl2008,JoachimiEtAl2013b}. 

\cite{TennetiEtAl2014b} extended the work of \cite{TennetiEtAl2014} to measure intrinsic alignment 2-point statistics in the MassiveBlack-II simulation, finding results qualitatively in agreement with recent measurements by \cite{JoachimiMandelbaumAbdallaBridle2011}. They measured both ED position angle statistics, which cross-correlates the underlying large-scale density field with the position angle of a halo, and the projected correlation function $w_{g+}(r_p)$, described above in Eq. \ref{projected_corr}. They were able to measure 2-point intrinsic alignment statistics that are qualitatively consistent with the magnitude and redshift/luminosity dependence of direct measurements of intrinsic alignment in surveys. Using the parameterization of Eq. \ref{eq:NLA}, they found results that qualitatively agree with the measurements of \cite{JoachimiMandelbaumAbdallaBridle2011}, but with smaller luminosity dependence that may be due to differences in the galaxy samples considered. Blue galaxies were found to have stronger misalignments than red galaxies, leading to a suppressed $w_{g+}$. Radial alignment of satellites within host halos was detected, and a scale-dependent bias due to the 1-halo term was identified that is not captured by the ad hoc nonlinear alignment model. Finally, there are indications that the amplitude of the alignment is decreased by a factor of 5-18 in galaxies that are consistent with the sample to be observed by LSST, relative to measured value for LRGs.

Based on these results, there appears to be a growing consensus that precisely how one measures intrinsic alignment of galaxies in simulations has a strong effect on the resulting large-scale intrinsic alignment correlations measured. This alignment changes when one considers the inner or outer portions of a dark matter halo, which are not typically ideal tracers of the observed, baryonic component of the galaxy shape, and further large, high-resolution hydrodynamical simulations will likely be necessary to measure directly the galaxy alignment in order to place more accurate constraints on how the intrinsic alignment of galaxies may be correlated on larger scales over the volumes and depths in redshift of planned weak lensing surveys. Pursuit of a purely halo-oriented model, or even a semi-analytical model based on simulation and observational constraints, may also be challenging if spin (mis-)alignments with filaments and sheets are indeed strong enough to produce measurable correlations in future surveys. The effects of baryonic infall and mergers on galaxy (and intrinsic alignment) evolution, as informed by hydrodynamic simulations, may also play a large part in the future evolution of models of intrinsic alignment.

\section{Mitigation of intrinsic alignment in weak gravitational lensing surveys}\label{mitigation}

The study of the intrinsic alignment of galaxies has been primarily motivated in recent years as a means to improve the potential of planned weak lensing surveys, which can be substantially impacted by the unmitigated, correlated intrinsic alignment signal. This correlated intrinsic alignment acts as a primary physical systematic to the weak lensing signal, biasing the measured power. A great deal of effort has thus gone into developing methods to estimate and isolate the impact of intrinsic alignment on the observed lensing power spectrum and bispectrum. These methods include directly applying parameterized models of galaxy intrinsic alignment, but also focus on using a variety of physical properties of the intrinsic alignment signal in order to isolate it from the true lensing portion of the observed spectrum or bispectrum.

\subsection{Marginalization over parameterized intrinsic alignment models}\label{margin}

The most direct method for separating the effects of galaxy intrinsic alignment and weak gravitational lensing on the correlated shapes of galaxies in large-scale surveys is through marginalization over some parameterized model of intrinsic alignment. This model can be taken to be some physically motivated model, like those described in Sec.~\ref{models}, or can be some more general parameterization of the intrinsic alignment signal. The most popular physical model is currently the linear alignment model of \cite{HirataSeljak2004,HirataSeljak2010} (see Sec.~\ref{la}), which relates the intrinsic alignment signal to the underlying tidal field, but recent work has attempted to improve and build on limitations of the linear alignment model on small scales (see Secs. \ref{nla}-\ref{sam}). In more generalized models, the signal may be parameterized, for example, as a function of properties like redshift, physical separation, or galaxy type, and can be expanded around some fiducial physical model. In both cases, the parameters which define an intrinsic alignment model are constrained along with other cosmological or nuisance parameters. 

As an early example, \cite{KingSchneider2003} demonstrated that generic, parameterized template functions could be employed to simultaneously fit both lensing ($\xi^{L}$) and intrinsic alignment ($\xi^I$) components of the observed ellipticity correlation $\xi(\theta,z_i,z_j)=\xi^{L}(\theta,z_i,z_j)+\xi^{I}(\theta,z_i,z_j)$ with photometric redshift information. The signals are assumed to be composed of some template functions
\begin{align}
\xi^L(\theta,z_i,z_j)=&\sum_{n=1}^{N_L}a_n A_n(\theta,z_i,z_j)\\
\xi^I(\theta,z_i,z_j)=&\sum_{n=1}^{N_I}b_n B_n(\theta,z_i,z_j),
\end{align}
with amplitude $a_n$ and $b_n$ of the $n$th lensing and intrinsic alignment template function, respectively. They used $N_L=3$ lensing functions $A_n$ built from various lensing correlation functions from CDM cosmologies and $N_I=9$ intrinsic alignment functions $B_n$ with an assumed spatial intrinsic alignment correlation function parameterized as
\begin{align}
\eta(r,z)=(1+\bar{z})^{\alpha}\exp(-r/R),
\end{align}
where $r$ is the comoving separation in units of $h^{-1}$ Mpc, $R$ is a correlation length, and $\bar{z}$ is the mean redshift of the galaxy pair. Three values each of $\alpha$ and $R$ are chosen as template functions, which are fitted against the assumed intrinsic alignment models of \cite{HeavensRefregierHeymans2000}, given in Eq.~(\ref{eq:hrh}), and \cite{Jing2002}, given in Eq.~(\ref{eq:jing}) for some specific halo mass choice. These models assume only an $II$ component to the intrinsic alignment signal, but \cite{KingSchneider2003} showed that by fitting the observed signal using the template functions described above with information about photometric redshifts, the degeneracy between $\Omega_m$ and $\sigma_8$ could be relieved through tomography despite the presence of an intrinsic alignment contamination in the ellipticity correlation. This template fitting process was expanded by \cite{King2005} to include the $GI$ contribution, with similar results. 

A major limitation to the use of parameterized intrinsic alignment models or templates for addressing cosmological bias in weak lensing surveys is the resulting degradation in figures of merit as the number of parameters increase. Current physical models of intrinsic alignment that possess fewer parameters, however, suffer from an inability to accurately describe the intrinsic alignment correlations on small, nonlinear scales. While some improvements to this have been attempted (Secs. \ref{nla}-\ref{sam}), there remains no consensus physical model that describes well the intrinsic alignment signal on all scales from basic principles. The creation of such a model remains challenging, as the connection between galaxy formation and evolution and the intrinsic alignment signal is not sufficiently understood to present an accurate physical model on nonlinear scales. 

However, such a direct model-fitting approach remains viable so long as the assumed systematic residuals do not strongly impact survey constraints, such as for a small or shallow weak lensing survey. \cite{HuffEtAl2014}, for example, utilized the linear alignment model with a scaling matched to previous measurements of the $GI$ correlation (see Sec.~\ref{detections}) in the Sloan Digital Sky Survey to unbias shear measurements. Using the central model of \cite{HirataEtAl2007}, \cite{HuffEtAl2014} reduced the magnitude of the shear-shear correlation by 8\% to account for a negative $GI$ contribution, with an error of 50\% on this determination propagated through to parameter constraints.

It has since been shown, though, that the linear alignment model typically used to describe intrinsic alignment, may not perform well in a minimal parameter marginalization process, as employed for the recent CFHT Lensing Survey (CFHTLenS) by \cite{heymans}. They employed a single parameter scaling of the intrinsic alignment signal in the linear alignment model, such that $P^{II}\propto A^2$ and $P^{GI}\propto A$. Including the extra parameter $A$ led to a reduction in the constraining power on $\sigma_8 (\Omega_m/0.27)^{\alpha}$ of about 30\% \cite{heymans}. Best-fit values of $A$ for the full galaxy sample were shown to be negative at 1.4$\sigma$, which would indicate a positive correlation for $GI$, though not at strong statistical significance, despite separate fits for early- and late-type galaxies both indicating a positive or near-zero value for $A$. This conflict, if confirmed by future observations, could indicate that a single-parameter linear alignment model assumption is too simple to capture the intrinsic alignment signal in a large weak lensing survey with galaxy types that may have different driving mechanisms for their alignments. These concerns will become important for future survey shear measurements, where systematic uncertainties become larger than statistical errors. 

\begin{figure}
\center
\includegraphics[width=\columnwidth]{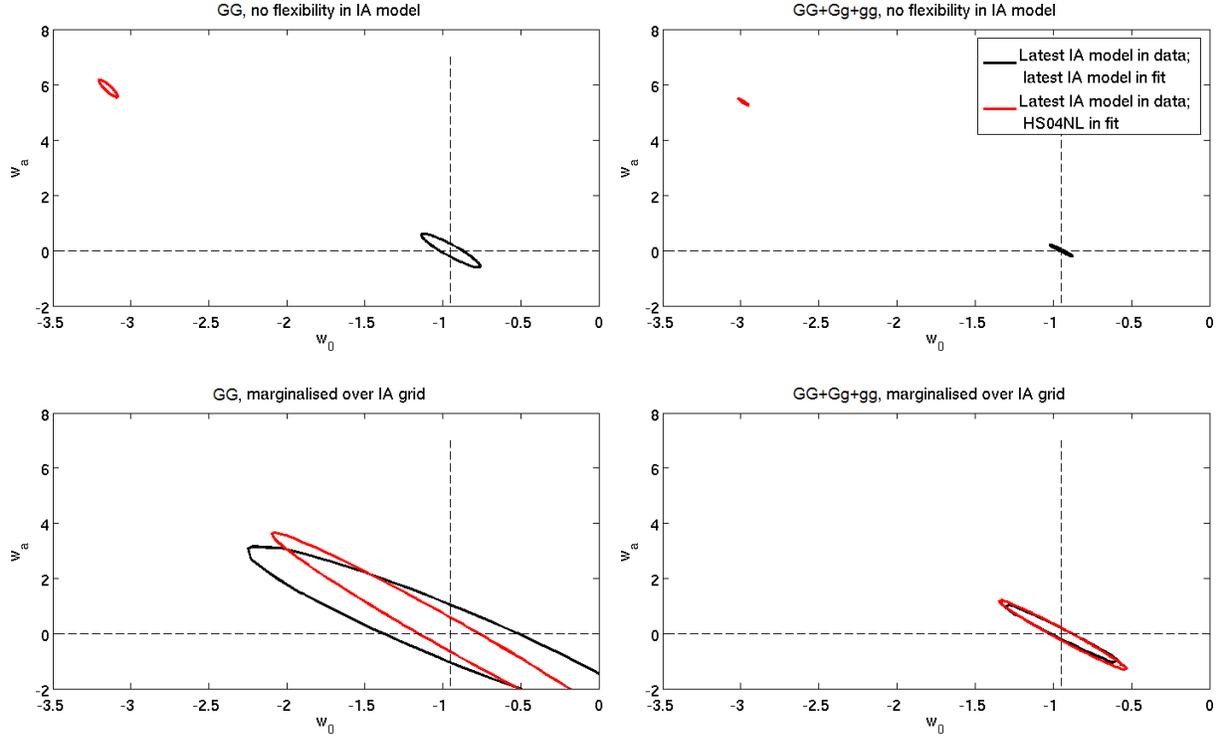}
\caption{
The effects of including an intrinsic alignment model in the data and either assuming the wrong model (described in Sec.~\ref{nla}) or marginalizing over a parameterization grid of intrinsic alignment models, like the processes described in Secs. \ref{margin} \& \ref{sc2} \protect\citep{BridleKing2007,JoachimiBridle2010}, for seven bins of both $k$ and $z$. Left panels include only shear-shear correlations. Right panels include shear--shear, shear--density, and density--density correlations. Top panels assume a wrong intrinsic alignment model. Bottom panels show the results of marginalization over the intrinsic alignment parameterization. Including position information as described in Sec.~\ref{sc2} nearly makes up for information loss due to the additional nuisance parameters. Source: Reproduced with permission from \protect\cite{KirkRassatHostBridle2011}, Oxford University Press on behalf of the Royal Astronomical Society.}\label{fig:mitig}
\end{figure}

A more general approach is that of \cite{BridleKing2007}, where a parameterized set of models is constructed which vary from a base physical model, in this case the ad hoc nonlinear alignment model, through a reasonable set of the parameter space. They choose a base parameterized model with arbitrary amplitude and redshift dependence, given for $X\in\{II,GI\}$ by
\begin{align}
P^{X}_{base}(k,\chi)=A_X\left(\frac{1+z}{1+z_0}\right)^{\gamma_X}P_{nl}^X(k,\chi),
\end{align}
where $P_{nl}^X$ is the nonlinear linear alignment model intrinsic alignment power spectra described in Sec.~\ref{nla}. This parameterization is generalized to include a scale-dependent function $Q^X(k;\chi)$
\begin{align}
P^{X}_{free}(k;\chi)=Q_X(k;\chi)P^{X}_{base}(k;\chi).
\end{align}
$Q^X$ is parameterized over $n$ bins of $k$ and $m$ bins of $z$ such that
\begin{align}
\ln{Q^X(k;\chi)}=&K Z B^X_{ij}+(1-K)ZB^X_{(i+1)j}+K(1-Z)B^X_{i(j+1)}\\
&+(1-K)(1-Z)B^X_{(i+1)(j+1)}\nonumber\\
K=&\frac{\ln{k}-\ln{k_i}}{\ln{k_{i+1}}-\ln{k_i}}\quad (k_i<k<k_{i+1})\\
Z=&\frac{\ln{1+z}-\ln{1+z_j}}{\ln{1+z_{j+1}}-\ln{1+z_j}}\quad (z_j<z<z_{j+1}).
\end{align}
$B^X_{ij}$ is a set of free parameters, and setting $B^X_{ij}=0$ is equivalent to $Q^X$ being unity. 

Using this parameterization, \cite{BridleKing2007} demonstrated the impact of the level of freedom in such parameterizations of the intrinsic alignment signal, as well as requirements on photometric redshift quality and the number of tomographic redshift bins. Parameterized models with more freedom necessarily require a larger number of parameters to capture the intrinsic alignment signal across a range of redshifts and scales, and thus degrade the figure of merit for constraints more strongly. Including galaxy position information as described below in Sec.~\ref{sc2} can maintain some of the original shear-shear constraining power, which is demonstrated in Fig.~\ref{fig:mitig}. This approach remains the most direct method for the mitigation of biases due to the intrinsic alignment signal, though determining the best strategy in terms of the competing benefit of using more sophisticated models with a larger number of free parameters versus the actual improvements gained by constraining the intrinsic alignment signal for removal is survey dependent. See also \cite{Bernstein2009,KitchingTaylor2011} for other discussions of marginalization over bias parameters, including intrinsic alignment.

Both approaches to employing intrinsic alignment models can be informed by ongoing measurements of the intrinsic alignment signal across galaxy samples (Sec.~\ref{detections}), by improvements in the methodology of studying galaxy alignment in simulations and its connection to galaxy properties (Sec.~\ref{sims}), and finally, by the inclusion of additional complementary survey measurements, either in the same survey or through overlapping probes, which we will discuss in the following sections.

\subsection{Redshift tomography, separation weighting, and sample limiting}\label{tomography}

\begin{figure}
\center
\includegraphics[width=.6\columnwidth]{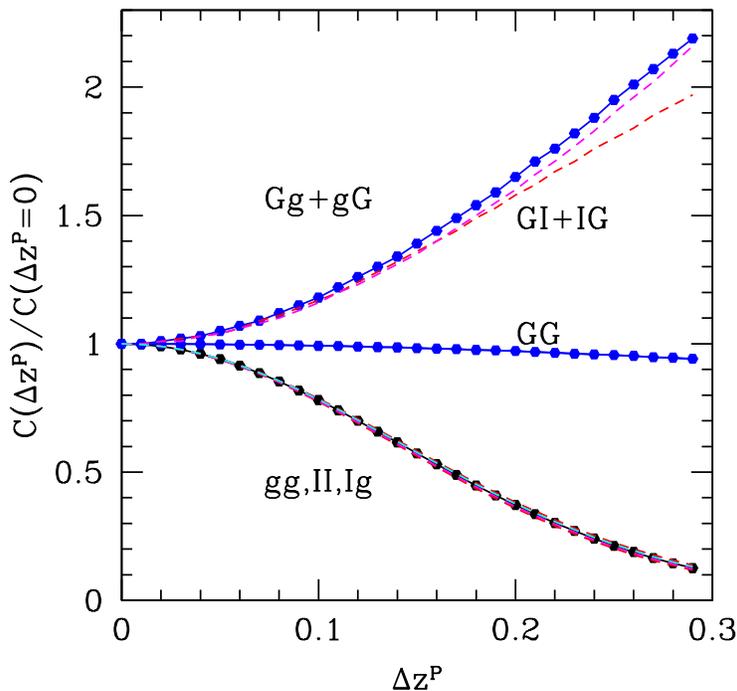}%
\caption{The redshift separation dependencies of the $GI$, $II$, and $GG$ spectra are shown, relative to the appropriate galaxy density cross-spectra. The $II$ correlation is diminished by about 75\% at $\Delta z^p=0.2$, while $GI$ grows by a similar fraction. The lensing signal $GG$ is relatively unaffected across this separation range. Both the halo intrinsic alignment model described in Sec.~\ref{haloia} and a toy model described in \protect\cite{Zhang2010b} are shown. Source: Reproduced with permission from \protect\cite{Zhang2010b}, Oxford University Press on behalf of the Royal Astronomical Society.}\label{fig:dzp}
\end{figure}

\begin{figure}
\center
\includegraphics[height=\textwidth,angle=270]{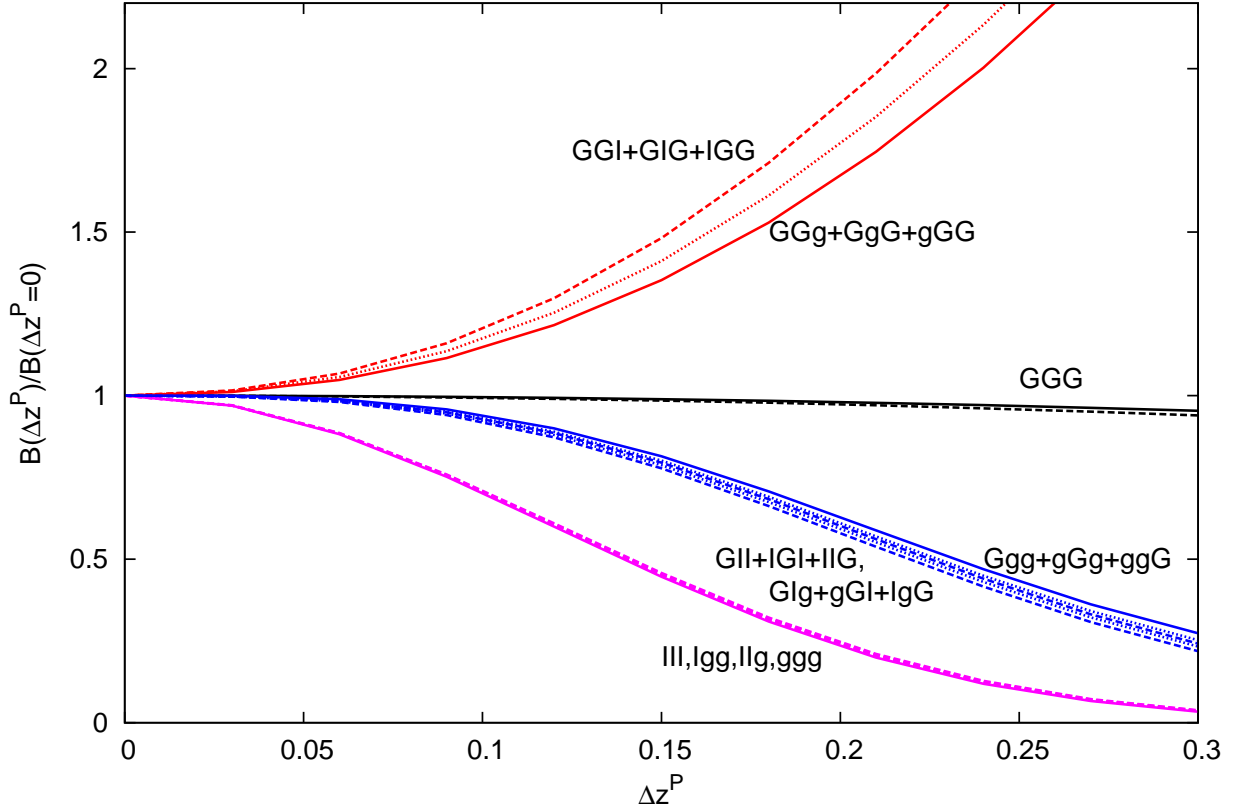}%
\caption{The galaxy separation dependencies of the $GGI$, $GII$, $III$, and $GGG$ bispectra are shown, relative to the appropriate galaxy density cross-bispectra. The $III$ correlation is diminished by about 75\% at $\Delta z^p=0.2$, which is similar to the impact on $II$, while $GGI$ is diminished by about 50\%. The $GGI$ correlation instead grows by about 75\%, similar to $GI$. The lensing signal $GGG$ is relatively unaffected across this separation range. Both the halo intrinsic alignment model described in Sec.~\ref{haloia} and a toy model described in \protect\cite{Zhang2010b} are shown. Source: Reproduced from \protect\cite{TroxelIshak2012b}.}\label{fig:dzp2}
\end{figure}

In order to avoid the necessity of modeling the intrinsic alignment signal and thus including additional parameters to constrain, one can employ certain physical properties of the intrinsic alignment signal to reduce its impact on weak lensing measurements. One such property is the strong separation dependence of the intrinsic ellipticity correlations. The separation dependence of the intrinsic alignment correlations are shown in Fig.~\ref{fig:dzp} for the power spectrum and in Fig.~\ref{fig:dzp2} for the bispectrum. The exclusion or down weighting of spatially close galaxies in the observed ellipticity correlation, proposed by \cite{CatelanKamionkowskiBlandford2001}, thus greatly reduces the magnitude of the intrinsic ellipticity correlation ($II$), but at the cost of reducing the available statistical power of the measured signal (increasing shape noise and cosmic variance). The $GI$ correlation is also unaffected, as it can occur over large redshift separations. The methodology of utilizing redshift information, particularly photometric redshifts, as a means to discriminate between physically close galaxies was explored by \cite{KingSchneider2002,Heavens2003,HeymansHeavens2003}. \cite{HeymansEtAl2004} applied this technique to results from the COMBO-17, Red-sequence Cluster, and VIRMOS-DESCART surveys, placing some early constraints on the magnitude of the intrinsic alignment signal. \cite{TakadaWhite2004} showed that employing a tomographic study of weak lensing using only cross-correlations between large redshift bins would increase errors on parameter constraints by about 10\% for five or more source redshift bins, while the $II$ contamination is typically rendered negligible. Tomography using photometric redshift bins does require accurate color information on galaxies, and some limitations of this are discussed by \cite{JainConnolyTakada2007} in light of intrinsic alignment contamination.

Another way to lower the impact of intrinsic alignment on shear measurements is to use complementary information on the intrinsic alignment signals (see Sec.~\ref{detections}) to limit the included galaxy sample by type, excluding those galaxies that contribute most strongly to the intrinsic alignment correlations. \cite{MandelbaumEtAl2006,HirataEtAl2007} first identified that Luminous Red Galaxies (LRGs) contribute most of the power of the $GI$ signal, and this has been confirmed by more recent studies. This would indicate that the exclusion of early type galaxies is one direct method to mitigate the impact of the $GI$ correlation on shear measurements. This was employed, for example, by the COSMOS tomographic study of shear in COSMOS by \cite{SchrabbackEtAl2010} along with exclusion of the redshift bin auto-correlations to decrease the impact of both $II$ and $GI$ effects. Most recently, CFHTLenS has confirmed this distinction between early- and late-type galaxies in a much larger galaxy sample, demonstrating that the intrinsic alignment signal from late-type galaxies is consistent with zero on large scales, while detecting a nonzero $GI$ signal from early type galaxies \citep{heymans}. \cite{JoachimiEtAl2013b} also demonstrated that removing 20\% of red foreground galaxies can suppress the intrinsic alignment contamination by up to a factor of two in a deep survey.

Whether using redshift tomography, an optimized down weighting of nearby galaxy pairs through redshift information, or sample limiting measurements to exclude galaxies which most strongly contribute to the intrinsic alignment signal, these mitigation schemes have become in some ways more feasible as a method of reducing the intrinsic ellipticity correlation component to the observed weak lensing signal, due to the very large number of galaxies and relative depth of recent and planned surveys. The loss of redshift auto-correlations or a subset of galaxy types, for example, is less catastrophic for a survey as the number of available photo-z bins increases due to better photo-z measurements, increased galaxy counts, and survey depths. The optimal methodology, however, remains survey dependent and requires careful analysis to balance information loss to the mitigation of intrinsic alignment bias with the improvement in accuracy of parameter determinations. Care must also be taken not to introduce additional biases due to galaxy sample cuts. With the identification of a long-range correlation between the intrinsic ellipticity and lensing of background galaxies by \cite{HirataSeljak2004}, redshift weighting can also only form part of a mitigation strategy for the intrinsic alignment signal, as correlations like $GI$ are unaffected. In fact, for deep lensing surveys the $GI$ signal grows to dominate the intrinsic ellipticity correlation ($II$), and one must employ different strategies for isolating its impact on the cosmic shear signal, like sample limiting, or others which we discuss further below.

\subsection{Nulling techniques}\label{nulling}

Since any galaxy can contribute to the $GI$ signal with multiple source galaxies at different redshifts, there is no direct method for excluding galaxy pairs that contribute to the total statistical $GI$ signal. However, one can construct more complicated procedures for weighting the cosmic shear signal by redshift to reduce the $GI$ contamination. This is possible because the intrinsic ellipticity-gravitational shear correlation has a unique geometry and redshift separation dependence compared to the cosmic shear signal. The nulling approach was discussed by \cite{HutererWhite2004} in the context of removing biases from small-scale physics, and developed by \cite{JoachimiSchneider2008,JoachimiSchneider2009} as a method for addressing the $GI$ intrinsic alignment contamination. It is referred to as nulling, since it nulls some component (e.g., $GI$) of the statistical signal. This is accomplished by constructing a new cosmic shear measure with a reweighted redshift distribution. This down weights the contribution by the $GI$ effect, and thus reduces its impact on cosmological constraints, though with some loss of statistical power associated with nulling a portion of the signal through downweighting. 

\begin{figure}
\center
\includegraphics[height=\textwidth,angle=270]{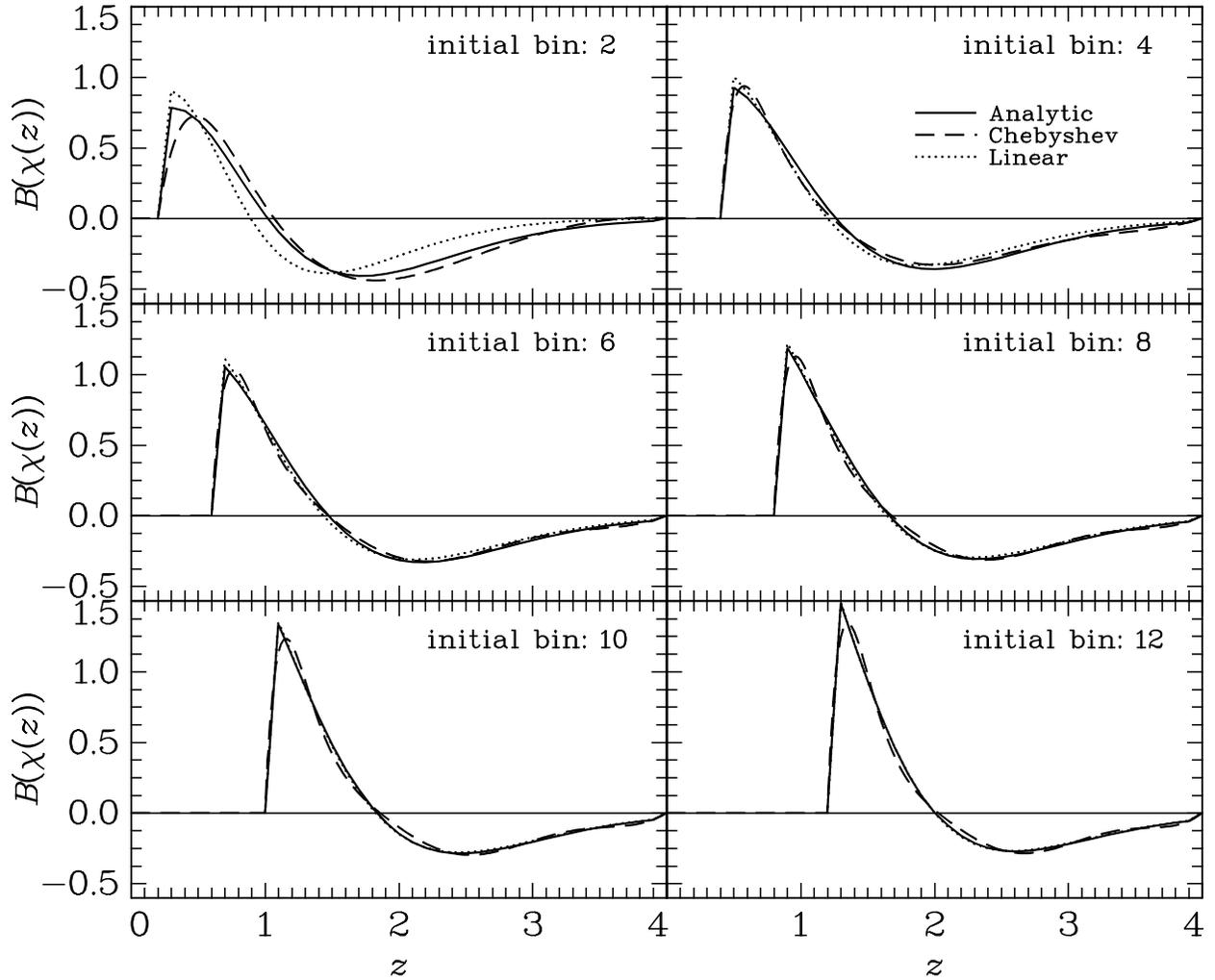}%
\caption{The weight functions $B_i(\chi)$, determined through three methods by \protect\cite{JoachimiSchneider2008}, for different initial bins $i$ and a total photo-z bin number $N_z=40$. Solid lines show a simplified analytical construction, dashed curves show a Chebyshev series construction, and dotted lines show a piecewise linear construction of the weight function. Source: Reproduced with permission from \protect\cite{JoachimiSchneider2008}. \copyright 2008 ESO.}\label{fig:nulling1}
\end{figure}

The nulling process assumes that a galaxy sample can be split into narrow redshift bins, such that the comoving galaxy distribution in bin $i$ can be written as 
\begin{equation}
n_i(\chi)\approx \delta^D(\chi-\chi(z_i)).
\end{equation}
The source of the $GI$ contamination to the shear spectrum comes from a correlation between the lensing of background galaxies in some redshift bin $j$ due to some matter distribution in a foreground redshift bin $i$ and the intrinsic ellipticity caused by the matter distribution's tidal field in bin $i$. Thus one can remove this component to the shear signal by not allowing a contribution to the convergence (or shear) due to matter in bin $i$. The standard projection for the convergence due to some distribution of galaxies was given in Eq.~(\ref{eq:keff2}), where in a spatially flat universe with comoving distance in units of $c/H_0$, it can be related to the density contrast by
\begin{align}
\kappa_i(\bm{\theta})=&\frac{3}{2}\Omega_m\int_0^{\chi_l} d\chi g_i(\chi)\frac{\chi}{a(\chi)}\delta(\chi\bm{\theta};\chi),
\end{align}
where
\begin{align}
g_i(\chi)=&\int_{\chi}^{\chi_l} d\chi' n_i(\chi')\left(1-\frac{\chi}{\chi'}\right).
\end{align}
Instead of the galaxy distribution, one can choose some arbitrary function $\hat{B}(\chi)$, such that
\begin{equation}
\hat{g}_i(\chi)=\int_{\chi}^{\chi_l}d\chi'\hat{B}_i(\chi')\left(1-\frac{\chi}{\chi'}\right).\label{eq:weight2}
\end{equation}
Then, in order to `null' the impact of some mass at comoving distance $\hat{\chi}$, one must simply choose some $\hat{B}(\chi)$ such that
\begin{equation}
\hat{g}_i(\chi)=\int_{\hat{\chi}}^{\chi_l}d\chi'\hat{B}_i(\chi')\left(1-\frac{\hat{\chi}}{\chi'}\right)=0,\label{eq:weight3}
\end{equation}
which renders the convergence due to such a mass zero. For the power spectrum, this weighting can be written 
\begin{align}
\hat{P}_i(\ell)\approx&\sum_{j=1}^{N_z}\hat{B}_i(\chi(z_j))P_{ij}(\ell)\chi'(z_j)\Delta z,
\end{align}
where $P_{ij}(\ell)$ is the 2D tomographic convergence power spectrum of Eq.~(\ref{eq:pstomo}) and $N_z$ is the total number of redshift bins.

\cite{JoachimiSchneider2008,JoachimiSchneider2009} describes the choice of $\hat{B}_i$ in order to remove the contribution of $GI$ to the cosmic shear information via the constraint in Eq.~(\ref{eq:weight3}) for some redshift bin $i$, while simultaneously maximizing the information content available to constrain cosmological parameters. This information optimization is designed to maximize the trace of the Fisher matrix for $N_{\ell}$ angular frequency bins of width $\Delta\ell_l$, $N_p$ parameters to be constrained, and survey area $A$. The power spectra in the trace of the Fisher matrix are evaluated for some fiducial parameters in the cosmological model. This trace is independent of the weighting amplitude, and thus a normalization of $\hat{B}$ can be pursued independently of this condition. In practice, the only non-zero $\hat{B}_j$ are for bins $i+1<j<N_z$. The redshift at which each bin is defined (e.g., central, median, or boundary values) is free to be chosen, with some choices producing improved reductions in the bias \citep{JoachimiSchneider2009}. The evaluation of $\hat{B}_i$ is discussed in detail in \cite{JoachimiSchneider2008,JoachimiSchneider2009}, and example functions are shown in Fig.~\ref{fig:nulling1}.

\begin{figure}
\center
\includegraphics[width=5.0in,height=5.0in]{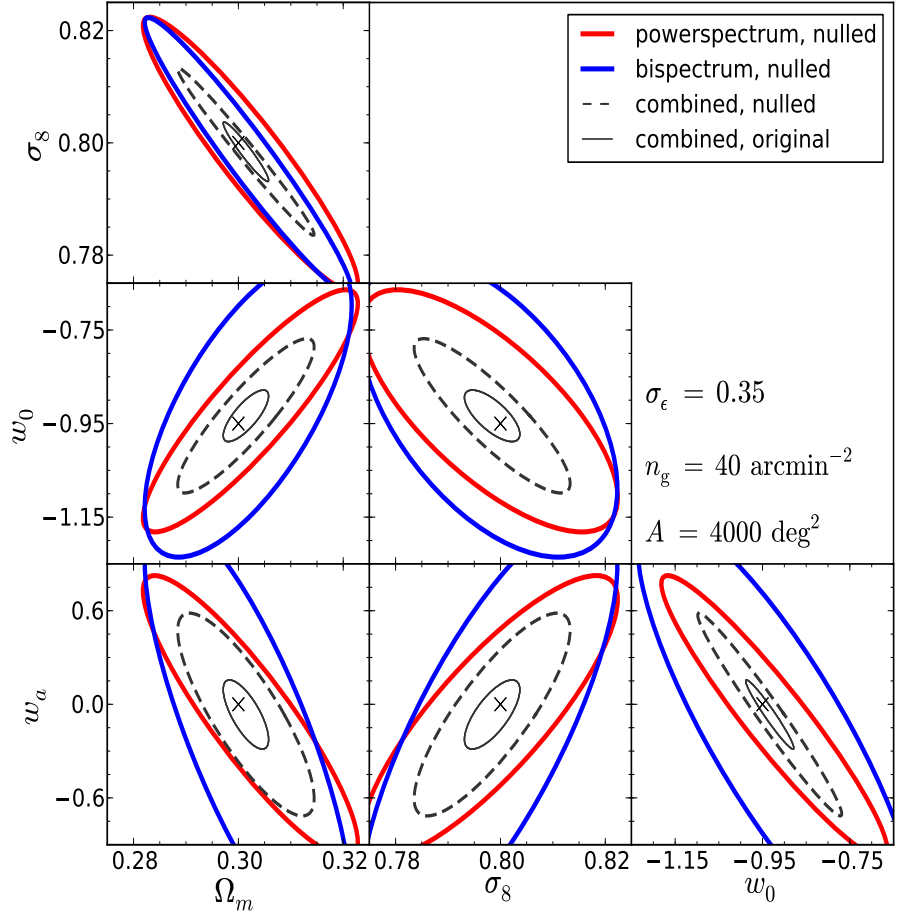}%
\caption{Resulting 1$\sigma$ parameter estimates for the given survey parameters after the nulling process using only the power spectrum, only the bispectrum, and a combination of both. This is compared to the original constraint from combining power spectrum and bispectrum. The nulling removes nearly all biases from the parameter estimates due to intrinsic alignment, at the cost of decreased statistical power. Source: Reproduced with permission from \protect\cite{ShiJoachimiSchneider2010}. \copyright 2010 ESO.}\label{fig:nulling2}
\end{figure}

With perfect redshift information, the nulling technique can totally remove the $GI$ portion of the lensing signal, removing the intrinsic alignment systematic error in parameter determinations. It simultaneously creates a moderate increase in the size of the associated confidence contours due to loss of statistical power. For 20 redshift bins, the increase in the 2D confidence region size is between 20\%-50\% \citep{JoachimiSchneider2008}. Most surveys, however, utilize uncertain photometric redshift information for galaxies, which conservatively limits the reduction in $GI$ by nulling to about a factor of 10 for 10 or more redshift bins, with a similar increase of up to 50\% in the size of confidence regions due to loss of statistical power \citep{JoachimiSchneider2009}.

The nulling technique has been expanded to the bispectrum for $GGI$ and $GII$ with similar results as for $GI$ by \cite{ShiJoachimiSchneider2010}. The nulling condition for the bispectrum of some mass at comoving distance $\chi_i$, for some weighting function $\hat{T}_{ij}(\chi_k)$, is now 
\begin{align}
\int_{\hat{\chi}}^{\chi_1}d\chi_k T_{ij}(\chi_k)\left(1-\frac{\chi_i}{\chi_k}\right)\chi_k'\Delta z_k=0.
\end{align}
The bispectrum estimator with this weighting function is then written
\begin{align}
\hat{B}_{ij}(\ell_1,\ell_2,\ell_3)\approx\sum_{k=i+1}^{N_z}\hat{T}_{ij}(\chi(z_k))B_{ijk}(\ell_1,\ell_2,\ell_3)\chi'(z_k)\Delta z_k.
\end{align}
The process for choosing the nulling weights $T_{ij}$ is described in detail by \cite{ShiJoachimiSchneider2010}. The 3-point nulling process is able to reduce the intrinsic alignment contamination by about factor of 10 for 10 redshift bins, with similar information loss as the 2-point nulling. The effect of nulling and combining the power spectrum and bispectrum on parameter constraints is shown in Fig.~\ref{fig:nulling2}.

The nulling technique is one option for mitigating the impact of the gravitational shear--intrinsic ellipticity correlation for surveys with the capability of utilizing many redshift bins with good photometric redshift information. It has no dependence on uncertain modeling of the intrinsic alignment signal, and uses purely geometric information obtained from a single survey to disentangle the impact of the $GI$ signal from cosmic shear. The technique was also proposed as a means to `boost' instead of null the intrinsic alignment signal \citep{JoachimiSchneider2010}, offering an alternative method of measuring the $GI$ correlation. It has also been proposed to null \citep{HeavensJoachimi2011} and boost \citep{Schneider2014} effects like magnification, which like intrinsic alignment can also bias the cosmic shear signal and provide alternative cosmological information.

\subsection{Self-calibration techniques}\label{sc}

Another alternative for the reduction and indirect measurement of the intrinsic alignment signal is the so-called `self-calibration' techniques, named because they calibrate the lensing signal by using a rescaling of complementary cross-correlations between various observables within a single survey. The self-calibration of systematic effects using the additional information gained from the gravitational shear-galaxy density cross-correlation and galaxy density-density correlation, in addition to the gravitational shear-shear correlation, has been discussed by several authors (e.g., \cite{BernsteinJain2004,HuJain2004,Zhan2006,Bernstein2009}), and was specifically examined as a means to self-calibrate the $GI$ signal by \cite{Zhang2010a,Zhang2010b}. The available information in a weak lensing survey relevant to the self-calibration, including the shear, density and intrinsic alignment components, can be written for $i<j$ as the following power spectra
\begin{align}
C^{(1)}_{ij}(\ell) =& C^{GG}_{ij}(\ell)+C^{IG}_{ij}(\ell)+C^{II}_{ij}(\ell),\nonumber\\
C^{(2)}_{ii}(\ell) =& C^{gG}_{ii}(\ell)+C^{gI}_{ii}(\ell),\nonumber\\
C^{(3)}_{ii}(\ell) =& C^{gg}_{ii}(\ell).\label{eq:2obs}
\end{align}
Unlike nulling, which uses a purely geometric approach to minimize the impact of $GI$, the self-calibration uses the relationship between the observed ellipticity correlations and cross-correlations of ellipticity and galaxy density to isolate the magnitude of the intrinsic alignment correlations. The self-calibration falls into two categories: the first of which self-calibrates the correlations between gravitational shear and intrinsic ellipticity from cross-correlations of photometric redshift bins, while the second self-calibrates the full set of intrinsic alignment correlations (or alternately their sum) within a single photometric redshift bin. 

\subsubsection{Self-calibration between photometric redshift bins}\label{selfcalibration1}

\begin{figure}
\center
\includegraphics[height=\textwidth,angle=270]{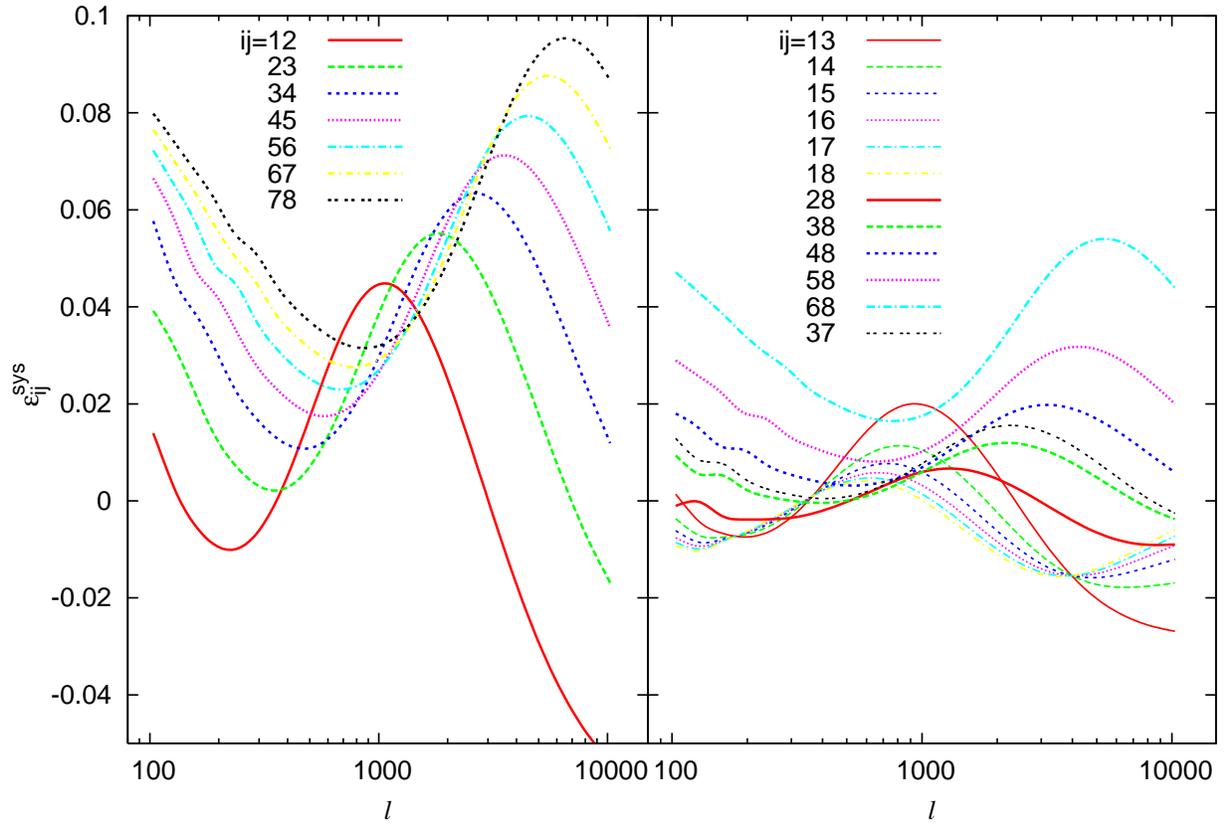}%
\caption{The accuracy of the scaling relationship in Eq.~(\ref{eq:2scale}) is shown for a variety of redshift bin combinations in a Stage IV survey. Equation (\ref{eq:2scale}) is accurate to within about 10\% for all redshift bin combinations, leading to a suppression of the $GI$ intrinsic alignment contamination to the power spectrum by a factor of 10 or more. Source: Reproduced from \protect\cite{TroxelIshak2012a}.}\label{fig:GIerror}
\end{figure}

The self-calibration of $GI$ in cross-correlations of different photometric redshift bins, $i<j$, where the $II$ signal is assumed to be negligible (e.g., Sec.~\ref{tomography}), requires no assumptions about any underlying intrinsic alignment model. It instead utilizes the ansatz of some deterministic galaxy bias $b_{i}$ relating the galaxy and matter densities in real space \citep{FryGaztanaga1993,Fry1994}, such that
\begin{align}
\delta_g(\bm{x})=b_{1}(\chi)\delta_m(\bm{x};\chi)+\frac{b_{2}(\chi)}{2}\delta_m(\bm{x};\chi)^2
\end{align}
to connect the galaxy density-intrinsic alignment ($gI$) and gravitational shear-intrinsic ellipticity ($GI$) correlations through some approximate scaling relationship. For the $GI$ self-calibration, the nonlinear bias is neglected. The redshift bins are assumed to be sufficiently narrow, like the nulling technique, so that one can safely make the approximation that the $GI$ and $gI$ spectra obey a simple scaling relation
\begin{align}
C^{IG}_{ij}(\ell)\approx \frac{W^G_{ij}}{b_{1}^i\Pi_{ii}}C^{Ig}_{ii}(\ell),\label{eq:2scale}
\end{align}
where $C^{IG}$ and $C^{Ig}$ are the 2D projected gravitational shear-intrinsic alignment and galaxy density-intrinsic alignment power spectra, respectively, and $b_{1}^i$ is the average galaxy bias within the $i$-th redshift bin. Finally,
\begin{align}
W^G_{ij}&=\int_0^{\chi_1}W_i(\chi)d\chi\\
\Pi_{ii}&=\int_0^{\chi'} n_i^2(\chi)d\chi.
\end{align}
This allows one to measure the intrinsic alignment in the galaxy shear--density cross-correlation ($C^{Ig}$) and infer the contamination $C^{IG}$ to the lensing signal using quantities measured from the survey. The inaccuracy of Eq. (\ref{eq:2scale}) is shown in Fig. \ref{fig:GIerror}, where the self-calibration is typically inaccurate by less than 10\%.

The $gI$ correlation is then isolated from the gravitational shear-galaxy density ($Gg$) correlation in the observed shear--density correlation, $C^{(2)}$, through an estimator that takes advantage of the different geometry dependence of the two correlations
\begin{equation}
\hat{C}^{Ig}_{ii}(\ell)=\frac{C^{(2)}_{ii}|_S(\ell)-Q^{GI}(\ell)C^{(2)}_{ii}(\ell)}{1-Q^{GI}(\ell)}.\label{eq:2cig}
\end{equation}
The suppression quotient $Q^{GI}(\ell)\equiv C^{gG}_{ii}|_S(\ell)/C^{gG}_{ii}(\ell)$ measures the relative suppression of the spectrum due to the geometry of the lensing kernel, where subscript `S' denotes a spectrum which measures only pairs such that the photometric redshifts $z^p_G<z^p_g$. $Q^{GI}$ is approximately equal to a factor $\bar{\eta}^{GI}_i=\eta^{GI}(\bar{z}_i)$, for mean bin redshift $\bar{z}_i$,
\begin{equation}
\eta^{GI}(z_L,z_g=z_L)=2\frac{\int_{i}dz^p_G\int_{i}dz^p_g\int_0^{\infty}dz_G W_L(z_L,z_G)p(z_G|z_G^p)p(z_g|z_g^p)S(z_G^p,z_g^p)n_i^p(z_G^p)n_i^p(z_g^p)}{\int_{i}dz^p_G\int_{i}dz^p_g\int_0^{\infty}dz_G W_L(z_L,z_G)p(z_G|z_G^p)p(z_g|z_g^p)n_i^p(z_G^p)n_i^p(z_g^p)},\label{eq:eta}
\end{equation}
where $\int_i\equiv\int_{\bar{z}_i-\Delta z_i/2}^{\bar{z}_i+\Delta z_i/2}$, $S(z,z')=1$ for $z<z'$ and $S=0$ otherwise. The case where $\eta,Q\rightarrow 0$ corresponds to the use of spectroscopic redshift information, where the selection rule completely removes the lensing signal.

The scaling relation in Eq.~(\ref{eq:2scale}) has been shown to be accurate to within 10\% for all but the lowest, adjacent photo-z bin combinations for a typical Stage IV weak lensing survey, which corresponds to a reduction in the magnitude of the $GI$ contaminant of a factor of 10 or more for various redshift bin pairs \citep{Zhang2010a,TroxelIshak2012a}. This is competitive with the photometric estimates of the nulling approach, but with separate benefits and challenges. The self-calibration is capable of simultaneously estimating the $GI$ signal, preserving it for use in other studies, and does not throw away significant statistical weight, which preserves the statistical power in parameter estimates. It instead uses an estimator to separate the lensing and intrinsic alignment components of the galaxy shear--density cross-correlation, but also preserves both for use in joint probe analysis. This will introduce an additional propagated error that must be considered if using the lensing component of the galaxy shear--density cross-correlation due to the imperfect separation. It also requires an additional measurement of the galaxy bias, though it has been shown that realistic projections for galaxy bias measurements in a Stage IV weak lensing survey are sufficient to render errors due to uncertainty in the galaxy bias negligible in the self-calibration \citep{Zhang2010a}. In both self-calibration and nulling techniques, cosmological priors are necessary, but the quality of constraints from complementary probes like the cosmic microwave background are sufficient. 

\begin{figure}
\center
\includegraphics[height=\textwidth,angle=270]{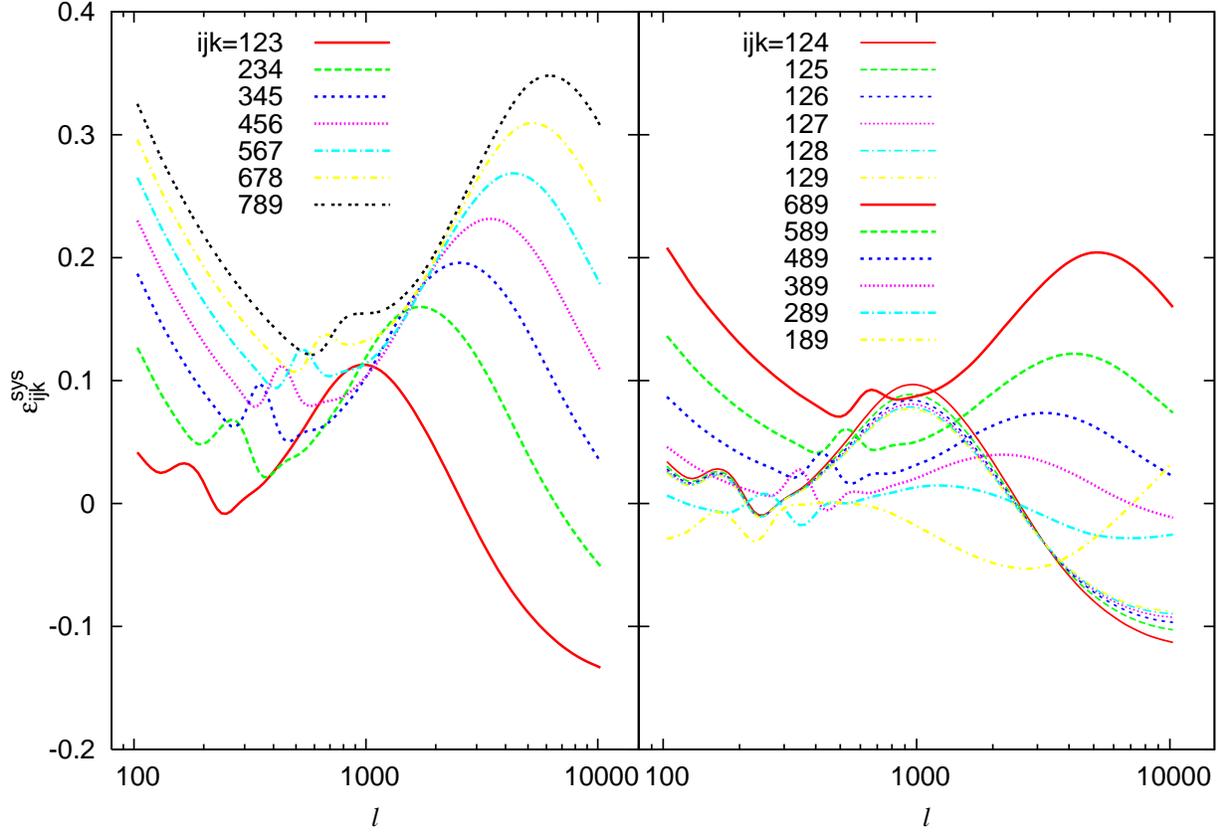}%
\caption{The accuracy of the scaling relationship in Eq.~(\ref{eq:3scale}) is shown for a variety of redshift bin combinations in a Stage IV survey. Equation (\ref{eq:3scale}) is accurate to within about 30\% for all redshift bin combinations, though most have an accuracy within 10\%, leading to a suppression of the $GGI$ intrinsic alignment contamination to the bispectrum by a factor of 3-10 or more. Source: Reproduced from \protect\cite{TroxelIshak2012a}.}\label{fig:GGIerror}
\end{figure}

The self-calibration technique for cross-correlations has recently been expanded to the bispectrum for the $GGI$ and $GII$ cross-correlations \citep{TroxelIshak2012c,TroxelIshak2012a}. In the case of the bispectrum, there are four observable (cross-)correlations between galaxy ellipticity and density for $i<j<k$
\begin{align}
B^{(1)}_{ijk}(\ell_1,\ell_2,\ell_3) =& B^{GGG}_{ijk}(\ell_1,\ell_2,\ell_3)+B^{IGG}_{ijk}(\ell_1,\ell_2,\ell_3)+B^{IIG}_{ijk}(\ell_1,\ell_2,\ell_3)+B^{III}_{ijk}(\ell_1,\ell_2,\ell_3),\nonumber\\
B^{(2)}_{iii}(\ell_1,\ell_2,\ell_3) =& B^{GGg}_{iii}(\ell_1,\ell_2,\ell_3)+B^{IGg}_{iii}(\ell_1,\ell_2,\ell_3)+B^{IIg}_{iii}(\ell_1,\ell_2,\ell_3),\nonumber\\
B^{(3)}_{iii}(\ell_1,\ell_2,\ell_3) =& B^{ggG}_{iii}(\ell_1,\ell_2,\ell_3)+B^{Igg}_{iii}(\ell_1,\ell_2,\ell_3),\nonumber\\
B^{(4)}_{iii}(\ell_1,\ell_2,\ell_3) =& B^{ggg}_{iii}(\ell_1,\ell_2,\ell_3).\label{eq:3obs}
\end{align}
$B^{III}_{ijk}$ is then negligible. The self-calibration then utilizes a scaling relationship between $B^{IGG}_{ijk}$ and $B^{Igg}_{ijk}$ in observables $B^{(1)}_{ijk}$ and $B^{(3)}_{iii}$ \citep{TroxelIshak2012a},
\begin{align}
B^{IGG}_{ijk}(\ell_1,\ell_2,\ell_3)\approx &\frac{W_{ijk}}{(b_{1}^i)^2\Pi_{iii}}B^{Igg}_{iii}(\ell_1,\ell_2,\ell_3)-\frac{b^i_2}{(b^i_1)^2}\frac{W_{ijk}}{\omega_{ii}\Pi_{ii}}\Big[C_{ii}^{Ig}(\ell_1)C_{ii}^{GG}(\ell_2)+C_{ii}^{GG}(\ell_2)C_{ii}^{Ig}(\ell_3)\label{eq:3scale}\\
&+\frac{\omega_{ii}}{b^i_1\Pi_{ii}}C^{II}_{ii}(\ell_1)C^{Ig}_{ii}(\ell_3)\Big]\nonumber
\end{align}
and between $B^{IIG}_{ijk}$ and $B^{IIg}_{ijk}$ in observables $B^{(1)}_{ijk}$ and $B^{(2)}_{iii}$ \citep{TroxelIshak2012c},
\begin{align}
B^{IIG}_{ijk}(\ell_1,\ell_2,\ell_3)\approx &\frac{W_{ijk}}{b_{1}^i\Pi_{iii}}B^{IIg}_{iii}(\ell_1,\ell_2,\ell_3)-\frac{b^i_2}{b^i_1}\frac{W_{ijk}}{\Pi^2_{ii}}\Big[C_{ii}^{II}(\ell_1)C_{ii}^{Ig}(\ell_2)+b^i_1C_{ii}^{Ig}(\ell_2)C_{ii}^{Ig}(\ell_3)\\
&+C^{II}_{ii}(\ell_1)C^{Ig}_{ii}(\ell_3)\Big].\nonumber
\end{align}
These use a similar process to the 2-point process described above for $GI$ to use the intrinsic alignment information in the galaxy shear--density--density and galaxy shear--shear--density spectra to infer the contamination to the lensing signal. These quantities depend on values for the spectrum extracted during the $GI$ self-calibration process. The inaccuracy of Eq. (\ref{eq:3scale}) is shown in Fig. \ref{fig:GGIerror}, where the self-calibration is typically inaccurate by less than 10\%, like the 2-point self-calibration, except for low redshift and spatially close tomographic bins, where the inaccuracy grows to less than about 30\%.

Similar estimators as for the $GI$ self-calibration can then be constructed for $B^{Igg}_{iii}$ and $B^{IIg}_{iii}$, respectively, as
\begin{align}
\hat{B}^{Igg}_{iii}(\ell_1,\ell_2,\ell_3)=&\frac{B^{(3)}_{ii}|_S(\ell_1,\ell_2,\ell_3)-Q^{GGI}(\ell_1,\ell_2,\ell_3)B^{(3)}_{ii}(\ell_1,\ell_2,\ell_3)}{1-Q^{GGI}(\ell_1,\ell_2,\ell_3)},\\
\hat{B}^{IIg}_{iii}(\ell_1,\ell_2,\ell_3)=&\frac{B^{(2)}_{ii}|_S(\ell_1,\ell_2,\ell_3)-Q^{GII}(\ell_1,\ell_2,\ell_3)B^{(2)}_{ii}(\ell_1,\ell_2,\ell_3)}{1-Q^{GII}(\ell_1,\ell_2,\ell_3)},
\end{align}
which again take advantage of the lensing geometry in the terms to isolate the signal of interest. The suppression quotients $Q_{GGI}$ and $Q_{GII}$ can again be approximated by some $\bar{\eta}^{GGI}_i$ and $\bar{\eta}^{GII}_i$, respectively, where
\begin{align}
\eta^{GGI}(z_L,z_g,z_{g'})=&3\frac{\int_{i}dz^p_G\int_{i}dz^p_g\int_{i}dz^p_{g'}\int_0^{\infty}dz_G W_L(z_L,z_G)S(z_G^p,z_g^p,z_{g'}^p)N^{p_{1}}_i}{\int_{\bar{z}_i-\Delta z_i/2}^{\bar{z}_i+\Delta z_i/2}dz^p_G\int_{i}dz^p_g\int_{i}dz^p_{g'}\int_0^{\infty}dz_G W_L(z_L,z_G)N^{p_{1}}_i},\\
\eta^{GII}(z_L,z_g)=&3\frac{\int_{i}dz^p_G\int_{i}dz^p_{G'}\int_{i}dz^p_{g}\int_0^{\infty}dz_G\int_0^{\infty}dz_{G'} W_L(z_L,z_G)W_L(z_L,z_{G'})S(z_G^p,z_{G'}^p,z_{g}^p)N^{p_{2}}_i}{\int_{\bar{z}_i-\Delta z_i/2}^{\bar{z}_i+\Delta z_i/2}dz^p_G\int_{i}dz^p_{G'}\int_{i}dz^p_{g}\int_0^{\infty}dz_G\int_0^{\infty}dz_{G'} W_L(z_L,z_G)W_L(z_L,z_{G'})N^{p_{2}}_i},
\end{align}
for $z_g=z_{g'}=z_L$ and
\begin{align}
N^{p_{1}}_i\equiv&p(z_G|z_G^p)p(z_g|z_g^p)p(z_{g'}|z_{g'}^p)n_i^p(z_G^p)n_i^p(z_g^p)n_i^p(z_{g'}^p)\\
N^{p_{2}}_i\equiv&p(z_G|z_G^p)p(z_{G'}|z_{G'}^p)p(z_{g}|z_{g}^p)n_i^p(z_G^p)n_i^p(z_{G'}^p)n_i^p(z_{g}^p).
\end{align}
The 3-point self-calibration techniques were shown to perform similarly to the $GI$ self-calibration, producing a reduction in $GGI$ and $GII$ by a factor of 3-10 or more, but with potentially non-negligible impact due to measurement errors in the nonlinear galaxy bias on some scales. 
 
\subsubsection{Self-calibration within a photometric redshift bin}\label{sc2}

The intrinsic alignment correlations can also be self-calibrated within each of several photometric redshift bins, where the impact of $II$ is still non-negligible \citep{Zhang2010b}. Instead of a single scaling relationship between the shear--shear correlation and the shear--density cross-correlation that is known, one instead constructs a set of scaling relationships between the corresponding cross-correlations of intrinsic alignment and shear and cross-correlations of shear and galaxy density, which share nearly identical redshift separation dependencies that are very different from that of the cosmic shear signal. This separation dependence is shown in Fig.~\ref{fig:dzp}. These scaling relationships,
\begin{align}
\frac{C^{GG}(\Delta z^p)}{C^{GG}(\Delta z^p=0)}\approx&1-f_{GG}(\Delta z^p)^2\\
C^{GI}(\Delta z^p)+C^{IG}(\Delta z^p)\approx&A_{GI}(C^{Gg}(\Delta z^p)+C^{gG}(\Delta z^p))\\
C^{gI}(\Delta z^p)\approx&A_{gI}C^{gg}(\Delta z^p)\\
C^{II}(\Delta z^p)\approx&A_{II}C^{gg}(\Delta z^p),
\end{align}
are parameterized for some $\ell$ and mean redshift $\bar{z}^p$ by the set of unknown scaling parameters $\{f_{GG},A_{GI},A_{Ig},$ $A_{II}\}$. These parameters can be marginalized over for measurements of the various shear and density (cross-) correlations between micro-bins of width $0.01$ with separation $\Delta z^p$ at redshift $\bar{z}^p$ within each larger photometric redshift bin.

This approach to self-calibrating the various intrinsic alignment signals (e.g., $II$ and $GI$) was shown to be safe against conservative estimates of photometric redshift errors and catastrophic outliers in a Stage IV weak lensing survey, and the impact due to intrinsic alignment on the redshift separation dependence of the observed shear correlation to be identifiable outside of expected survey noise \citep{Zhang2010a}. The process can be modified to measure only the total intrinsic alignment contamination, if one is primarily interested in the cosmic shear signal, or to isolate individual intrinsic alignment components, but with increased measurement errors. A comprehensive strategy for employing the process will necessarily be survey dependent.

\begin{figure}
\center
\includegraphics[width=.6\columnwidth]{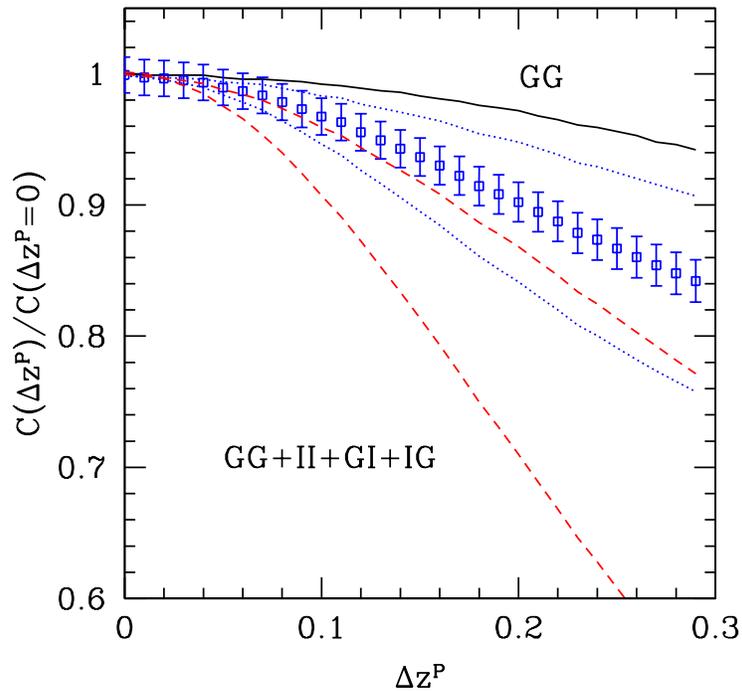}%
\caption{The total impact of the intrinsic alignment separation dependence relative to the pure $GG$ signal. The data points and error bars represent a fiducial intrinsic alignment contamination from the halo model discussed in Sec.~\ref{haloia} with error estimates for a Stage IV survey. Also shown is a $\pm50\%$ scaling of the intrinsic alignment signal for the fiducial (blue dotted lines) and a different toy intrinsic alignment model (red dashed lines), described in \protect\cite{Zhang2010b}. The deviation of the signal due to intrinsic alignment from the expected lensing result (black line) is larger than the survey error at separations greater than about $\Delta z^p=0.1$ for expected levels of intrinsic alignment contamination in the power spectrum. Source: Reproduced with permission from \protect\cite{Zhang2010b}, Oxford University Press on behalf of the Royal Astronomical Society.}\label{fig:dzpm}
\end{figure}

\begin{figure}
\center
\includegraphics[height=\textwidth,angle=270]{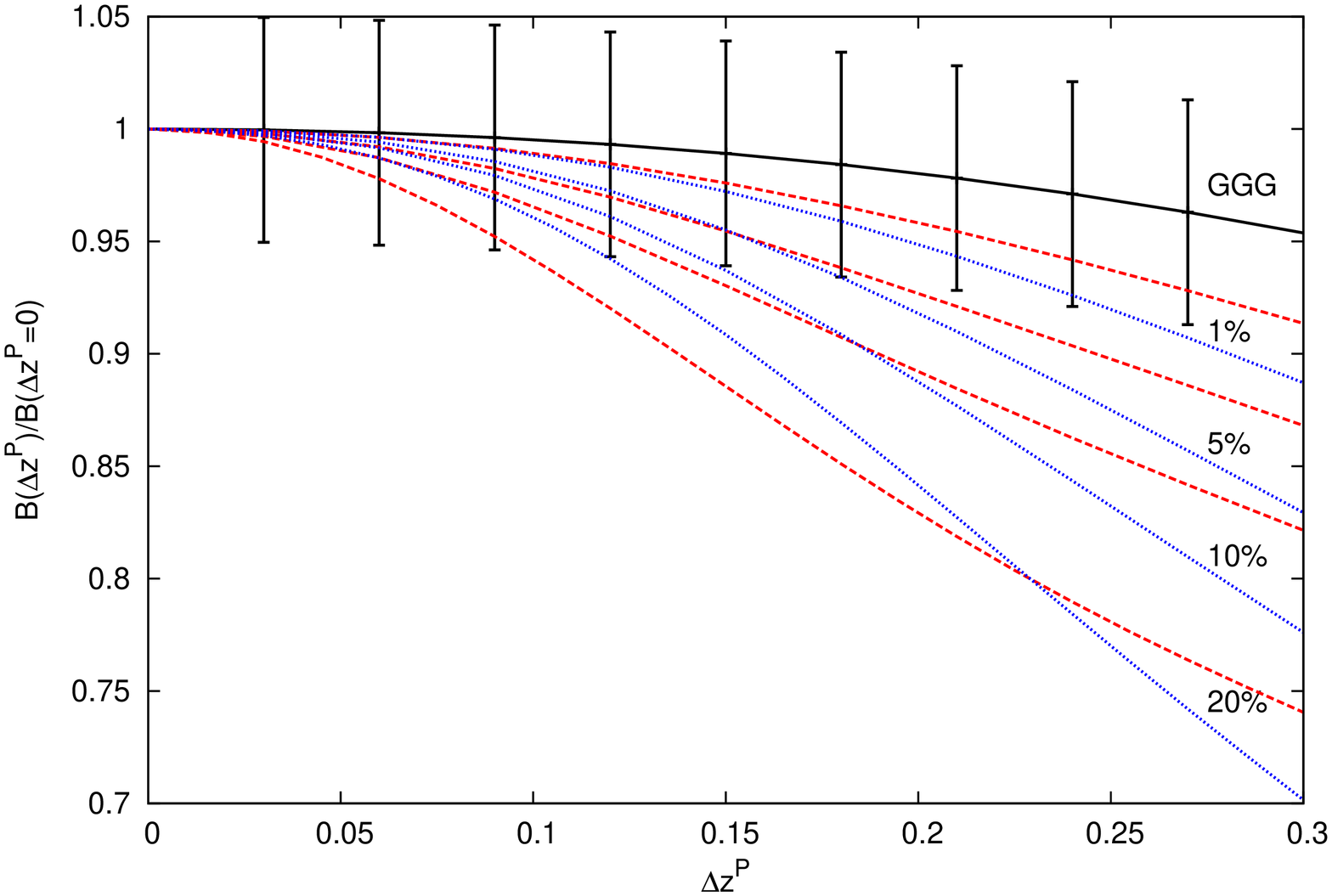}%
\caption{The total impact of the intrinsic alignment separation dependence relative to the pure $GGG$ signal. Both the halo intrinsic alignment model (dashed red lines) described in Sec.~\ref{haloia} and a toy model (dotted blue lines) described in \cite{Zhang2010b} are shown for a total contamination relative to the $GGG$ bispectrum of 1\%, 5\%, 10\%, and 20\%. The data points and error bars represent the expected value with error estimates of the $GGG$ bispectrum for a Stage IV survey. The deviation of the signal due to intrinsic alignment from the expected lensing result is larger than the survey error at separations greater than $\Delta z^p=0.15$ for expected levels of intrinsic alignment contamination in the bispectrum. Source: Reproduced from \protect\cite{TroxelIshak2012b}.}\label{fig:dzpm2}
\end{figure}

This self-calibration approach was recently expanded for the 3-point intrinsic alignment signals ($III$, $GII$, $GGI$) by \cite{TroxelIshak2012b}, but with strong constraints likely only possible on the total magnitude of the intrinsic alignment signal within a single photometric redshift bin ($GGI+GII+III$). In the 3-point case, the set of scaling relationships between the various bispectra can be expressed as
\begin{align}
\frac{B^{GGG}(\Delta z^p)}{B^{GGG}(\Delta z^p=0)}\approx&1-f_{GGG}(\Delta z^p)^2\\
B^{GGI}(\Delta z^p)+B^{GIG}(\Delta z^p)+B^{IGG}(\Delta z^p)\approx&A_{GGI}(B^{GGg}(\Delta z^p)+B^{GgG}(\Delta z^p)+B^{gGG}(\Delta z^p))\\
B^{GII}(\Delta z^p)+B^{IGI}(\Delta z^p)+B^{IIG}(\Delta z^p)\approx&A_{GII}(B^{Ggg}(\Delta z^p)+B^{gGg}(\Delta z^p)+B^{ggG}(\Delta z^p))\\
B^{Igg}(\Delta z^p)\approx&A_{Igg}C^{ggg}(\Delta z^p)\\
B^{IIg}(\Delta z^p)\approx&A_{IIg}C^{ggg}(\Delta z^p)\\
B^{III}(\Delta z^p)\approx&A_{III}C^{ggg}(\Delta z^p),
\end{align}
again parameterized for some $\ell$ and mean redshift $\bar{z}^p$ by the set of unknown scaling parameters $\{f_{GGG}, A_{GGI},$ $A_{GII}, A_{Igg}, A_{IIg}, A_{III}\}$. For the bispectrum, the choice of redshift separation is indistinct, and \cite{TroxelIshak2012b} discussed the relative performance of several choices for the meaning of $\Delta z^p$ with respect to triangle shape and scale. As with the 2-point redshift separation dependencies of $GI$ and $II$, the total impact of the intrinsic alignment redshift separation dependence on the bispectrum was shown to be detectable outside of the expected survey measurement errors in the shear bispectrum. This is also true for expected levels of intrinsic alignment contamination, but with much lower expected signal-to-noise possible than for the spectrum. It is likely that only the total sum $B^{GGI}+B^{GII}+B^{III}$ will possibly be measurable in this way for the bispectrum.

The ability to isolate the intrinsic alignment component(s) of the signal depends at the very least on being able to measure the redshift separation dependence of the signal at a sufficient number of separations or micro-bins, as shown in Figs. \ref{fig:dzpm} \& \ref{fig:dzpm2}. For the spectrum, this requires more than four data points of the expected signal lie outside the survey error on a fiducial lensing-only curve. This is clearly the case for Fig. \ref{fig:dzpm}. For the bispectrum, more than six data points are required to lie outside the fiducial lensing signal and survey noise estimates. 

A similar process that does not require direct measurement of the spectra specifically as a function of redshift separation, but employs a very similar strategy to constrain the intrinsic alignment components of the spectra, was also employed by \cite{JoachimiBridle2010,KirkBridleSchneider2010} as part of simultaneously marginalizing over the presence of both intrinsic alignment and magnification in the observed correlations to constrain cosmological parameters. \cite{JoachimiBridle2010}, for example, introduce instead a set of bias parameters to build the set of relationships with respect to the matter power spectrum:
\begin{align}
C_{ij}^{GG}(\ell)=&\int_0^{\chi}d\chi'\frac{W_i(\chi')W_j(\chi')}{\chi'^2}P_{\delta}(\frac{\ell}{\chi'},\chi')\\
C_{ij}^{IG}(\ell)=&\int_0^{\chi}d\chi'\frac{f_i(\chi')W_j(\chi')}{\chi'^2}b_I(\frac{\ell}{\chi'},\chi')r_I(\frac{\ell}{\chi'},\chi')P_{\delta}(\frac{\ell}{\chi'},\chi')\\
C_{ij}^{II}(\ell)=&\int_0^{\chi}d\chi'\frac{f_i(\chi')f_j(\chi')}{\chi'^2}b_I^2(\frac{\ell}{\chi'},\chi')P_{\delta}(\frac{\ell}{\chi'},\chi')\\
C_{ij}^{gG}(\ell)=&\int_0^{\chi}d\chi'\frac{f_i(\chi')W_j(\chi')}{\chi'^2}b_g(\frac{\ell}{\chi'},\chi')r_g(\frac{\ell}{\chi'},\chi')P_{\delta}(\frac{\ell}{\chi'},\chi')\\
C_{ij}^{gI}(\ell)=&\int_0^{\chi}d\chi'\frac{f_i(\chi')f_j(\chi')}{\chi'^2}b_g(\frac{\ell}{\chi'},\chi')r_g(\frac{\ell}{\chi'},\chi')b_I(\frac{\ell}{\chi'},\chi')r_I(\frac{\ell}{\chi'},\chi')P_{\delta}(\frac{\ell}{\chi'},\chi')\\
C_{ij}^{gg}(\ell)=&\int_0^{\chi}d\chi'\frac{f_i(\chi')f_j(\chi')}{\chi'^2}b_g^2(\frac{\ell}{\chi'},\chi')P_{\delta}(\frac{\ell}{\chi'},\chi')\\
C_{ij}^{mG}(\ell)=&2(\alpha_i-1)C_{ij}^{GG}(\ell)\\
C_{ij}^{mI}(\ell)=&2(\alpha_i-1)C_{ij}^{IG}(\ell)\\
C_{ij}^{gm}(\ell)=&2(\alpha_j-1)C_{ij}^{gG}(\ell)\\
C_{ij}^{mm}(\ell)=&4(\alpha_i-1)(\alpha_j-1)C_{ij}^{GG}(\ell),
\end{align}
where we have shown the flat ($k=0$) case for simplicity. The quantity $\alpha_i$ can be constrained separately in a survey, leaving a minimal set of bias parameters $b_x\in\{b_I,r_I,b_g,r_g\}$. Each of these terms were in turn given a general parameterization
\begin{align}
b_x=A_x Q_x(k,z(\chi))b_x^{fid}(k,\chi).
\end{align}
This kind of marginalization process was shown to be successful by \cite{JoachimiBridle2010,KirkBridleSchneider2010}, despite the necessary additional parameters, due to the inclusion of the additional information in the various correlations between galaxy ellipticity and density, which mitigates the loss of statistical information due to the extra parameters marginalized over. 

One of the unresolved challenges to employing any of these self-calibration approaches is the tendency in recent years to propose the galaxy shear--density cross-correlation as a solution for the calibration of a range of systematics, including for example intrinsic alignment and photometric redshifts, while also using it to constrain cosmology. The degree to which the simultaneous implementations of these various self-calibration techniques will challenge the total information content in the shear--density cross-correlation is thus far unexplored, but may be mediated by the inclusion of additional complementary data sets.

\subsection{Calibration of the intrinsic alignment signal with complementary data sets}\label{cmb}

\begin{figure}
\center
\includegraphics[width=.8\columnwidth]{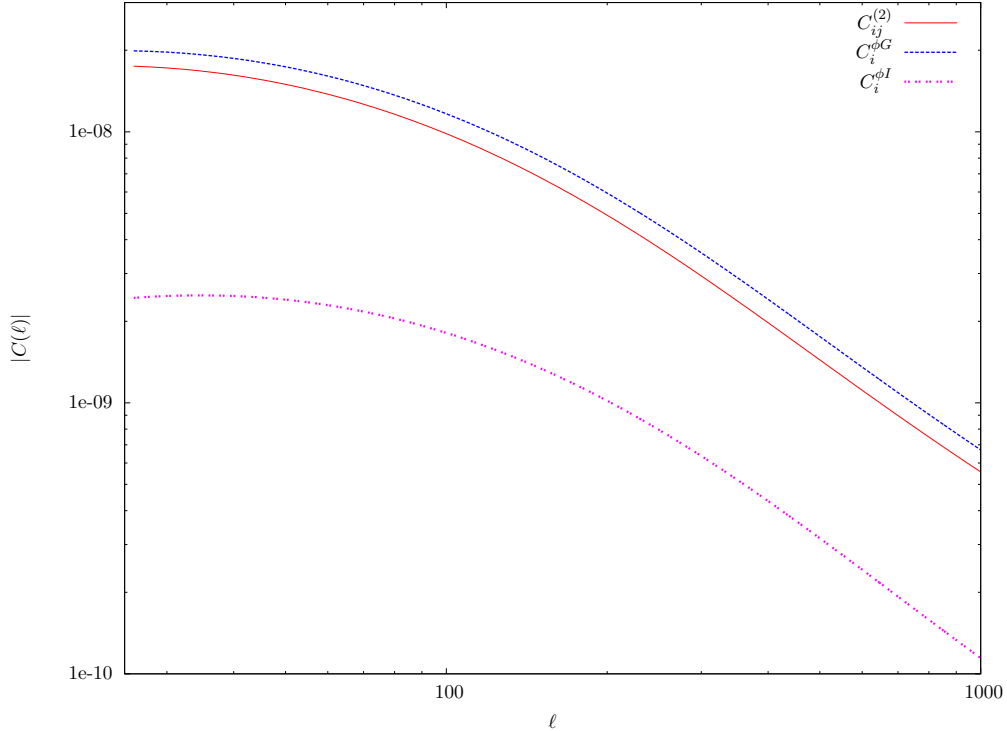}%
\caption{The $\phi I$ and $\phi G$ power spectra, compared to the total observed CMB lensing--galaxy lensing signal. The magnitude of the $\phi I$ signal is about 15\% that of the $\phi G$ signal and negative. This is compared to a contamination in galaxy lensing ($GG$) by $GI$ of about 10\%. Source: Reproduced from \protect\cite{TroxelIshak2014}.}\label{fig:cmbia}
\end{figure}

Gravitational lensing of the CMB and its cross-correlation with galaxy lensing \citep{cmb1,cmb2,cmb3,cmb4,cmb7,cmb5,cmb6,HuOkamoto2002,cmbl1,cmbl3,cmbl4,polarbear2013,cmbl2} have been suggested as a method for calibrating multiplicative biases in galaxy lensing \citep{Vallinotto2012,DasEtAl2013}, which are difficult to constrain and degenerate with the growth function (e.g., \cite{HutererEtAl2006,AmaraEtAl2008}). The CMB lensing signal is unaffected by galaxy intrinsic alignment, considered to be an additive bias to the lensing signal, and offers a separate measure of lensing by large-scale structure in the universe. The cross-correlation of CMB lensing and galaxy lensing was recently detected for the first time by \cite{hand}, and does include an intrinsic alignment correlation like $GI$, which was labeled $\phi I$ by \cite{TroxelIshak2014}. In the same way that the intrinsic ellipticity of a foreground galaxy is correlated with the lensing of a background galaxy to form the $GI$ cross-correlation, a foreground galaxy can also be correlated with the lensing deflection induced in the CMB temperature fluctuations or the polarization signal to form the $\phi I$ cross-correlation. This was first commented on by \cite{HirataSeljak2004}, and \cite{HallTaylor2014,TroxelIshak2014,KitchingEtAl2014} recently provided estimates of its impact on the lensing signal. This $\phi I$ cross-spectrum can be represented analytically as part of the observed signal, along with the CMB lensing--galaxy lensing cross-spectrum ($\phi G$)
\begin{align}
C^{(obs)}_{i}(\ell)=&C^{\phi G}_{i}(\ell)+C^{\phi I}_{i}(\ell).
\end{align}
Using variants of the linear alignment models that agree with low-z intrinsic alignment measurements (e.g., Sec \ref{nla}), \cite{HallTaylor2014,TroxelIshak2014} have separately determined that the magnitude of the $\phi I$ signal is about 15\% that of the $\phi G$ signal, shown in Fig.~\ref{fig:cmbia}, making its fractional impact about 50\% stronger than that of the $GI$ contaminant in galaxy lensing when calculated in the same way. \cite{KitchingEtAl2014} further investigated the impact on cosmological constraints using both galaxy lensing and CMB lensing in a full 3D analysis, finding also that constraints on the amplitude of intrinsic alignment models can be improved by a factor of 2 by including CMB lensing in the analysis with complementary improvements in constraints on cosmological parameters. 

It was shown by \cite{TroxelIshak2014} that this $\phi I$ cross-correlation can be connected to the $GI$ and $gI$ cross-correlations in various redshift bins from a galaxy lensing survey through scaling relationships, which are constructed similarly to those in Sec.~\ref{selfcalibration1}. This gives
\begin{align}
C^{I\phi}_i(\ell)\approx &\frac{W^{\phi}_i}{W^G_{ij}}C^{IG}_{ij}(\ell) \label{eq:scalegiphii}\\
C^{I\phi}_i(\ell)\approx &\frac{W^{\phi}_i}{b_i \Pi_{ii}}C^{Ig}_{ii}(\ell),\label{eq:scaleigphii}
\end{align}
where $b_{1}^i$, $W^G_{ij}$, and $\Pi_{ii}$ were described in Sec.~\ref{selfcalibration1}, and  
\begin{align}
W^{\phi}_i=&\int_0^{\chi}d\chi'W^{\phi}(\chi')n_{i}(\chi')\\
W^{\phi}(\chi)=&\frac{3}{2}\Omega_m(1+z)\chi(1-\frac{\chi}{\chi^{1}}).
\end{align}
Unlike the self-calibration process for $GI$, the scaling factor $W^{\phi}_i/W^G_{ij}$ does not require explicit cosmological priors to evaluate or a measurement of the galaxy bias, while the accuracy of the scaling relationship in Eq.~(\ref{eq:scaleigphii}) is more accurate than Eqs. (\ref{eq:2scale}) \& (\ref{eq:scalegiphii}), typically to within 5\% in all redshift bins for a Stage IV lensing survey. 

Using the additional information in the cross-correlation between CMB lensing and galaxy lensing, though it requires overlapping measurements of lensing in both galaxy and CMB surveys, provides an additional means to isolate information on the intrinsic alignment signal to ellipticity-ellipticity and ellipticity-density measurements, and is particularly suited for higher redshift intrinsic alignment information compared to the lensing signal, which has a peak efficiency at lower redshift. Combining these approaches to simultaneously utilize both the $gI$ and $\phi I$ cross-correlations to successfully calibrate the $GI$ cross-correlation, either as presented in Eqs. (\ref{eq:scalegiphii}) \& (\ref{eq:scaleigphii}) or as part of an approach like \cite{JoachimiBridle2010,KirkBridleSchneider2010}, will be one focus in comprehensive approaches to mitigating the effects of intrinsic alignment on weak lensing survey science. 

\subsection{Other paths toward mitigating intrinsic alignment}\label{otherm}

\subsubsection{E- and B-mode decomposition}\label{ebmode}

A basic test for the presence of intrinsic alignment in the observed shear signal (or other systematic effects in the data) is a non-zero B-mode component of the ellipticity correlation (e.g., \cite{PenLeeSeljak2000,CrittendenNatarajanPenTheuns2001,CrittendenNatarajanPenTheuns2002}). The observed correlation can be decomposed into E- and B-mode components, where the E-mode component of the signal is curl-free. The true lensing signal is necessarily curl-free, since the shear is related to the gradient of the potential 
\begin{align}
\gamma_{ij}(x)=(\partial_i\partial_j-\frac{1}{2}\delta_{ij}\nabla^2)\psi(x),
\end{align}
except on small scales where source redshift clustering has an impact \citep{SchneiderEtAl2002}. The intrinsic alignment signal, by contrast, possesses both E- and B-mode components, and thus the non-zero detection of a B-mode signal could be used to place constraints on the contamination of the observed lensing signal by intrinsic alignment (e.g., \cite{HeymansHeavens2003}), or as a diagnostic to test the removal of intrinsic alignment contributions to the observed ellipticity correlation through some of the methods discussed above. This is limited in usefulness to considerations of the $II$ correlation in most cases, however, since the B-mode component of the intrinsic alignment signal should not correlate with the E-mode lensing signal in the $GI$ correlation. Leading order terms from linear alignment models also predict a zero B-mode contribution, however, and so the E/B-mode decomposition may have limited use as a diagnostic for the presence of intrinsic alignment. Other works that discuss the decomposition of the lensing signal into E- and B-mode components, and its impact on constraints of systematics like intrinsic alignment include, for example, \cite{HoekstraEtAl2002,SchneiderKilbinger2007,SchneiderEiflerKrause2010,Becker2013}.

\subsubsection{Measurements of intrinsic alignment that are insensitive to gravitational lensing}

There have also been recent suggestions for utilizing novel approaches for distinguishing the intrinsic shape of a galaxy, which would in principle allow for a weak lensing detection without either intrinsic alignment bias or significant shape noise. One of these approaches by \cite{BrownBattye2011} seeks to utilize radio weak lensing measurements to identify the orientation of a galaxy's polarized emission, which is both related to the intrinsic alignment of the galaxy and unaffected by gravitational lensing. Combining polarization measurements with shape determinations then allows one to disentangle intrinsic alignment from cosmic shear. This can also be extended to optical polarization \citep{AuditSimmons1999}. Both, however, would require better knowledge of the polarization emission properties of source galaxies, and the effects, for example, of intervening magnetic fields in rotating the polarization plane \citep{BrownBattye2011} in a process which would inject an additional position-dependent systematic similar to intrinsic alignment.

A similar approach by \cite{HuffEtAl2013} promotes a smaller, purely spectroscopic survey with overlap of larger photometric surveys that will enable the measurement of the resolved rotation velocity of disk galaxies. Combined with a Tully--Fisher relationship that connects the circular rotation velocity with luminosity \citep{TullyFisher1977}, the inclination of the disks could be identified, and thus the intrinsic shape of the galaxy prior to gravitational lensing. Like using polarization orientation, this could significantly increase statistical power in surveys by limiting random shape noise. It would also minimize potential bias from intrinsic alignment on weak lensing measurements. \cite{HuffEtAl2013} claimed that such a `Stage III' spectroscopic survey can be competitive with best-case estimates of planned Stage IV photometric surveys.

\section{Summary and future outlook}\label{summary}

The potential and importance of the weak gravitational lensing of galaxies, in particular by large-scale structure (cosmic shear), and its cross-correlations with galaxy positions and other probes of structure to constrain parameters and models of cosmology has become clear in the past decades. This includes the mapping of the distribution and evolution of structure (composed both of baryonic and dark matter), the study of the nature and evolution of cosmic acceleration (`dark energy'), and testing theories of gravity at cosmological scales. We have reviewed the theory and formalisms related to applying measurements of weak lensing shear or convergence to the study of cosmology and discussed some of the challenges involved in realizing the potential of this powerful cosmic probe in Secs. \ref{intro}-\ref{formalisms}. One of the most serious physical systematics of the weak lensing of galaxies, which can introduce large biases in cosmological information and is in some ways considered a barrier to the full use of weak lensing, is the intrinsic alignment of the galaxies (i.e., their shapes and orientations before being lensed).

We have explored the intrinsic alignment of galaxies in Secs.~\ref{backIA}-\ref{mitigation} and reviewed in Sec.~\ref{backIA} a great deal of work that has gone into characterizing and modeling how they can be correlated with each other ($II$ or $III$ correlations) and with the large-scale tidal field due to linear density perturbations ($GI$, $GGI$, or $GII$ correlations). These correlations of intrinsic alignment can heavily contaminate the cosmic shear signal, even across the large volumes of planned Stage IV surveys, due to the (anti-)correlation of the intrinsic alignment with the lensing of background galaxies. It is vital for the success of large weak lensing survey science that we successfully isolate the effects of intrinsic alignment from weak gravitational lensing.

We provided a summary of strategies to characterize and model the intrinsic alignment signal on both large and small scales in Sec.~\ref{models}. Recent attempts to incorporate intrinsic alignment into a halo model formalism (Sec.~\ref{haloia}) or to compile a semi-analytical description of the intrinsic alignment signals from measurements in large N-body dark matter simulations and smaller hydrodynamical simulations (Sec.~\ref{sam}) have met with some success in describing the properties of the intrinsic alignment signal on smaller scales to match with simulation and survey detections. The resulting impact of intrinsic alignment on cosmological information due to various models of the intrinsic alignment correlations were briefly discussed in Sec.~\ref{impacts}.

In the past decade, as the size and capability of our surveys have improved, the correlated intrinsic alignment of galaxies has been directly measured in a variety of data sets. Strategies for making these measurements and results were reviewed in Sec.~\ref{detections}. These measurements have provided needed empirical constraints on the properties of the intrinsic alignment signal, including its magnitude, dependency on galaxy types and luminosities, and evolution in time. These constraints in turn inform better models of the intrinsic alignment signal (e.g., Sec.~\ref{haloia} \& \ref{sam}), and suggest methods to mitigate the impact of the intrinsic alignment signal on weak lensing science (Sec.~\ref{mitigation}). 

A complementary approach to directly or indirectly measuring the intrinsic alignment of galaxies in surveys is to characterize the intrinsic alignment through studies of cosmological simulations. The use of simulations to study intrinsic alignment was discussed in Sec.~\ref{sims}, where we present a variety of results that help explain how galaxies are aligned on small and large scales. The use of simulations to study intrinsic alignment, however, is still limited by our computational ability to produce hydrodynamical simulations of large enough volume from which the true correlations of the shapes of galaxies can be directly measured. We have thus far been limited primarily to the use of dark matter halos as tracers of the galaxy shape in cosmological scale dark matter only N-body simulations (Sec.~\ref{dmonly}) for measuring intrinsic alignment correlations. These approaches are limited, however, by how well dark matter halos can be used as tracers of the baryonic galaxy shape (Sec.~\ref{misalignment}). Work is now being done to use smaller hydrodynamical simulations (Sec.~\ref{hydro}) to inform how this (mis-)alignment with dark matter halos can be characterized and incorporated into semi-analytical models (e.g. Sec.~\ref{sam}), which use simulations to inform model building. 

While understanding the intrinsic alignment of galaxies through model fitting is ideal, in order to use the intrinsic alignment signal as a direct probe for structure formation in the universe, this is currently limited by our understanding of how galaxies become aligned or misaligned over time in large-scale structure and halos. We also yet lack the ability to produce large-scale hydrodynamical simulations of sufficient size to study intrinsic alignment correlations directly in our cosmological models. Another approach to the large impact of intrinsic alignment in weak lensing surveys is then to design techniques to simply mitigate the impact of intrinsic alignment on the weak lensing signal, without needing exact knowledge of intrinsic alignment modeling.

We reviewed the development of a series of mitigation strategies for weak lensing surveys in Sec.~\ref{mitigation}. These range from methods that simultaneously fit the intrinsic alignment signal to a parameterized model or template (Sec.~\ref{margin}), those that remove the intrinsic alignment-contaminated information in a survey (Secs. \ref{tomography}-\ref{nulling}), and others that indirectly isolate and preserve the intrinsic alignment signal based on complementary information in other cross-correlations with the cosmic shear signal (Secs. \ref{sc}-\ref{cmb}). Some new proposals even outline methods to measure the intrinsic alignment of galaxies independently of the lensing signal (Sec.~\ref{otherm}).

In sum, the intrinsic alignment of galaxies remains a difficult, and as yet unresolved challenge for the use of weak lensing in large surveys as a truly precise (and more importantly, accurate) cosmological probe. However, a great deal of progress has been made toward understanding and mitigating its influence on cosmological constraints. With the analysis of ongoing Stage III surveys and the development of pipelines for planned Stage IV surveys, it is a challenge that our field has met with enthusiasm. A clever combination of the methods and measurements presented thus far will provide a solid base upon which future models and analysis methods can successfully resolve the challenge that the intrinsic alignment of galaxies poses to weak gravitational lensing science.

\section*{Acknowledgments}
We thank Matthias Bartelmann, Jonathan Blazek, Sarah Bridle, Michael Brown, Rupert Croft, Catherine Heymans, Alina Kiessling, Martin Kilbinger, Lindsay King, Donnacha Kirk, Rachel Mandelbaum, Austin Peel, Michael Schneider, Tim Schrabback, Masahiro Takada, Ludovic Van Waerbeke, and Joseph Zuntz for reading or providing useful comments on early versions of the manuscript. We are also very grateful for the thorough and helpful comments of the referee, and also wish to thank Jonathan Blazek, Benjamin Joachimi, Donnacha Kirk, Rachel Mandelbaum, Michael Schneider, Xun Shi, and Pengjie Zhang for figures reproduced from previous works. MI acknowledges that this work was supported in part by the National Science Foundation under grant AST-1109667. MT acknowledges support by the NASA/TSGC graduate fellowship program, and is very grateful for the hospitality of the Center for Particle Cosmology at the University of Pennsylvania, where part of this work was written.

\newpage
\section*{References}

\bibliographystyle{model2-names} 
\bibliography{PhysReptResub4weps}

\end{document}